\documentclass[12pt,preprintnumbers,superscriptaddress,amsmath,amssymb,nofootinbib]{revtex4}

\usepackage{graphicx}
\usepackage{dcolumn}
\usepackage{bm}
\usepackage{bbm}
\usepackage{amssymb}
\usepackage{amsmath}
\usepackage{epsfig}    
\usepackage{color}
\usepackage{slashed}
\usepackage{cancel}

\usepackage[colorlinks,
            linkcolor=black,
            anchorcolor=black,
            citecolor=black
            ]{hyperref}

\def\cb{c_\beta}

\def\l{\lambda}

\def\sb{s_\beta}

\begin{document} 

\title{One-loop renormalized Higgs vertices in Georgi-Machacek model}

\author{Cheng-Wei Chiang}
\email[e-mail: ]{chengwei@phys.ntu.edu.tw}
\affiliation{Department of Physics and Center for Theoretical Physics, 
National Taiwan University, Taipei, Taiwan 10617, R.O.C.}
\affiliation{Institute of Physics, Academia Sinica, Taipei, Taiwan 11529, R.O.C.}
\affiliation{Department of Physics and Center of High Energy and High Field Physics, National Central University, Chungli, Taiwan 32001, R.O.C.}

\author{An-Li Kuo}
\email[e-mail: ]{101222028@cc.ncu.edu.tw}
\affiliation{Department of Physics and Center of High Energy and High Field Physics, National Central University, Chungli, Taiwan 32001, R.O.C.}
             
\author{Kei Yagyu}
\email[e-mail: ]{yagyu@st.seikei.ac.jp}
\affiliation{Seikei University, Musashino, Tokyo 180-8633, Japan}

\begin{abstract}

We compute renormalized vertices of the 125 GeV Higgs boson $h$ with the weak gauge bosons ($hVV$), fermions ($hf\bar{f}$) and itself ($hhh$) in the Georgi-Machacek model at one-loop level. 
The renormalization is performed based on the on-shell scheme with the use of the minimal subtraction scheme only for the $hhh$ vertex. 
We explicitly show the gauge dependence in the counterterms of the scalar mixing parameters in the general $R_\xi$ gauge, 
and that the dependence can be removed by using the pinch technique in physical scattering processes. 
We then discuss the possible allowed deviations in these one-loop corrected Higgs couplings from the standard model predictions by scanning model parameters under the constraints of perturbative unitarity and vacuum stability as well as those from experimental data. 

\end{abstract}
\maketitle


\newpage
\section{Introduction}

Discovery of the 125 GeV Higgs boson at the CERN Large Hadron Collider (LHC) had completed the particle spectrum of the Standard Model (SM).  
This, however, does not necessarily mean that the SM is the ultimate theory describing elementary particle physics, because of theoretically unsatisfactory issues, such as the gauge hierarchy problem and unexplained phenomena of neutrino mass, dark matter and baryon asymmetry of the Universe. 
These problems are expected to be solved in new physics (NP) beyond the SM at or above the TeV scale. 
In NP models, the Higgs sector is often extended from the minimal form assumed in the SM, and its properties strongly depend on the NP scenario. 
Therefore, a determination of the structure of the Higgs sector using experimental data is important to narrow down possible NP models.

There are basically two ways to identify an extended Higgs sector: the direct search and the indirect search.  The former approach is to discover additional Higgs bosons, while the latter is to find deviations in various observables related to the discovered Higgs boson $(h)$ from the SM predictions.
So far, no additional Higgs boson has been discovered at the LHC, and this situation makes the indirect search attractive. 
Currently the Higgs boson couplings are measured with insufficient accuracies at the LHC, {\it e.g.}, a 10\% level uncertainty in the $hVV$ ($V=W^\pm,Z$) couplings~\cite{LHC1}.
They are expected to be measured with much better accuracies in future collider experiments, such as the high-luminosity LHC and $e^+e^-$ colliders, 
where they can be determined to the percent or sub-percent level~\cite{Dawson:2013bba,Fujii:2015jha}.

In order to make a sensible comparison with such precision measurements, one needs to reduce theoretical uncertainties in models with an extended Higgs sector.  In particular, radiative corrections to the Higgs boson couplings should be taken into account. 
One-loop corrections to various Higgs boson couplings have been studied in several models with a simple Higgs extension, {\it e.g.}, 
models with an additional isospin singlet scalar field, the Higgs singlet model (HSM)~\cite{HSM1,HSM2,HSM3,HSM4}; 
a doublet scalar field, two-Higgs doublet models (THDMs)~\cite{THDM1,THDM2,THDM3,THDM4}, and a complex triplet field, the Higgs triplet model (HTM)~\cite{HTM1,HTM2}. 
Recently, a numerical tool {\tt H-COUP}~\cite{HCOUP} has been constructed to compute various $h$ couplings at one-loop level in the HSM and THDMs without any gauge dependence~\cite{Yagyu-Gauge}.

In this paper, we investigate one-loop corrections to the Higgs boson couplings in the Georgi-Machacek (GM) model~\cite{Georgi:1985nv,Chanowitz:1985ug}
which has the capacity of providing Majorana mass to left-handed neutrinos through the type-II seesaw mechanism as in the HTM~\cite{Konetschny:1977bn,Schechter:1980gr,Cheng:1980qt}. 
This model realizes the minimal Higgs sector containing isospin triplet scalar fields while having 
an approximate custodial symmetry
\footnote{The custodial symmetry is actually broken explicitly by the $U(1)_Y$ gauge coupling as 
it also happens in the SM. } in the scalar sector. 
Thanks to this custodial symmetry, the vacuum expectation value (VEV) of the Higgs triplet fields is allowed to be sizeable while keeping the electroweak rho parameter $\rho = 1$ at tree level, 
a desirable property that is consistent with experimental observations~\cite{pdg}.
This property can provide phenomenologically interesting predictions.  For example, 
the $hVV$ couplings can be larger than their SM values, leading to enhanced weak gauge boson scattering processes via the SM-like and exotic Higgs bosons~\cite{Chiang:2013rua,Chiang:2015kka}.
It has also been shown that with an ${\cal O}(10)$-GeV triplet VEV, the model allows a sufficiently strong first-order phase transition to facilitate successful electroweak baryogenesis~\cite{Chiang:2014hia}.

In our earlier work~\cite{GM_lett}, the one-loop corrected $hVV$ couplings had been calculated in the GM model without presenting the details. 
Besides, the Yukawa couplings ($hf\bar{f}$) and the Higgs self-coupling ($hhh$) were not computed in that work. 
In this paper, we present in detail our computations of all these $h$ couplings ({\it i.e.}, $hVV$, $hf\bar{f}$ and $hhh$) at one-loop level. 
We apply the on-shell renormalization scheme to our calculation, where counterterms for scalar mixing parameters remain gauge-dependent as been generally shown for models with mixing among scalar fields~\cite{Yamada,Espinosa}. 
We discuss how such a gauge dependence can be removed by using the pinch technique for physical processes.

The structure of this paper is as follows.  
In Section~\ref{sec:model}, we briefly review the GM model, separately discussing the scalar potential, the scalar kinetic energies, and the Yukawa interactions.  
We also discuss the decoupling limit of the model.  
Section~\ref{sec:reno} is devoted to the discussions of renormalization in each of the gauge, fermion, and scalar sectors.  
We introduce the necessary counterterms and renormalization conditions to determine these counterterms.  
Section~\ref{sec:gauge} takes special care of the gauge dependence issue in some scalar 2-point functions.  
We will adopt the general $R_\xi$ gauge in the computations, and employ the pinch technique to remove the gauge dependence in physical scattering processes.  
In Section~\ref{sec:reno-vertex}, we derive the renormalized $hVV$, $hf\bar{f}$ and $hhh$ vertices. 
For the $hVV$ and $hf\bar{f}$ vertices, we further define the form factors of these vertices.
We then show the simple plots for the renormalized scale factors for $hVV$, $hf\bar{f}$ and $hhh$ couplings normalized to their SM predictions and discuss their momentum dependence.
Section~\ref{sec:constraints} discusses and lists theoretical and experimental constraints to be imposed in the parameter scan of the model.  
Section~\ref{sec:numana} presents the numerical result for the renormalized scale factors by scanning model parameters under the both theoretical and experimental constraints.   
Section~\ref{sec:summary} summarizes our findings in this work.  
Appendices~\ref{sec:mass} and \ref{sec:hcoup} give explicit formulas of the masses and interactions of the Higgs bosons in the model, respectively. 
In Appendix~\ref{sec:1pi}, the loop functions are defined, and explicit formulae for contributions from 1PI diagrams that appear in our calculations are presented.

\section{The Model \label{sec:model}}

The scalar sector of the GM model is composed of a weak isospin doublet field $\phi$ with hypercharge $Y=1/2$ and weak isospin triplet fields $\chi$ and $\xi$ with $Y=1$ and $Y=0$, respectively.
These scalar fields can be expressed in the $SU(2)_L\times SU(2)_R$ bi-doublet $(\Phi)$ and bi-triplet $(\Delta)$ forms as:
\begin{align}
\Phi \equiv (\phi^c,\phi) = \left(
\begin{array}{cc}
\phi^{0*} & \phi^+ \\
-\phi^- & \phi^0
\end{array}\right),\quad 
\Delta\equiv(\chi^c,\xi,\chi)= \left(
\begin{array}{ccc}
\chi^{0*} & \xi^+ & \chi^{++} \\
-\chi^- & \xi^0 & \chi^{+} \\
\chi^{--} & -\xi^- & \chi^{0} 
\end{array}\right), \label{eq:Higgs_matrices}
\end{align}
where $\phi^c = i\tau^2 \phi^*$ and $\chi^c = C_3\chi^*$ are the charge-conjugated $\phi$ and $\chi$ fields, respectively. 
The matrix $C_3$ is given by 
\begin{align}
C_3 = \begin{pmatrix}
0 &0&1\\
0&-1&0\\
1&0&0
\end{pmatrix}. 
\end{align}
The neutral components are parameterized by 
\begin{align}
\phi^0&=\frac{1}{\sqrt{2}}(\phi_r+v_\phi+i\phi_i), \quad 
\chi^0=\frac{1}{\sqrt{2}}(\chi_r+i\chi_i)+v_\chi,\quad \xi^0=\xi_r+v_\xi, \label{eq:neutral}
\end{align}
where $v_\phi$, $v_\chi$ and $v_\xi$ are the VEVs of $\phi^0$, $\chi^0$ and $\xi^0$, respectively. 
For later convenience, we re-express the two triplet VEVs by 
\begin{align}
v_\chi = v_\Delta,\quad v_\xi = v_\Delta + \nu.
\end{align}
The $\nu$ parameter describes the deviation from alignment in the triplet VEVs, {\it i.e.}, $\langle\Delta\rangle = v_\Delta {\mathbbm 1}_{3\times 3}$.

In the following subsections, we first discuss the scalar potential and explain the necessity of introducing $SU(2)_L\times SU(2)_R$ breaking terms in order to make the model consistent at loop levels. 
We then give the Lagrangian of the scalar kinetic terms and the Yukawa interactions. 
Finally, we discuss the decoupling property of the GM model.

\subsection{Scalar potential}

The $SU(2)_L\times U(1)_Y$ gauge-invariant scalar potential can be expressed as follows:
\begin{align}
V = V_{\text{cust}}(\Phi,\Delta) + V_{\cancel{\text{cust}}}(\phi,\chi,\xi), 
\end{align}
where $V_{\text{cust}}$ and $V_{\cancel{\text{cust}}}$ are respectively given as a function of $\{ \Phi,\Delta \}$ and $\{ \phi,\chi,\xi \}$
\footnote{Given the scalar fields in the model, the most general $SU(2)_L\times U(1)_Y$ gauge-invariant scalar potential 
has 14 real and 2 complex parameters.  Imposing the global $SU(2)_L\times SU(2)_R$ symmetry renders relations among the parameters and results in the custodial symmetric potential given in Eq.~(\ref{eq:pot}) described by 9 real parameters, as shown in Ref.~\cite{Blasi}. }. 
$V_{\cancel{\text{cust}}}$ is defined such that when it is vanishing, the potential has the most general global $SU(2)_L\times SU(2)_R$ symmetry which is spontaneously broken down to the diagonal part $SU(2)_V$, the so-called custodial symmetry, 
under the assumption of vacuum alignment: $v_\chi = v_\xi$ or, equivalently, $\nu =0$. 
In this configuration, the electroweak rho parameter $\rho$ is predicted to be $1$ at tree level as we will see in the next subsection.

Nonetheless, even if we take $V_{\cancel{\text{cust}}} = 0$ at tree level, $V_{\cancel{\text{cust}}}$ generally re-appears at loop levels due to, {\it e.g.}, hypercharge gauge boson loops as a consequence of $SU(2)_L\times SU(2)_R$ breaking effects in the kinetic term.   
In addition, such loop contributions contain ultra-violet (UV) divergences which cannot be cancelled by counterterms associated with the $V_{\text{cust}}$ part alone. 
Therefore, in order to make the model consistent at loop levels, we need to introduce custodial symmetry breaking terms from the beginning. 
The simplest choice to make our calculations of renormalized  vertices for the discovered Higgs boson consistent is to introduce
\begin{align}
V_{\cancel{\text{cust}}} = \frac{m_\xi^2}{2} \xi^\dagger\, \xi, \label{custb}
\end{align}
where $\xi = (\xi^+,\xi^0,-\xi^-)^T$. 
The other possible terms for $V_{\cancel{\text{cust}}}$ can be important for the computation of one-loop corrections to physical quantities related to the extra Higgs bosons, but are not our concerns here.

Explicitly, the most general custodial symmetric potential is given by 
\begin{align}
V_{\text{cust}}
=&
m_\Phi^2\text{tr}(\Phi^\dagger\Phi)+m_\Delta^2\text{tr}(\Delta^\dagger\Delta)
+\lambda_1[\text{tr}(\Phi^\dagger\Phi)]^2+\lambda_2[\text{tr}(\Delta^\dagger\Delta)]^2
+\lambda_3\text{tr}[(\Delta^\dagger\Delta)^2]\notag\\
&+\lambda_4\text{tr}(\Phi^\dagger\Phi)\text{tr}(\Delta^\dagger\Delta)
+\lambda_5\text{tr}\left(\Phi^\dagger\frac{\tau^a}{2}\Phi\frac{\tau^b}{2}\right)
\text{tr}(\Delta^\dagger t^a\Delta t^b)\notag\\
&+\mu_1\text{tr}\left(\Phi^\dagger \frac{\tau^a}{2}\Phi\frac{\tau^b}{2}\right)(P^\dagger \Delta P)^{ab}
+\mu_2\text{tr}\left(\Delta^\dagger t^a\Delta t^b\right)(P^\dagger \Delta P)^{ab}, \label{eq:pot}
\end{align}
where $\tau^a/2$ and $t^a$ ($a=1,2,3$) are the $2\times 2$ and $3\times 3$ representations of the $SU(2)$ generators, respectively. 
The matrix $P$ gives the similarity transformation $P (-i\epsilon^a)P^\dagger = t^a$ with $\epsilon^a$ being 
the adjoint representation of the $SU(2)$ generators, and is given as
\begin{align}
P=\left(
\begin{array}{ccc}
-1/\sqrt{2} & i/\sqrt{2} & 0 \\
0 & 0 & 1 \\
1/\sqrt{2} & i/\sqrt{2} & 0
\end{array}\right). 
\end{align}

To obtain the mass eigenvalues for the physical Higgs bosons, one imposes the tadpole conditions at tree level: 
\begin{align}
\left.\frac{\partial V}{\partial \phi_r}\right|_0 =
\left.\frac{\partial V}{\partial \chi_r}\right|_0 = 
\left.\frac{\partial V}{\partial \xi_r}\right|_0 = 0. 
\end{align}
Using the above three equations, one can re-write the three mass parameters $m_\Phi^2$, $m_\Delta^2$ and $m_\xi^2$ in terms of the other parameters in the scalar potential. 
We note that in the limit of $\nu = 0$, $m_\xi^2$ also vanishes and the tadpole conditions for $\chi_r$ and $\xi_r$ become identical as a consequence of restoring the custodial symmetry at tree level.
Detailed analytic expressions for the physical Higgs bosons and their squared masses are presented in Appendix~\ref{sec:mass} for the general $\nu \neq 0$ case.

We here highlight some important properties of the mass spectrum in the $\nu = 0$ limit.  
The mass eigenstates of Higgs bosons can be classified under the custodial
$SU(2)_V$ symmetry into one 5-plet $(H_5^{\pm\pm},H_5^\pm,H_5^0)$, one 3-plet $(H_3^\pm,H_3^0)$ 
and two singlets $H_1$ and $h$ with $h$ being identified with the observed 125 GeV Higgs boson in our work. 
The Higgs bosons belonging to the same $SU(2)_V$ multiplet are degenerate in mass, as seen in Eq.~(\ref{massH}). 
Taking $\nu \to 0$, various mixing angles defined in Eq.~(\ref{eigen2}) among the scalar bosons reduce to
\begin{align}
\begin{split}
&\tan\beta \equiv \tan\beta_{\text{odd}} = \tan\beta_1^{\pm}  = \frac{v_\phi}{2\sqrt{2}v_\Delta},  \\
&\tan \beta_2^\pm = \tan\gamma = \tan\alpha_{1,2} = 0, \\
& \tan\alpha \equiv \tan\alpha_3. 
\end{split} \label{cust2}
\end{align}
Therefore, the rotation matrices to separate the Nambu-Goldstone (NG) bosons from the physical CP-odd and singly-charged Higgs bosons become the same. 
This also shows the recovery of the custodial symmetry. 
Consequently, all the potential parameters can be expressed in terms of the following 9 parameters:
\begin{align}
m_{H_5}^2,~m_{H_3}^2,~m_{H_1}^2,~m_h^2,~\mu_1,~\mu_2,~v,~\beta,~\alpha.  
\label{para}
\end{align}

\subsection{Kinetic terms}

The kinetic terms of the scalar fields are given by
\begin{align}
\mathcal{L}_{\text{kin}}&=\frac{1}{2}\text{tr}(D_\mu \Phi)^\dagger (D^\mu \Phi)
+\frac{1}{2}\text{tr}(D_\mu \Delta)^\dagger (D^\mu \Delta), \label{lkin}
\end{align}
where the covariant derivatives
\begin{align}
\begin{split}
D_\mu \Phi   &=\partial_\mu\Phi -ig\frac{\tau^a}{2}W_\mu^a\Phi + ig'B_\mu \Phi\frac{\tau^3}{2},\\
D_\mu \Delta &=\partial_\mu\Delta -igt^aW_\mu^a\Delta + ig'B_\mu \Delta t^3. \label{cov}
\end{split}
\end{align}
 
The weak gauge boson masses are calculated to be 
\begin{align}
m_W^2 = \frac{g^2}{4}(v_\phi^2+8v_\Delta^2 + \bar{\nu}^2),\quad 
m_Z^2 = \frac{g^2_Z}{4}(v_\phi^2+8v_\Delta^2), \label{mvsq}
\end{align}
where $\bar{\nu} \equiv 2\sqrt{\nu(2v_\Delta+\nu)}$ and $g_Z \equiv \sqrt{g^2+g^{\prime 2}}$. 
As in the SM, the electroweak symmetry breaking $SU(2)_L\times U(1)_Y \to U(1)_{\text{EM}}$ 
forces the following relation among the gauge couplings:
\begin{align}
e (=\sqrt{4\pi \alpha_{\text{EM}}}) = gs_W^{} = g^\prime c_W, 
\end{align}
where 
$\alpha_{\text{EM}}$ is the fine structure constant and $c_W$ $(s_W)$ is the cosine (sine) of the weak mixing angle $\theta_W$. 
Using these relations, we can also write $g_Z^{} = g/c_W^{}$. 
From Eq.~(\ref{mvsq}), we can identify the VEV $v$, which is related to the Fermi's decay constant $G_F$ by $v = (\sqrt{2}G_F)^{-1/2}$, as  
\begin{align}
v^2 = v_\phi^2+8v_\Delta^2 + \bar{\nu}^2. \label{vevs}
\end{align}
The tree-level rho parameter is then given by
\begin{align}
\rho_\text{tree} \equiv \frac{m_W^2}{m_Z^2c_W^2} = \frac{v^2}{v^2-\bar{\nu}^2}. 
\end{align}
Therefore, a non-zero $\nu$ would make the rho parameter deviate from unity at tree level. 
This implies that unlike in the SM, the electroweak sector is now empirically fixed by four independent parameters, {\it e.g.}, the set of $\{ m_W^{},m_Z^{},\alpha_{\text{EM}}, \nu \}$.  
In fact, the necessity of four input parameters in the electroweak sector generally appears in models with $\rho_{\text{tree}}\neq 1$~\cite{Blank,triplet-ew}. 
In terms of these four parameters, $s_W^2$ and $v^2$ are given by:  
\begin{align}
s^2_W = 1 -\frac{m_W^2}{m_Z^2}\left(1 - \frac{\bar{\nu}^2}{v^2} \right), \quad
v^2   = \frac{m_W^2 s_W^2}{\pi\alpha_{\text{EM}}}. \label{vsq}
\end{align}

The 3- and 4-point interaction terms of Higgs bosons to gauge bosons are also obtained from Eq.~(\ref{cov}), with their expressions in the $\nu \to 0$ limit given in Appendix~\ref{sec:hcoup}. 
We here list several remarkable features regarding the gauge interactions of the Higgs bosons in the model:
\begin{enumerate}
\item 
The SM-like Higgs boson couplings $hWW$ and $hZZ$ can be larger than the SM predictions at tree level. This does not happen in models constructed with only singlet and/or doublet scalars. 
\item 
The 5-plet Higgs bosons have the scalar-gauge-gauge type interactions, while the 3-plet Higgs bosons do not, as seen in Eq.~(\ref{sgg}).  The 3-plet Higgs bosons are thus said to be gauge-phobic.
\item 
The $H_5^0WW$ coupling normalized by the SM $hWW$ coupling defined as $c_{H_5WW}$ is different from that associated with the $Z$ boson ({\it i.e.}, $c_{H_5ZZ}^{}$).  
In particular, $c_{H_5ZZ} / c_{H_5WW} = -2$, as seen in Eq.~(\ref{c5_1}).  This property is not seen in the corresponding couplings of $h$ and $H_1$. 
\end{enumerate}

\subsection{Yukawa interactions}

The Yukawa Lagrangian for the third-generation fermions is given by 
\begin{align}
{\cal L}_Y = -y_t\bar{Q}_L^3 \, \phi^c\, t_R^{} -y_b\bar{Q}_L^3 \,\phi\, b_R^{} -y_\tau\bar{L}_L^3 \,\phi\, \tau_R^{} +\text{H.c.}, \label{yuk}
\end{align}
where $Q_L^3 = (t,b)_L^T$ and $L_L^3 = (\nu_\tau,\tau)_L^T$.  
The Yukawa interactions for the other SM fermions can be simply obtained by generalizing the above Yukawa couplings to $3\times 3$ Yukawa matrices. 
In the $\nu \to 0$ limit, fermion masses are obtained as $m_f = y_f v s_\beta/\sqrt{2}$ for $f \in \{ t,b,\tau \}$. 
We note that there is another type of Yukawa interaction terms for the $\chi$ field, which is expressed as
\begin{align}
\overline{L_L^c}i\tau^2 \chi L_L+\text{H.c.},
\end{align}
and gives Majorana mass to left-handed neutrinos. 
Typically, the size of this Yukawa coupling is expected to be as small as ${\cal O}(10^{-9}\text{--}10^{-10})$ for $v_\Delta^{} = {\cal O}(1)$ GeV to reproduce the observed neutrino oscillations.  Thus, these interactions do not play any important role in the following discussions and are ignored throughout this paper.

The interaction terms for the physical Higgs bosons are given in Appendix~\ref{sec:hcoup} in the $\nu \to 0$ limit. 
It should be noted that the 5-plet Higgs bosons do not couple to fermions and are thus fermio-phobic, while the 3-plet Higgs bosons do. 
In fact, the structure of the Yukawa couplings of the 3-plet Higgs bosons is the same as that of the CP-odd and charged Higgs bosons in the Type-I THDM.

\subsection{Decoupling limit}

In this subsection, we briefly discuss the decoupling limit in the GM model.  As clarified in Ref.~\cite{Logan}, the decoupling limit can be realized by taking $m_\Delta^{}$ to  infinity, with all the extra Higgs boson masses also going to infinity and the SM predictions being reproduced.

In order to clearly see how the decoupling limit works, we expand physical parameters in the decoupling regime ({\it i.e.}, $m_\Delta^{} \gg v$) in powers of $1/m_\Delta^2$. 
The masses of extra Higgs bosons are expanded as 
\begin{align}
\begin{split}
m_{H_5}&=\frac{m_\Delta}{4\sqrt2}\left[8+\frac{v^2}{m_\Delta^2}(4\l_4-\l_5)\right]+{\cal O}\left(\frac{v^4}{m_\Delta^3}\right)~,\\
m_{H_3}&=\frac{m_\Delta}{4\sqrt2}\left[8+\frac{v^2}{m_\Delta^2}(4\l_4+\l_5)\right]+{\cal O}\left(\frac{v^4}{m_\Delta^3}\right)~,\\
m_{H_1}&=\frac{m_\Delta}{4\sqrt2}\left[8+\frac{v^2}{m_\Delta^2}(4\l_4+2\l_5)\right]+{\cal O}\left(\frac{v^4}{m_\Delta^3}\right)~.\\
\end{split}
\label{eq:mHdecoupling}
\end{align}
With these mass parameters growing virtually linearly with $m_\Delta^{}$, these extra Higgs bosons are decoupled from the theory in the $m_\Delta\to\infty$ limit. 
It is also seen that the differences among these mass parameters are suppressed by ${\cal O}(1/m_\Delta^{})$ or higher.  
Keeping terms up to order $1/m_\Delta^{}$, we obtain a relation among these mass parameters~\cite{Chiang:2012cn}
\begin{align}
m_{H_1} = \frac{3}{2}m_{H_3}-\frac{1}{2}m_{H_5}~.
\label{eq:massrelation}
\end{align}
On the other hand, the mass of the SM-like Higgs boson $h$ is mainly given by the $\lambda_1$ term as in the SM: 
\begin{align}
\l_1=\frac{1}{8}\left\{\frac{m_h^2}{v^2}+\frac{3\mu_1^2}{8m_\Delta^2}+\frac{\mu_1^2}{16m_\Delta^4}\left[5m_h^2-9v^2(2\l_4+\l_5)\right]\right\}+{\cal O}\left(\frac{v^6}{m_\Delta^6}\right). 
\end{align}

Next, we check the decoupling behavior of the couplings associated with the SM-like Higgs boson $h$. 
At tree level, the $h$ couplings are modified from the SM predictions due to the mixing between the CP-even Higgs bosons and the VEVs. 
The former and the latter are respectively parameterized by $\alpha$ and $v_\Delta^{}$ (or $\beta$). 
Expanding in powers of $1/m_\Delta^2$, we obtain
\begin{align}
\begin{split}
\sin\alpha= &-\frac{\sqrt3v\left|\mu_1\right|}{8m_\Delta^2}\left\{2+\frac{1}{m_\Delta^2}\left[m_h^2-2v^2(2\l_4+\l_5)\right]\right\}+{\cal O}\left(\frac{v^6}{m_\Delta^6}\right)~,\\
v_\Delta= &\frac{v^2\left|\mu_1\right|}{16m_\Delta^2}\left[2-\frac{v^2}{m_\Delta^2}(2\l_4+\l_5)\right] +{\cal O}\left(\frac{v^5}{m_\Delta^4}\right). 
\end{split}
\label{eq:vevdecoupling}
\end{align}
As expected, both of these parameters approach zero in the limit of $m_\Delta\to\infty$ with $\mu_1$ taken to be finite. 
The decoupling behavior of the $h$ couplings can be shown more directly by expanding 
the normalized $hVV$ $(c_{hVV}^{})$, $hff$ $(c_{hff})$ and $hhh$ $(c_{hhh})$ couplings by their SM values as 
\begin{align}
\begin{split}
c_{hVV}^{}  & = 1+\frac{3v^2}{32}\frac{\mu_1^2}{m_\Delta^4}+{\cal O}\left(\frac{v^6}{m_\Delta^6}\right)~,\\
c_{hff}^{}  & = 1-\frac{v^2}{32}\frac{\mu_1^2}{m_\Delta^4}+{\cal O}\left(\frac{v^6}{m_\Delta^6}\right)~,\\
c_{hhh}^{}  & = 1+\frac{v^2\mu_1^2}{32m_\Delta^4}\left[\frac{12v^2}{m_h^2}(2\l_4+\l_5)-7\right]+{\cal O}\left(\frac{v^6}{m_\Delta^6}\right)~.
\end{split}
\end{align}
As expected, they all become one in the decoupling limit.

\section{Renormalization\label{sec:reno}}

In this section, we discuss the renormalization prescription to obtain finite one-loop corrected Higgs boson couplings. 
Our renormalization is based on the on-shell scheme, 
where counterterms are introduced to cancel the radiative corrections to the mass parameters (as well as wave functions) for various fields on their mass shells.

In our calculation, unrenormalized one-loop contributions to 2-point and 3-point functions are constructed in the so-called tadpole scheme~\cite{TP1,TP2} as 
\begin{align}
\begin{split}
&\Pi_{AB}(p^2) = \Pi_{AB}^{\text{1PI}}(p^2) + \Pi_{AB}^{\text{Tad}}~~(\text{for 2-point functions}), \\
&\Gamma_{ABC}(p_1^2,p_2^2,q^2) = \Gamma_{ABC}^{\text{1PI}}(p_1^2,p_2^2,q^2) + \Gamma_{ABC}^{\text{Tad}}~~(\text{for 3-point functions}),
\end{split}
 \label{Xpoint-def}
\end{align}
where $A$, $B$, and $C$ refer to particles on the external legs.
The first and second terms on the right-hand sides denote the contributions from 1-particle irreducible (1PI) and tadpole inserted diagrams, respectively. 
Obviously, there is no momentum dependence in the tadpole inserted contributions ($\Pi_{AB}^{\text{Tad}}$ and $\Gamma_{ABC}^{\text{Tad}}$). 
For later convenience, we define the derivative
\begin{align}
\Pi_{AB}'(m^2) \equiv \frac{d}{dp^2}\Pi_{AB}(p^2)\Big|_{p^2 = m^2}
\end{align}
for a generic 2-point function.
We assume that effects of custodial $SU(2)_V$ symmetry breaking are introduced at the one-loop level; 
namely, we take the $SU(2)_V$ breaking parameter $\nu = 0$ at tree level. 
Therefore, in the calculations of one-loop diagrams, we can make use of the tree-level properties discussed in the previous section,
such as a degenerate mass for the Higgs bosons belonging to the same $SU(2)_V$ multiplet, because including deviations from the tree-level properties would be of higher order corrections.

In the subsequent subsections, we discuss the renormalization of the parameters in the gauge sector, the fermion sector and the scalar sector in order.

\subsection{Gauge sector \label{sec:reno-gauge}}

We shift the following electroweak parameters and the field wave functions of $SU(2)_L$ and $U(1)_Y$ gauge bosons denoted by $W_\mu^a$ ($a=1,2,3$) and $B_\mu$ as: 
\begin{align}
\begin{split}
m_W^2 &\to m_W^2+\delta m_W^2,\quad
m_Z^2 \to m_Z^2+\delta m_Z^2,\quad 
\alpha_{\text{EM}} \to \alpha_{\text{EM}} + \delta\alpha_{\text{EM}},\quad 
\nu \to 0+\delta \nu, \\
W_\mu^a &\to \left(1+\frac{1}{2}\delta Z_W \right)W_\mu^a, \quad 
B_\mu   \to \left(1+\frac{1}{2}\delta Z_B \right)B_\mu, 
\end{split}
\label{shift}
\end{align}
in which we have introduced 6 counterterms. 
Using Eq.~(\ref{vsq}), the counterterms $\delta v$ and $\delta s_W^2$ are given by
\begin{align}
\begin{split}
\delta s_W^2 &= -\frac{m_W^2}{m_Z^2}\left(\frac{\delta m_W^2}{m_W^2}-\frac{\delta m_Z^2}{m_Z^2}-\frac{8v_\Delta\delta\nu}{v^2}\right),  \\
\frac{\delta v}{v} & = \frac{1}{2}\left[\left(1-\frac{c_W^2}{s_W^2} \right)\frac{\delta m_W^2}{m_W^2} + \frac{c_W^2}{s_W^2}\frac{\delta m_Z^2}{m_Z^2} 
 - \frac{\delta \alpha_{\text{EM}}}{\alpha_{\text{EM}}} + \frac{c_W^2}{s_W^2}\frac{8v_\Delta\delta\nu}{v^2}\right]. 
 \end{split}
 \label{delsw}
\end{align}
Furthermore, it is convenient to define the following counterterms for the wave functions of the physical $Z$ boson and photon fields:
\begin{align}
\left(\begin{array}{c}
\delta Z_Z\\
\delta Z_\gamma
\end{array}\right)
&=\left(
\begin{array}{cc}
c_W^2 & s_W^2\\
s_W^2 & c_W^2
\end{array}\right)
\left(\begin{array}{c}
\delta Z_W\\
\delta Z_B
\end{array}\right),  \quad 
\delta Z_{Z\gamma }=\frac{c_Ws_W}{c_W^2-s_W^2}(\delta Z_Z-\delta Z_\gamma). \label{wave}
\end{align}
The renormalized gauge boson 2-point functions $\hat{\Pi}_{XY}$, ($XY =WW,\,ZZ,\,Z\gamma,\,\gamma\gamma$) can then be defined as follows:
\begin{align}
\begin{split}
\hat{\Pi}_{WW}(p^2)&=\Pi_{WW}(p^2)-\delta m_W^2+\delta Z_{W}(p^2-m_W^2),\\
\hat{\Pi}_{ZZ}(p^2)&=\Pi_{ZZ}(p^2)-\delta m_Z^2+\delta Z_{Z}(p^2-m_Z^2),\\
\hat{\Pi}_{Z\gamma}(p^2)&=\Pi_{ Z\gamma}(p^2)+\delta Z_{Z\gamma}\left(p^2-\frac{m_Z^2}{2}\right)-m_Z^2\frac{\delta s_W^2}{2s_Wc_W}, \\
\hat{\Pi}_{\gamma\gamma}(p^2)&=\Pi_{\gamma\gamma}(p^2)+p^2\delta Z_{\gamma},
\end{split}
\end{align}
with $\Pi_{XY}^{}$ being the nurenormalized 2-point functions defined in Eq.~(\ref{Xpoint-def}).

In order to determine the counterterms in Eq.~(\ref{shift}), 
we impose the following five on-shell conditions, which are the same as those used in the SM~\cite{Hollik}: 
\begin{align}
\begin{split}
&\text{Re}\,\hat{\Pi}_{WW}(m_W^2)=0,\quad \text{Re}\,\hat{\Pi}_{ZZ}(m_Z^2) = 0,\quad
\hat{\Pi}_{\gamma\gamma}'(0) =0,\quad
\hat{\Pi}_{Z\gamma }(0)=0,\\
&\hat{\Gamma}_\mu^{\gamma ee } (q^2=0,p_1\hspace{-3.4mm}/\hspace{2mm}=p_2\hspace{-3.4mm}/\hspace{2mm}=m_e) =ie\gamma_\mu, 
\end{split}
\label{rc1}
\end{align}
where $\hat{\Gamma}_\mu^{\gamma ee }$ is the renormalized photon-electron-positron vertex. 
From them, the five counterterms are determined as follows:
\begin{align}
\begin{split}
\delta m_W^2 &= \text{Re}\,\Pi_{WW}(m_W^2),\quad  \delta m_Z^2 = \text{Re}\,\Pi_{ZZ}(m_Z^2),\quad 
\frac{\delta\alpha_{\text{EM}}}{\alpha_{\text{EM}}} = \Pi_{\gamma\gamma}'(0) +\frac{2s_W^{}}{c_W^{}}\frac{\Pi_{Z\gamma}(0)}{m_Z^2},\\
\delta Z_\gamma &= -\Pi_{\gamma\gamma}'(0) , \quad \delta Z_{Z\gamma} = \frac{2}{m_Z^2}\Pi_{Z\gamma}(0)+\frac{\delta s_W^2}{s_Wc_W}. 
\end{split}
\end{align}
Using Eq.~(\ref{wave}), one then finds
\begin{align}
\begin{split}
\delta Z_Z &= -\Pi_{\gamma\gamma}'(0)+\frac{2(c_W^2-s_W^2)}{c_Ws_W}\frac{\Pi_{ Z\gamma}(0)}{m_Z^2}
+\frac{c_W^2-s_W^2}{c_W^2}\frac{\delta s_W^2}{s_W^2},\\
\delta Z_W &=  -\Pi_{\gamma\gamma}'(0)+\frac{2c_W}{s_W}\frac{\Pi_{ Z\gamma}(0)}{m_Z^2}
+\frac{\delta s_W^2}{s_W^2}.
\end{split}
 \end{align}
As explained in Section~\ref{sec:model}, there is one additional counterterm $\delta \nu$ in the GM model.  
Therefore, we need another condition to fix it.  Following the earlier work in Ref.~\cite{GM_lett}, we demand that 
the electroweak oblique $T$ parameter, $T \equiv T_{\text{GM}}-T_{\text{SM}}$ with $T_{\text{GM}}$ and $T_{\text{SM}}$ being respectively the $T$ parameter calculated in the GM model and the SM, be equal to its experimental value:
\begin{align}
T = T_{\text{exp}}, 
\end{align}
where
\begin{align}
\alpha_{\text{EM}}\,T  
&=\frac{\Delta\Pi_{ZZ}(0)}{m_Z^2}-\frac{\Delta\Pi_{WW}(0)}{m_W^2}+\frac{8v_\Delta\delta\nu}{v^2}, 
\label{eq:Tpara}
\end{align}
with $\Delta \Pi_{VV} \equiv  \Pi_{VV}|_{\text{GM}}- \Pi_{VV}|_{\text{SM}}$.
Therefore, $\delta \nu$ is determined as 
\begin{align}
\delta \nu  = \frac{v^2}{8v_\Delta^{}}\left[ \frac{\Delta\Pi_{WW}(0)}{m_W^2}-\frac{\Delta \Pi_{ZZ}(0)}{m_Z^2} + \alpha_{\text{EM}}\,T_{\text{exp}} \right]. 
\end{align}
We will set $T_{\text{exp}} = 0$ in the discussion of numerical analyses. 

\subsection{Fermion sector}

The renormalization for the fermion sector can be done in the same way as in the SM. 
Left-handed and right-handed fermions ($\psi_L$ and $\psi_R$) and their masses $m_f$ are shifted as 
\begin{align}
\psi_{L/R} \to \left(1+ \frac{1}{2}\delta Z_{L/R}^f\right)\psi_{L/R},\quad m_f \to m_f + \delta m_f. 
\end{align}
Following Ref.~\cite{Hollik}, these counterterms are given by 
\begin{align}
\begin{split}
  \delta m_f^{} & = m_f \left[
  \Pi_{ff,V}^\textrm{}(m_f^2) + \Pi_{ff,S}(m_f^2) \right],\\
  \delta Z_V^f &  \left(\equiv \frac{\delta Z_L^f+\delta Z_R^f}{2} \right) =  - \Pi_{ff,V}(m_f^2) -2m_f^2\left[\Pi_{ff,V}'(m_f^2) + \Pi_{ff,S}'(m_f^2) \right], 
  \end{split}
\end{align}
where $\Pi_{ff,V}$ and $\Pi_{ff,S}$ are the vector and scalar parts of the fermion 2-point functions defined in Eq.~(\ref{piffdef}) at the one-loop level, respectively.  
Although another independent wave function renormalization factor $\delta Z_A^f = (\delta Z_L^f-\delta Z_R^f)/2$ can be constructed, it does not appear in 
subsequent discussions.

\subsection{Scalar sector}

Finally, we discuss the renormalization of parameters in the scalar potential. 
In particular, we concentrate on the neutral scalar part, because the charged scalar states are not relevant for the discussions of the renormalized Higgs boson vertices in Section~\ref{sec:reno-vertex}. 
We shift the parameters defined in Eq.~(\ref{para}) as follows:
\begin{align}
\begin{split}
&(m_{H_5}^2,m_{H_3}^2,m_{H_1}^2,m_{h}^2) \to (m_{H_5}^2,m_{H_3}^2,m_{H_1}^2,m_{h}^2) + (\delta m_{H_5}^2,\delta m_{H_3}^2,\delta m_{H_1}^2,\delta m_{h}^2), \\
&\mu_i \to \mu_{1,2} + \delta \mu_{1,2},~~\beta \to \beta + \delta \beta,~~ \alpha \to \alpha + \delta \alpha,~~ \alpha_{1,2} \to 0 + \delta\alpha_{1,2},  \label{counterterms}
\end{split}
\end{align}
where the shifts for $v$ and $\nu$ are already done in Section~\ref{sec:reno-gauge}. 
We here also shift the mixing angles $\alpha_1$ and $\alpha_2$ which become zero at tree level due to the custodial symmetry
\footnote{In our choice of the scalar potential given in Eqs.~(\ref{custb}) and (\ref{eq:pot}), 
the counterterms $\delta \alpha_{1,2}$ can be expressed by the other counterterms given in Eq.~(\ref{counterterms}).  }. 
We note that there are also counterterms for the three tadpoles of $\phi_r$, $\chi_r$ and $\xi_r$.  But these counterterms should be zero in the tadpole scheme~\cite{TP2}, as their contributions are already included in the tadpole inserted diagrams in Eq.~(\ref{Xpoint-def}). 
The wave functions for the CP-odd and CP-even Higgs bosons are then shifted as follows: 
\begin{align}
\begin{split}
&\begin{pmatrix}
G^0 \\
H_3^0
\end{pmatrix}
\to  \left[1 + \frac{1}{2}\begin{pmatrix}
\delta Z_{G^0} & \delta Z_{G^0H_3^0} + 2\delta \beta \\
\delta Z_{H_3^0G^0} - 2\delta \beta & \delta Z_{H_3^0} 
\end{pmatrix}\right]
\begin{pmatrix}
G^0 \\
H_3^0
\end{pmatrix}, \\
&\begin{pmatrix}
H_1\\
h \\
H_5^0
\end{pmatrix}
\to  \left[1 + \frac{1}{2}\begin{pmatrix}
\delta Z_{H_1} & \delta Z_{H_1h} + 2\delta \alpha & \delta Z_{H_1H_5^0}+ 2\delta \alpha_2 \\
\delta Z_{hH_1}-2\delta \alpha & \delta Z_h & \delta Z_{hH_5^0}+2\delta \alpha_1\\
\delta Z_{H_5^0H_1}-2\delta \alpha_2&\delta Z_{H_5^0h}-2\delta \alpha_1 & \delta Z_{H_5^0}
\end{pmatrix}\right]
\begin{pmatrix}
H_1\\
h \\
H_5^0
\end{pmatrix}, 
\end{split}
\end{align}
where $\delta Z_{ij} = \delta Z_{ji}$. 
%
%

The renormalized 2-point functions for the neutral scalar fields are given by 
\begin{align}
\begin{split}
\hat{\Pi}_{SS'}(p^2) & = \Pi_{SS'}(p^2) + \frac{1}{2}(2p^2-m_{S}^2-m_{S'}^2)\delta Z_{SS'} - (m_S^2 - m_{S'}^2) \delta \theta_{SS'}, 
\end{split}
\end{align}
where $S,~S'\in \{ H_5^0,~H_1,~h,~H_3^0,~G^0 \}$, $\delta Z_{SS}=\delta Z_S$ and $(m_{H_5^0}^2,m_{H_3^0}^2,m_{G^0}^2) = (m_{H_5}^2,m_{H_3}^2,0)$.
In addition, $\delta \theta_{SS'}$ is $\delta \alpha$, 
$\delta \alpha_1$,  $\delta \alpha_2$ and $\delta \beta$ for $(S,S') = (H_1,h)$,
$(h,H_5^0)$, $(H_1,H_5^0)$ and $(G^0,H_3^0)$, respectively, and $\delta \theta_{SS'} = -\delta \theta_{S'S}$. 
To determine these counterterms, we impose the following on-shell conditions: 
\begin{align}
\begin{split}
&\text{Re}\,\hat{\Pi}_{SS}(m_S^2) = 0, \quad \text{Re}\,\hat{\Pi}_{SS}'(m_S^2) =0 ,\\
&\text{Re}\,\hat{\Pi}_{SS'}(m_S^2) =\text{Re}\,\hat{\Pi}_{SS'}(m_{S'}^2)= 0~~\mbox{for } S \neq S'. 
\end{split}
\end{align}
Counterterms are then determined as 
\begin{align}
\delta m_S^2 &= \text{Re}\,\Pi_{SS}(m_{S}^2), \quad \delta Z_{S}  = -\text{Re}\,\Pi_{SS}'(m_{S}^2), 
\end{align}
and 
\begin{align}
\begin{split}
\delta \alpha    &=  \frac{1}{2(m_{H_1}^2-m_h^2)}\text{Re}\left[\Pi_{H_1 h}(m_{H_1}^2)+\Pi_{H_1 h}(m_{h}^2) \right], \\
\delta \alpha_1  &=  \frac{1}{2(m_{h}^2-m_{H_5}^2)}\text{Re}\left[\Pi_{H_5^0 h}(m_{h}^2)+\Pi_{H_5^0 h}(m_{H_5}^2) \right], \\
\delta \alpha_2  &=  \frac{1}{2(m_{H_1}^2-m_{H_5}^2)}\text{Re}\left[\Pi_{H_1 H_5^0}(m_{H_1}^2)+\Pi_{H_1H_5^0}(m_{H_5}^2) \right], \\
\delta \beta     &=  -\frac{1}{2m_{H_3}^2}\text{Re}\left[\Pi_{G^0H_3^0}(0)+\Pi_{G^0H_3^0}(m_{H_3}^2) \right], \label{delta_beta}\\
\delta Z_{H_1 h}  &=  -\frac{1}{m_{H_1}^2-m_h^2}\text{Re}\left[\Pi_{H_1 h}(m_{H_1}^2)-\Pi_{H_1 h}(m_h^2) \right], \\
\delta Z_{hH_5^0 }  &=  -\frac{1}{m_{h}^2-m_{H_5}^2}\text{Re}\left[\Pi_{H_5^0 h}(m_{h}^2)-\Pi_{H_5^0 h}(m_{H_5}^2) \right], \\
\delta Z_{H_1H_5^0 } &=  -\frac{1}{m_{H_1}^2-m_{H_5}^2}\text{Re}\left[\Pi_{H_1 H_5^0}(m_{H_1}^2)-\Pi_{H_1 H_5^0}(m_{H_5}^2) \right], \\
\delta Z_{G^0H_3^0} &=  \frac{1}{m_{H_3}^2}\text{Re}\left[\Pi_{G^0 H_3^0}(0)-\Pi_{G^0 H_3^0}(m_{H_3}^2) \right]. 
\end{split}
\end{align}

There are still two counterterms $\delta\mu_{1,2}$ that are not fixed by the above conditions. 
These counterterms appear in the renormalized $hhh$ vertex, and we will discuss how to determine these counterterms in Section~\ref{sec:hhh}.

\section{Gauge Dependence \label{sec:gauge}}

In the previous section, we have determined all the counterterms by imposing the on-shell renormalization conditions except for $\delta \mu_{1,2}$.
As a result, they can be expressed in terms of 2-point functions defined in Eq.~(\ref{Xpoint-def}). 
However, it has been known that there remains gauge dependence in the counterterms for the mixing angles, {\it e.g.}, $\delta \beta$ and $\delta \alpha$, in the on-shell scheme as  
it can be proved using the Nielsen identity~\cite{NI}.

In this section, we first show the gauge dependence in the scalar 2-point functions, particularly for the CP-even and CP-odd scalar bosons. 
In order to manifestly show the gauge dependence, we perform the calculation in the general $R_\xi$ gauge, where 
the propagator of a gauge boson $V(=W,Z)$ is expressed using the gauge parameter $\xi_V$ as 
\begin{align}
\frac{-i}{p^2-m_V^2}\left[g^{\mu\nu} - (1-\xi_V)\frac{p^\mu p^\nu}{p^2-\xi_V m_V^2}\right].
\end{align}
We then discuss how one can remove such gauge dependence by employing the pinch technique~\cite{Cornwall:1981zr,Cornwall:1989gv}.

\subsection{CP-even part}

First, we show explicitly the cancellation of the gauge dependence in the mixing of CP-even Higgs bosons. 
Here, we only show the $\xi_W$-dependent part because the $\xi_Z$ part can be simply obtained by the replacements of $(g,W,G^\pm) \to (g_Z^{}/2,Z,G^0)$.  
The 2-point functions for $\phi_1$--$\phi_2$ ($\phi_{1,2} \in \{ h, H_1 \}$) are expressed as 
\begin{align}
&\Pi_{\phi_1\phi_2}(q^2)  = \Pi_{\phi_1\phi_2}(q^2)\Big|_{\xi = 1}
+\Pi_{\phi_1\phi_2}(q^2)\Big|_{\text{G.D.}},  \label{pi12r}
\end{align}
where the first term on the right-hand side corresponds to the result calculated in the 't~Hooft-Feynman gauge.
%
On the other hand, the second term in Eq.~\eqref{pi12r} depends on the gauge parameter and is explicitly given by
\begin{align}
\Pi_{\phi_1\phi_2}(q^2)\Big|_{\text{G.D.}} &= (1-\xi_W)\frac{g^2}{64\pi^2}\Big[  c_{\phi_1 VV}^{}c_{\phi_2 VV}^{}(q^4-m_{\phi_1}^2m_{\phi_2}^2)C_0(q^2;W,G^\pm) \notag\\
&\quad + 2c_{\phi_1 H_3V}c_{\phi_2 H_3V}f(q^2;m_{\phi_1},m_{\phi_2},m_{H_3})C_0(q^2;W,G^\pm,H_3^\pm)\notag\\
&\quad -(c_{\phi_1VV}c_{\phi_2VV} + c_{\phi_1H_3V}c_{\phi_2H_3V})(2q^2 -m_{\phi_1}^2 -m_{\phi_2}^2)B_0(0;m_W^{},m_{G^\pm}^{}) \Big], \label{pi12}
\end{align}
where we have introduced
\begin{align}
\begin{split}
C_0(q^2;A,B)   &\equiv \frac{1}{m_A^2-m_B^2}[B_0(q^2;m_A,m_A)-B_0(q^2;m_B,m_B)], \\
C_0(q^2;A,B,C) &\equiv \frac{1}{m_A^2-m_B^2}[B_0(q^2;m_A,m_C)-B_0(q^2;m_B,m_C)], \\
f(q^2;m_A,m_B,m_C) &\equiv  (q^2-m_C^2)^2-(m_C^2-m_A^2)(m_C^2-m_B^2).  
\end{split}
\end{align}
The function $B_0$ is the Passarino-Veltman's scalar 2-point function~\cite{PV} defined in Section~\ref{sec:lf}.
We see that for $\phi_1 = \phi_2 (\equiv \phi)$, $\Pi_{\phi_1\phi_2}\big|_{\text{G.D.}}$ vanishes at $q^2 = m_\phi^2$ as it is expected by the Nielsen identity, so that 
the counterterms for the mass parameters ($\delta m_h^2$ and $\delta m_{H_1}^2$) are gauge-independent. 
For $\phi_1 \neq \phi_2$, $\Pi_{\phi_1\phi_2}\big|_{\text{G.D.}}$ does not vanish either at $q^2 = m_{\phi_1}^2$ or $q^2 = m_{\phi_2}^2$, and thus
the gauge dependence shows up in $\delta \alpha$.

In order to remove such gauge dependence, 
one can add pinch-term contributions to the above 2-point functions.  
Pinch-terms are the ``propagator-like'' part of vertex and box-diagram corrections to a 2-to-2 fermion scattering process ({\it i.e.}, $f\bar{f}' \to f\bar{f}'$ with $f$ and $f'$ being SM fermions), 
and can be extracted by cancelling internal fermion propagators using a contracted loop momentum in the numerator.   
Here the extracted pinch-terms do not depend on the choice of the external fermions. 
Since the pinch-terms also depend on the gauge choice, they can be expressed in a way similar to Eq.~(\ref{pi12r}) as 
\begin{align}
\Pi_{\phi_1\phi_2}^{\text{PT}}(q^2) &= \Pi_{\phi_1\phi_2}^{\text{PT}}(q^2)\Big|_{\xi = 1} + \Pi_{\phi_1\phi_2}^{\text{PT}}(q^2)\Big|_{\text{G.D.}}, \label{pi12pt}
\end{align}
where 
\begin{align}
\Pi_{\phi_1\phi_2}^{\text{PT}}(q^2)\Big|_{\xi = 1} 
=& 
-\frac{g^2}{16\pi^2}\left(q^2-\frac{m_{\phi_1}^2}{2}-\frac{m_{\phi_2}^2}{2}\right) \notag\\
      & \times \left[c_{\phi_1 VV}c_{\phi_2 VV}B_0(q^2;m_W,m_W)+c_{\phi_1 H_3V}c_{\phi_2 H_3V}B_0(q^2;m_{H_3},m_W) \right].  \label{pipipi}
\end{align}
One can verify that the second term of Eq.~(\ref{pi12pt}) satisfies the property $\Pi_{\phi_1\phi_2}^{\text{PT}}(q^2)\Big|_{\text{G.D.}} = -\Pi_{\phi_1\phi_2}(q^2)\Big|_{\text{G.D.}}$.  
Therefore, the 2-point functions with the pinch-terms $\overline{\Pi}_{\phi_1\phi_2} \equiv \Pi_{\phi_1\phi_2} + \Pi_{\phi_1\phi_2}^{\text{PT}}$ are gauge-independent, and one should consider the gauge-independent counterterm $\delta \bar{\alpha}$ defined by  
\begin{align}
\delta \bar{\alpha}    &=  \frac{1}{2(m_{H_1}^2-m_h^2)}\text{Re}\left[\overline{\Pi}_{H_1 h}(m_{H_1}^2) + \overline{\Pi}_{H_1 h}(m_{h}^2) \right],
\label{alpbar}
\end{align}
instead of $\delta\alpha$, as we will do in the following discussions.

\subsection{CP-odd part}

Next, we discuss the gauge dependence of the 2-point functions for the CP-odd scalar bosons $A_1$--$A_2$ ($A_{1,2} \in \{ G^0, H_3^0 \}$). 
Similar to Eq.~(\ref{pi12}), we have 
\begin{align}
\Pi_{A_1A_2}(q^2)  = \Pi_{A_1A_2}(q^2)\Big|_{\xi = 1}  +   \Pi_{A_1A_2}(q^2)\Big|_{\text{G.D.}}. 
\end{align}
The gauge-dependent part is expressed as 
\begin{align}
& \!\!\!\! 
\Pi_{A_1A_2}(q^2)\Big|_{\text{G.D.}} \notag\\
=& 
(1-\xi_W^{})\frac{g^2}{32\pi^2} \Bigg\{\zeta_{A_1}\zeta_{A_2}f(q^2;m_{A_1},m_{A_2},m_{H_5})C_0(q^2;W,G^\pm,H_5^\pm) \notag\\
&\quad\quad
- \zeta_{A_1}\zeta_{A_2}\left(q^2-\frac{m_{A_1}^2}{2}-\frac{m_{A_2}^2}{2} \right)B_0(0;m_W,m_{G^\pm})\notag\\
&\quad\quad
+\delta_{A_1 H_3^0}\delta_{A_2 H_3^0}(q^2-m_{H_3}^2)
\left[ (q^2-m_{H_3}^2)C_0(q^2;W,G^\pm,H_3^\pm) -B_0(0;m_W,m_{G^\pm}) \right] \Bigg\}\notag\\
&+(1-\xi_Z)\frac{g_Z^2}{64\pi^2}\sum_{\varphi=h,H_1,H_5^0} c_{\varphi A_1Z}c_{\varphi A_2Z} \notag\\
&\quad 
\times\left[f(q^2;m_{A_1},m_{A_2},m_{\varphi})C_0(q^2;Z,G^0,\varphi)- \left(q^2-\frac{m_{A_1}^2}{2}-\frac{m_{A_2}^2}{2} \right)B_0(0;m_Z,m_{G^0})\right], \label{pi5}
\end{align}
where $(m_{A_i},\zeta_{A_i},c_{\varphi A_1Z})= (m_{H_3},-s_\beta,-c_{\varphi H_3 Z})$ for $A_i = H_3^0$ and 
$(0,c_\beta,c_{\varphi Z Z})$ for $A_i = G^0$, and $c_{\varphi H_3 Z}$ and $c_{\varphi ZZ}$ are given in Eqs.~(\ref{c5_1}) and (\ref{c5_2}). 
Again, if $A_1 \neq A_2$, $\Pi_{A_1A_2}\big|_{\text{G.D.}}$ does not vanish at either $q^2 = m_{A_1}^2$ or $m_{A_2}^2$, and the counterterm $\delta\beta$ determined by Eq.~(\ref{delta_beta}) remains gauge-dependent.

Analogous to the CP-even case, we can add the pinch-terms extracted from a $f\bar{f'} \to f\bar{f}'$ process:
\begin{align}
\Pi_{A_1A_2}^{\text{PT}}(q^2) & = \Pi_{A_1A_2}^{\text{PT}}(q^2)\Big|_{\xi = 1}  + \Pi_{A_1A_2}^{\text{PT}}(q^2)\Big|_{\text{G.D.}}, 
\end{align}
where the gauge-independent part 
\begin{align}
\Pi_{A_1A_2}^{\text{PT}}(q^2)\Big|_{\xi = 1} 
=& 
-\delta_{A_1H_3^0}\delta_{A_2H_3^0}\frac{g^2}{16\pi^2}(q^2-m_{H_3}^2)B_0(q^2;m_W,m_{H_3})\notag\\
&-\frac{g_Z^2}{32\pi^2}\left(q^2-\frac{m_{A_1}^2}{2}-\frac{m_{A_2}^2}{2} \right)\sum_{\phi=h,H_1}c_{\phi A_1V}c_{\phi A_2V}B_0(q^2;m_Z,m_\phi), 
\end{align}
while the gauge-dependent part 
\begin{align}
\Pi_{A_1A_2}^{\text{PT}}(q^2)\Big|_{\text{G.D.}}
=
& - \delta_{A_1 H_3^0}\delta_{A_2 H_3^0}(1-\xi_W)\frac{g^2}{32\pi^2}(q^2-m_{H_3}^2)
\notag\\
& \qquad \left[ (q^2-m_{H_3}^2)C_0(q^2;W,G^\pm,H_3^\pm)-B_0(0;m_W,m_{G^\pm}) \right]
\notag\\
&-(1-\xi_Z)\frac{g_Z^2}{64\pi^2}
\Bigg\{\sum_{\phi = h,H}c_{\phi A_1Z}c_{\phi A_2Z}
\Big[f(q^2;m_{A_1},m_{A_2},m_{\phi})C_0(q^2;Z,G^0,\phi) \notag\\
&\qquad\qquad\qquad\qquad\qquad\qquad
-\left(q^2-\frac{m_{A_1}^2}{2}-\frac{m_{A_2}^2}{2}\right)B_0(0;m_Z,m_{G^0}) \Big]
\notag\\
&\qquad\qquad\qquad\qquad
-\frac{5}{3}\zeta_{A_1}\zeta_{A_2}\left(q^2-\frac{m_{A_1}^2}{2}-\frac{m_{A_2}^2}{2}\right)B_0(0;m_Z,m_{G^0})\Bigg\}. 
\end{align}
Analogous to Eq.~(\ref{alpbar}), one should consider the pinch term-included counterterm $\delta \bar{\beta}$ defined by  
\begin{align}
\delta \bar{\beta}    &=  -\frac{1}{2m_{H_3}^2}\text{Re}\left[\overline{\Pi}_{G^0 H_3^0}(0) + \overline{\Pi}_{G^0 H_3^0}(m_{H_3}^2) \right], \label{pinched-beta}
\end{align}
where $\overline{\Pi}_{A_1A_2} \equiv \Pi_{A_1A_2} + \Pi_{A_1A_2}^{\text{PT}}$.
It is intriguing to note that $\Pi_{A_1A_2}^{\text{PT}}\big|_{\text{G.D.}} \not= -\Pi_{A_1A_2}\big|_{\text{G.D.}}$ in this case; that is, $\delta \bar{\beta}$ still has explicit gauge dependence. 
In fact, the 5-plet Higgs boson loop contributions to $\Pi_{A_1A_2}\big|_{\text{G.D.}}$, the terms proportional to $\zeta_{A_1}^{}\zeta_{A_2}^{}$ and $c_{H_5^0 A_1Z}^{}c_{H_5^0 A_2Z}^{}$ in Eq.~(\ref{pi5}),  
are not cancelled by the pinch-terms because of the fermio-phobic nature of the 5-plet Higgs bosons.  
Therefore, even after the pinch-terms are included, gauge dependence still remains in the 2-point functions for the CP-odd scalar bosons. 
This does not happen, for example, in the THDMs, where the gauge dependence in the 2-point functions for the CP-odd Higgs bosons does cancel by adding the pinch-terms, as shown in Refs.~\cite{TP2,Yagyu-Gauge}.

\begin{figure}[t]
\centering
\includegraphics[scale=0.2]{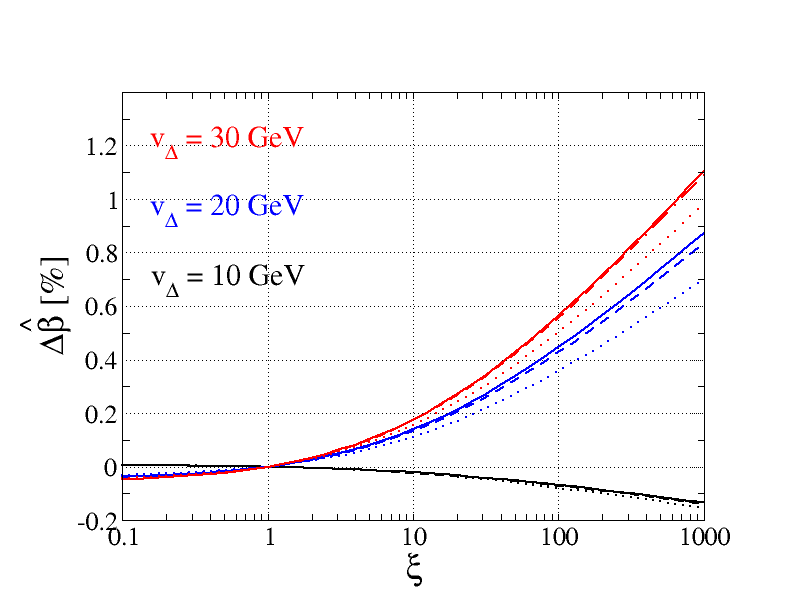}\hspace{-5mm}
\includegraphics[scale=0.2]{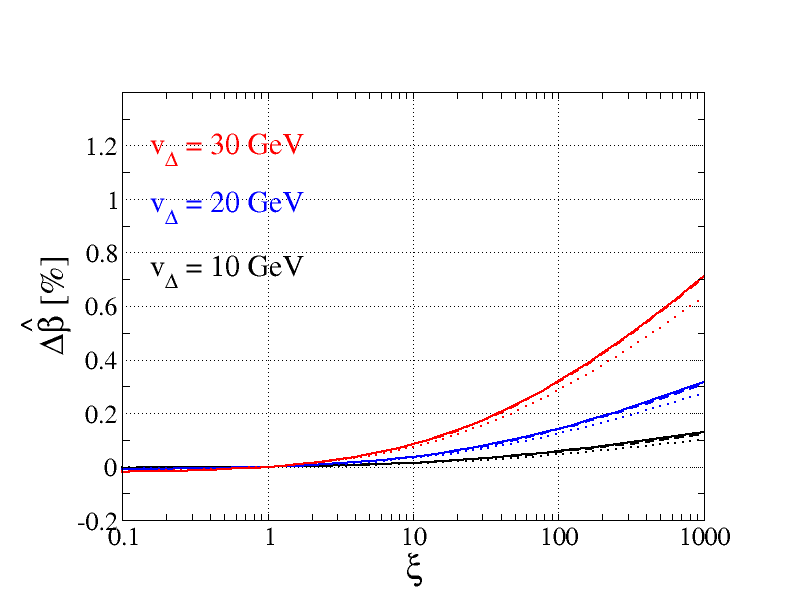}\hspace{-5mm}
\includegraphics[scale=0.2]{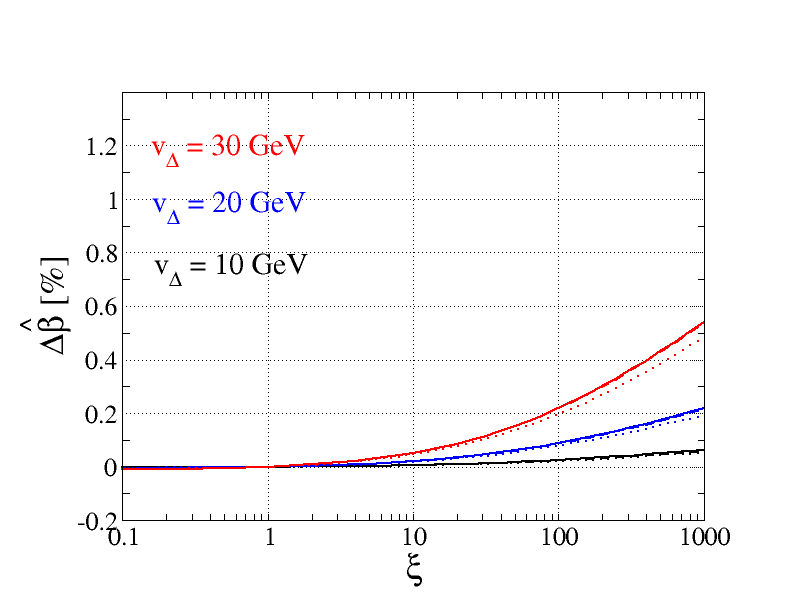}
\caption{Gauge dependence of the renormalized mixing angle $\hat{\beta}$ in the case of $\mu_1 = -100$ GeV and 
$\mu_2=0$ with $\xi_W = \xi_Z = \xi$, where $\Delta \hat{\beta}$ is defined in Eq.~(\ref{delbb}).  The left, middle and right plots show the case with $m_{H_5}=m_{H_3}=m_{H_1}=300$, 500 and 700~GeV, respectively. 
For each plot, the black, blue and red curves show respectively the cases of $v_\Delta^{} = 10$, 20 and 30~GeV, 
while the solid, dashed and dotted curves denote respectively the case with $\alpha = 0^\circ$, $-10^\circ$ and $-20^\circ$. }
\label{delb}
\end{figure}

To see the $\xi$ dependence, we introduce 
\begin{align}
\Delta \hat{\beta} 
\equiv 
\frac{\left. \hat{\beta} \right|_{\xi} - \left. \hat{\beta} \right|_{\xi=1}}{\left. \hat{\beta} \right|_{\xi = 1}},
\quad
\hat{\beta} = \beta + \left. \delta \bar{\beta} \right|_{\text{fin}},  
\label{delbb}
\end{align}
where $\left. \delta \bar{\beta} \right|_{\text{fin}}$ is the finite part of $\delta \bar{\beta}$. 
In Fig.~\ref{delb}, we show how the renormalized mixing angle $\hat{\beta}$ depends on the choice of the gauge parameter $\xi$ ($=\xi_W^{} = \xi_Z^{}$). 
We see that the gauge dependence of $\hat{\beta}$ becomes larger for larger values of $\xi$ and/or $v_\Delta^{}$, but it is at most about 1\% or smaller when $\xi \leq 10^3$. 
Therefore, in the numerical evaluation for the renormalized Higgs boson vertices, the actual effect from the gauge dependence in $\delta \bar{\beta}$ is negligibly small.  Moreover, the modifications cause by varying $\alpha$ 
between $0^\circ$ and $-20^\circ$, a range of phenomenological interest, are much minor.
In the following discussion, we will use $\delta \bar{\beta}$ instead of $\delta\beta$.

Before closing this section, we would like to remark that the gauge dependence of the 2-point function for the CP-odd Higgs bosons is cancelled if we add up all the contributions to the $f\bar{f}' \to f\bar{f}'$ scattering amplitude
from the $H_3^0$--$H_3^0$, $H_3^0$--$G^0$ and $G^0$--$G^0$ mediators:
\begin{align}
{\cal M} &= 
-\frac{m_fm_{f'}}{v^2}(\bar{f}\gamma_5 f)(\bar{f}'\gamma_5 f')\notag\\
&\quad
\times \left[
\left(\frac{\cot\beta}{q^2-m_{H_3}^2}\right)^2\overline{\Pi}_{H_3^0H_3^0}(q^2)
+\frac{2\cot\beta}{q^2(q^2-m_{H_3}^2)}\overline{\Pi}_{H_3^0G^0}(q^2)  + \frac{1}{q^4}\overline{\Pi}_{G^0G^0}(q^2) \right]. 
\end{align}
One can explicitly verify that the gauge dependence in the above expression is exactly cancelled among the three terms in the square brackets.

\section{Renormalized Higgs Vertices  \label{sec:reno-vertex}}

In this section, we compute renormalized $hV^\mu V^\nu$ ($V=W,Z$), $hf\bar{f}$ and $hhh$ vertices based on the on-shell scheme discussed in Section~\ref{sec:reno}.  We note that the on-shell conditions are insufficient 
to fix all the counterterms appearing in the renormalized $hhh$ vertex.  Therefore, we have to introduce an additional condition, the minimal subtraction (MS) scheme, to be discussed in Section~\ref{sec:hhh}. 
All the analytic expressions for the 1PI diagram contributions (variables labeled with the superscript ``1PI'') to 1-, 2- and 3-point functions are given in Appendix~\ref{sec:1pi}.

Hereafter, we use the shorthand notation for the trigonometric function as $s_\theta = \sin\theta$, $c_\theta = \cos\theta$ and $t_\theta = \tan\theta$. 

\subsection{Renormalized $hVV$ vertex}

The renormalized $hV^\mu V^\nu$ vertices can be generally decomposed as:
\begin{align}
\hat{\Gamma}_{hVV}^{\mu\nu}(p_1^2,p_2^2,q^2)=\hat{\Gamma}_{hVV}^1g^{\mu\nu}+\hat{\Gamma}_{hVV}^2\frac{p_1^\nu p_2^\mu}{m_V^2}
+i\hat{\Gamma}_{hVV}^3\epsilon^{\mu\nu\rho\sigma}\frac{p_{1\rho} p_{2\sigma}}{m_V^2}, \label{form_factor}
\end{align}
where $p_{1,2}^\mu$ and $q^\mu (=p_1^\mu+p_2^\mu$) are the incoming momenta of the gauge bosons and the outgoing momentum of $h$, respectively. 
Each of the renormalized form factors $\hat \Gamma^i_{hVV}$ can be further decomposed into four parts as:
\begin{align}
\hat{\Gamma}_{hVV}^i(p_1^2,p_2^2,q^2)&=\Gamma^{i,\text{tree}}_{hVV} + \delta \Gamma^i_{hVV}+\Gamma^{i,\text{1PI}}_{hVV}(p_1^2,p_2^2,q^2)
+T_{hVV}^{i},\quad (i=1,2,3), 
\label{reno-hvv}
\end{align}
where $\Gamma^{i,\text{tree}}_{hVV}$, $\delta \Gamma_{hVV}^i$, $\Gamma^{i,\text{1PI}}_{hVV}$ and $T_{hVV}^{i}$
denote the contributions from tree-level diagrams, counterterms, 1PI diagrams and tadpoles, respectively. 
We note that the tadpole part $\Pi_{AB}^{\text {Tad}}$ in the counterterms is here grouped into $T_{hVV}^{i}$.  According to Eq.~(\ref{Xpoint-def}),
\begin{align}
T_{hVV}^{i} = \delta \Gamma^i_{hVV}  (\text{tadpole part}) + \Gamma^{i,\text{Tad}}_{hVV}. 
\end{align}
Each term in Eq.~(\ref{reno-hvv}) is given as follows:
\begin{align}
\begin{split}
\Gamma_{hVV}^{1,\text{tree}} & = \frac{2m_V^2}{v}c_{hVV}^{}, \\
\delta \Gamma_{hVV}^1 & =  \frac{2m_V^2}{v}\Bigg[c_{hVV}^{}\left(\frac{\delta m_V^2}{m_V^2}-\frac{\delta v}{v} +\delta Z_V +\frac{1}{2}\delta Z_h\right) \\
&\hspace{15mm}  - c_{hH_3V}\delta \bar{\beta} +C_\nu^V \frac{\delta\nu}{v}+\frac{c_{H_1VV}}{2}\delta Z_{H_1 h}+\frac{c_{H_5VV}}{2}\delta Z_{H_5h}\Bigg],  \\
T_{hVV}^{1,\text{Tad}} &=\frac{2m_V^2}{v^2m_{H_3}^2}c_{hH_3V}\left(T_h^{\text{1PI}}\,c_{hH_3V}^{} + T_{H_1}^{\text{1PI}}c_{H_1H_3V}^{} + T_{H_5^0}^{\text{1PI}}c_{H_5H_3Z}\right) , 
\end{split}   \label{reno-hvv2}
\end{align}
where $C_\nu^W = \sqrt{2}s_\beta c_{hH_3V}$ and $C_\nu^Z = \sqrt{2}c_\beta c_{hVV}$. 
For $i=2,3$, we have $\Gamma_{hVV}^{i,\text{tree}} = \delta \Gamma_{hVV}^{i}=T_{hVV}^{i} = 0$. 
As mentioned in the previous section, we adopt the pinched counterterm $\delta \bar{\beta}$ defined in Eq.~(\ref{pinched-beta}) instead of $\delta \beta$ given in Eq.~(\ref{delta_beta}).

\subsection{Renormalized $hf\bar{f}$ vertex \label{sec:hff}}

The renormalized $hf\bar{f}$ vertices can be expressed in terms of eight form factors as follows:
\begin{align}
\hat{\Gamma}_{hff}(p_1^2,p_2^2,q^2)&=
\hat{\Gamma}_{hff}^S+\gamma_5 \hat{\Gamma}_{hff}^P+p_1\hspace{-3.5mm}/\hspace{2mm}\hat{\Gamma}_{hff}^{V_1}
+p_2\hspace{-3.5mm}/\hspace{2mm}\hat{\Gamma}_{hff}^{V_2}\notag\\
&\quad +p_1\hspace{-3.5mm}/\hspace{2mm}\gamma_5 \hat{\Gamma}_{hff}^{A_1}
+p_2\hspace{-3.5mm}/\hspace{2mm}\gamma_5\hat{\Gamma}_{hff}^{A_2}
+p_1\hspace{-3.5mm}/\hspace{2mm}p_2\hspace{-3.5mm}/\hspace{2mm}\hat{\Gamma}_{hff}^{T}
+p_1\hspace{-3.5mm}/\hspace{2mm}p_2\hspace{-3.5mm}/\hspace{2mm}\gamma_5\hat{\Gamma}_{hff}^{PT},  \label{form_factor2}
\end{align}
where $p_{1,2}^\mu$ and $q^\mu (=p_1^\mu+p_2^\mu$) 
are the incoming momenta of the fermions and the outgoing momentum of $h$, respectively. 
Analogous to the renormalized $hV^\mu V^\nu$ vertices, each of the renormalized form factors is further decomposed into the following four parts:
\begin{align}
\hat{\Gamma}_{hff}^i(p_1^2,p_2^2,q^2)&=\Gamma^{i,\text{tree}}_{hff}
+\delta \Gamma^i_{hff}+\Gamma^{i,\text{1PI}}_{hff} +T^i_{hff},
\label{reno-hff}
\end{align}
where $i=S,P,V_1,V_2,A_1,A_2,T,PT$.
We note that the tadpole term cannot be inserted to the tree-level $hf\bar{f}$ diagram; that is, $\Gamma_{hff}^{\text{Tad}} = 0$. 
Hence, the tadpole contribution $T^i_{hff}$ is obtained only from the corresponding counterterm. 
Each term of Eq.~(\ref{reno-hff}), except for the 1PI part, is given as follows:
\begin{align}
\begin{split}
\Gamma^{S,\text{tree}}_{hff} &= -\frac{m_f}{v}c_{hff}, \\
\delta \Gamma_{hff}^S    &= -\frac{m_f}{v}c_{hff}\left(
\frac{\delta m_f}{m_f}+ \delta Z_V^f -\frac{\delta v}{v} -\cot\beta \delta \bar{\beta} 
+\sqrt{2}c_\beta \frac{\delta \nu}{v}  + \frac{\delta Z_h}{2}  + \frac{c_{Hff}}{2c_{hff}}\delta Z_{H_1h} \right), \\
T_{hff}^S & = -\frac{m_f}{v}c_{hff}\frac{\cot\beta}{m_{H_3}^2v} \left(c_{hH_3V} T_h^{\text{1PI}} + c_{H_1H_3V} T_{H_1^0}^{\text{1PI}} +\frac{2\sqrt{3}}{3}s_\beta T_{H_5^0}^{\text{1PI}} \right),
\end{split}
\end{align}
with $\Gamma^{i,\text{tree}}_{hff} = \delta \Gamma_{hff}^{i} = T_{hff}^{i} = 0 ~~\text{for}~~i\neq S$. 
As for the renormalized $hV^\mu V^\nu$ vertex, we also use the pinched counterterm $\delta \bar{\beta}$ in the contribution to the $hf\bar{f}$ vertex.

\subsection{Renormalized $hhh$ vertex  \label{sec:hhh} }

Finally, we compute the renormalized $hhh$ vertex which is trivial in the Lorentz structure as it is a scalar vertex. 
Analogous to the $hV^\mu V^\nu $ and $hf\bar{f}$ vertices, the renormalized $hhh$ vertex can be expressed as  
\begin{align}
\hat{\Gamma}_{hhh}(p_1^2,p_2^2,q^2)
&=\Gamma_{hhh}^{\text{tree}} + \delta\Gamma_{hhh} + \Gamma_{hhh}^{\text{1PI}} + T_{hhh}, 
\end{align}
where $p_{1,2}^\mu$  and $q^\mu (=p_1^\mu+p_2^\mu$) are the incoming and outgoing momenta for the Higgs boson, respectively. 
Each of the contributions is given as follows:
\begin{align}
\begin{split}
\Gamma_{hhh}^{\text{tree}} 
=& 
3!\lambda_{hhh} =  -\frac{3m_h^2}{v}\left(\frac{c_\alpha^3}{s_\beta}-\frac{2\sqrt{6}}{3}\frac{s_\alpha^3}{c_\beta} \right)
 +\mu_1s_\alpha^2 t_\beta (3\sqrt{2}c_\alpha +2\sqrt{3}s_\alpha t_\beta) -2\sqrt{3}\mu_2 s_\alpha^3, \\
\delta \Gamma_{hhh} 
=&
3! \left[\delta \lambda_{hhh} + \frac{3}{2}\lambda_{hhh}\delta Z_h 
+ \frac{1}{2}\lambda_{H_1hh}(\delta Z_{H_1h} + 2\delta\bar{\alpha}) \right], \\
T_{hhh}
=& 3! \left\{
C(m_h^2)\frac{m_h^2}{v} \left[
2\sum_{\phi=h,H_1}(1+2\delta_{\phi h})
\frac{\lambda_{\phi hh}T_\phi^{\text{1PI}}}{m_h^2m_\phi^2} 
+ \sum_{\varphi=h,H_1,H_5^0} c_{\varphi WW}\frac{T_\varphi^{\text{1PI}}}{vm_\varphi^2}
\right] \right. \\
& \quad\quad
+\sum_{\phi=h,H_1} \left[c_{hH_3V}\frac{s_{2\alpha}}{s_{2\beta}} \lambda_{\phi Hh}\frac{T_\phi^{\text{1PI}}}{vm_\phi^2}
  +(1+3\delta_{\phi h}) \lambda_{\phi hhh}\frac{T_\phi^{\text{1PI}}}{m_\phi^2} \right] 
   \\
& \quad\quad
\left.
-C(\delta \beta)\sum_{\varphi=h,H_1,H_5^0} c_{\varphi H_3 Z}
\frac{m_\varphi^2-m_{H_3}^2}{m_{H_3}^2}\frac{T_\varphi^{\text{1PI}}}{m_\varphi^2}
+ \frac{3}{\sqrt{6}} C(\delta\nu) \frac{T_{H_5^0}^{\text{1PI}}}{m_{H_5}^2} \right\}, 
\end{split}
\label{ren_hhh}
\end{align}
where $\lambda_{\phi_i\phi_j\phi_k}$ and $\lambda_{\phi_i\phi_j\phi_k\phi_l}$ are defined in Eq.~(\ref{self}), and 
\begin{align}
\begin{split}
C(\delta m_h^2)
&= -\frac{1}{3s_{2\beta}}(3c_\alpha^3 c_\beta  - 2\sqrt{6}s_\alpha^3 s_\beta) , \\
C(\delta \beta)
&= \left(\frac{c_\alpha^3}{2t_\beta s_\beta} 
+ \frac{\sqrt{6}}{3}s_\alpha^3 \frac{t_\beta}{c_\beta} \right)\frac{m_h^2}{v^2} 
+ s_\alpha^2\frac{3\sqrt{2}c_\alpha + 4\sqrt{3}s_\alpha t_\beta}{6c_\beta^2}\frac{\mu_1}{v}, \\
C(\delta \nu)  &=\left[ \frac{2\sqrt{3}}{9}s_\alpha^3\left(3-\frac{2}{c_{\beta}^2}\right) -\frac{\sqrt{2}c_\alpha^3}{2t_\beta} \right]\frac{m_h^2}{v^2} 
- \frac{2c_\alpha s_\alpha^2 s_\beta}{9c_\beta^2} (3 + 2\sqrt{6}t_\alpha t_\beta) \frac{\mu_1}{v}.  \label{cdelta}
\end{split}
\end{align}
The counterterm $\delta\lambda_{hhh}$ is expressed as 
\begin{align}
\delta \lambda_{hhh} 
=& 
C(\delta m_h^2)\frac{m_h^2}{v}\left(\frac{\delta m_h^2}{m_h^2} -\frac{\delta v}{v}\right)
+\left(\frac{c_{hH_3V}^{}}{2}\frac{s_{2\alpha}}{s_{2\beta}}\frac{m_{H_1}^2 - m_h^2}{v}-\lambda_{H_1hh}\right)\delta \bar{\alpha}\notag\\
&
+ C(\delta \beta)v\delta \bar{\beta} + C(\delta \nu) \delta\nu 
 + \frac{s_\alpha^2 t_\beta}{6} (3\sqrt{2}c_\alpha + 2\sqrt{3}s_\alpha t_\beta)\delta \mu_1  - \frac{s_\alpha^3}{\sqrt{3}}\delta \mu_2 . \label{delhhh}
\end{align}
Notice here that $\delta \alpha$ and $\delta \beta$ are correctly replaced by the corresponding pinched counterterms $\delta \bar{\alpha}$ and $\delta \bar{\beta}$.

In Eq.~(\ref{delhhh}), the counterterms $\delta\mu_1$ and $\delta\mu_2$ show up and cannot be individually determined by applying the on-shell scheme, as alluded to in the beginning of the section. 
A similar situation also happens in THDMs~\cite{THDM1} and the HSM~\cite{HSM3}. 
In this paper, we apply the MS scheme to fix the combination of $\delta \mu_1$ and $\delta \mu_2$, where these counterterms are determined so as to cancel only the UV divergent part 
$\Delta_{\text{div}}$  of $\delta\Gamma_{hhh}$ (without the $\delta\mu_1$ and $\delta\mu_2$ terms), $\Gamma_{hhh}^{\text{1PI}}$ and $T_{hhh}$. 
Here, $\Delta_{\text{div}} \equiv 1/\epsilon + \ln4\pi-\gamma_E+\ln\mu^2$ with $\mu$ and $\epsilon$ being defined in Appendix~\ref{sec:lf} and $\gamma_E$ being the Euler-Mascheroni constant. 
The same method has also been applied to fix the counterterm for the $hhh$ vertex in THDMs~\cite{THDM1,THDM4} and that in the HSM~{\cite{HSM3}}.  
%

\subsection{Renormalized Higgs boson couplings  \label{sec:hcoup}}

We can now calculate the renormalized Higgs boson vertices ({\it i.e.}, $hVV$, $hf\bar{f}$ and $hhh$ vertices) from the discussions in the previous subsections. 
We here define the renormalized scale factors $\hat{\kappa}_X^{}$ for the Higgs boson couplings, which are convenient to discuss the deviation in the couplings from the SM predictions, as follows:
\begin{align}
\begin{split}
\hat\kappa_V(p^2) &\equiv 
\frac{\hat\Gamma_{hVV}^1(m_V^2,p^2,m_h^2)_\text{GM}}{\hat\Gamma_{hVV}^1(m_V^2,p^2,m_h^2)_\text{SM}}~,
\\
\hat\kappa_t(p^2) &\equiv 
\frac{\hat\Gamma_{htt}^S(m_t^2,p^2,m_h^2)_\text{GM}}{\hat\Gamma_{htt}^S(m_t^2,p^2,m_h^2)_\text{SM}}~,
\\
\hat\kappa_h(p^2) &\equiv
\frac{\hat\Gamma_{hhh}(m_h^2,m_h^2,p^2)_\text{GM}}{\hat\Gamma_{hhh}(m_h^2,m_h^2,p^2)_\text{SM}}~,
\end{split} \label{kappa-def1}
\end{align}
where $p^2$ denotes the squared momentum for the off-shell particle, namely, $V^*$, $t^*$ and $h^*$ for the $hVV$, $ht\bar{t}$ and $hhh$ couplings, respectively. 
For the $hb\bar{b}$ and $h\tau^+\tau^-$ couplings, we define their renormalized scale factors without the momentum dependence since the on-shell decays $h \to b\bar{b}$ and $h \to \tau^+\tau^-$
are allowed: 
\begin{align}
\hat\kappa_{f} \equiv\frac{\hat\Gamma_{hff}^S(m_{f}^2,m_{f}^2,m_h^2)_\text{GM}}{\hat\Gamma_{hff}^S(m_{f}^2,m_{f}^2,m_h^2)_\text{SM}}~~~\text{for}~~~f = b,\tau. \label{kappa-def2}
\end{align}
For the numerical evaluation of these scale factors, we use the following SM input parameters~\cite{pdg}:
\begin{align}
G_F &= 1.166379\times10^{-5} \ \text{GeV}^{-2}~, &
\alpha_\text{EM}^{-1} &= 137.035999~, &
\Delta\alpha_\text{EM} &= 0.06635 ~,
\notag\\
m_Z &= 91.1876 \ \text{GeV}~, &
m_h &= 125 \ \text{GeV} ~, &
m_t &= 173.21 \ \text{GeV}~,
\\
m_b &= 4.66 \ \text{GeV}~, &
m_c &= 1.275 \ \text{GeV}~, &
m_\tau &= 1.77684 \ \text{GeV}~,
\notag
\end{align}
where all the other quarks and leptons are assumed to be massless. 

\begin{figure}[t]
\centering
\includegraphics[scale=0.3]{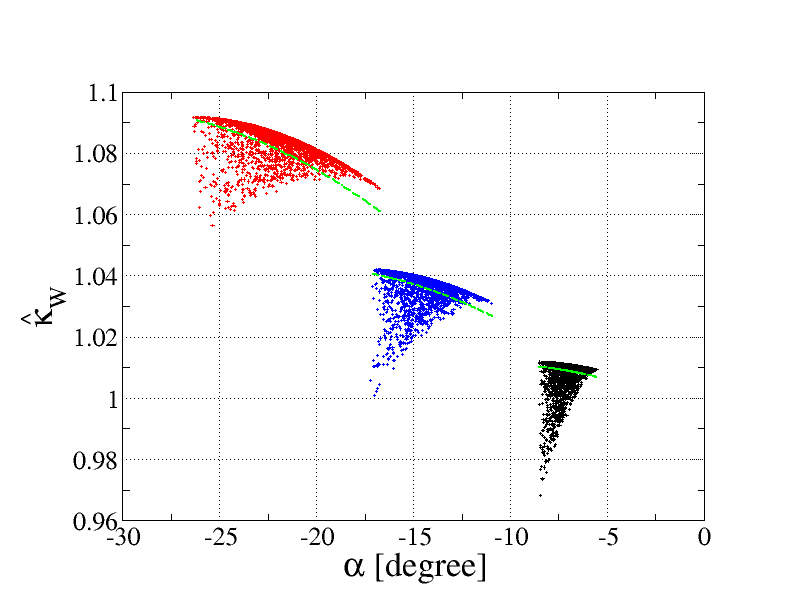}\hspace{-5mm}
\includegraphics[scale=0.3]{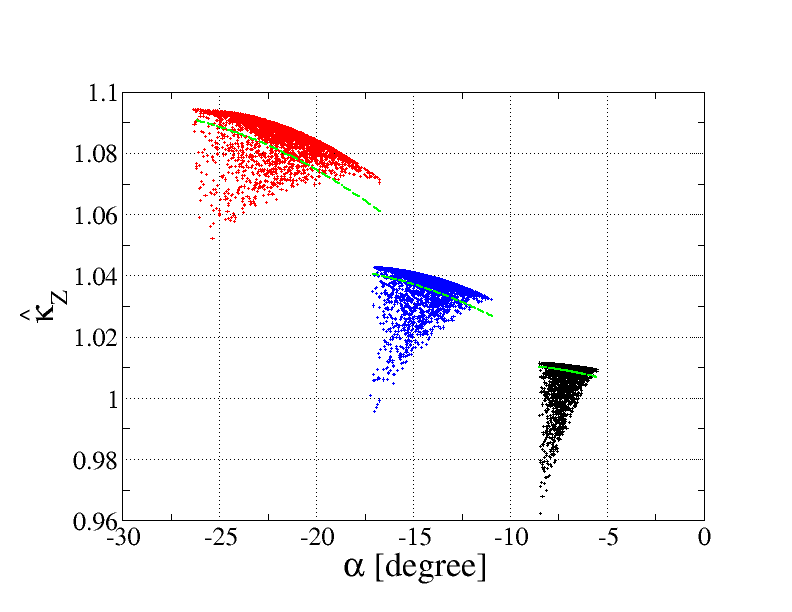} \\
\includegraphics[scale=0.3]{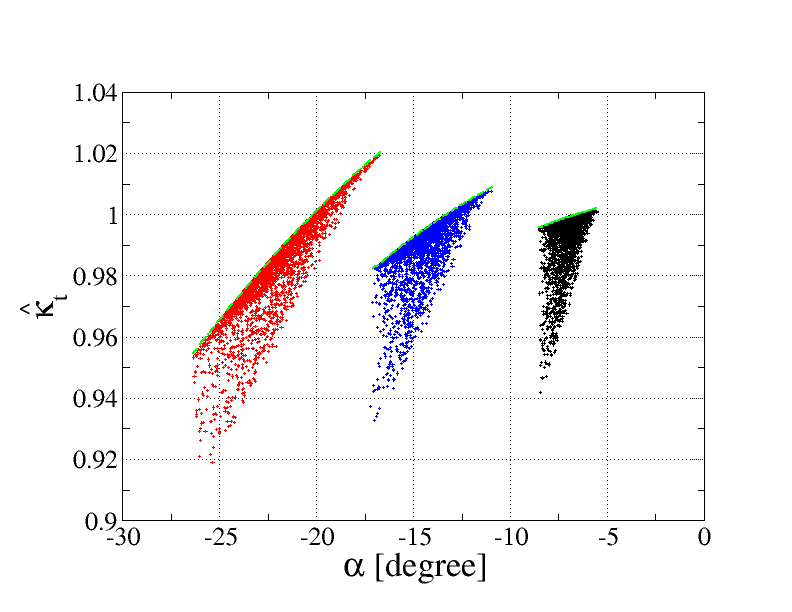}\hspace{-5mm}
\includegraphics[scale=0.3]{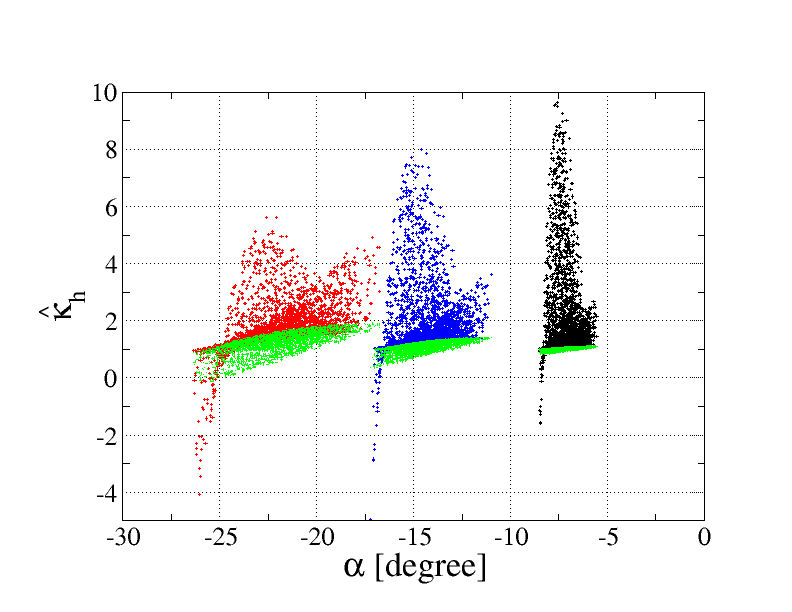} 
\caption{Renormalized scale factors $\hat{\kappa}_W$ (upper left), 
$\hat{\kappa}_Z$ (upper right),
$\hat{\kappa}_t$ (lower left) and 
$\hat{\kappa}_h$ (lower left) as functions of $\alpha$ for $m_{H_5}^{} =m_{H_3}^{} = m_{H_1}^{} =400$~GeV, where the squared momentum is taken to be (250 GeV)$^2$ for $\hat{\kappa}_W$ and $\hat{\kappa}_Z$ and (500 GeV)$^2$ for $\hat{\kappa}_t$ and $\hat{\kappa}_h$.  In these plots, $v_\Delta^{}$ is fixed to be 10~GeV (black), 20~GeV (blue) and 30~GeV (red), while $\mu_1$ and $\mu_2$ are scanned. 
The tree-level predictions are also shown by the dashed green curves (for $\hat{\kappa}_W$, $\hat{\kappa}_Z$) and $\hat{\kappa}_t$  and green dots (for $\hat{\kappa}_h$). 
We only show the points allowed by the constraints of perturbative unitarity and vacuum stability, to be described in Section~\ref{sec:theory}. }
\label{kappas}
\end{figure}

In order to understand the typical behaviors of the renormalized scale factors, 
we show $\hat{\kappa}_{W}$, $\hat{\kappa}_{Z}$, $\hat{\kappa}_{t}$ and $\hat{\kappa}_{h}$ as functions of the mixing angle $\alpha$ in Fig.~\ref{kappas}. 
In these plots,  the squared momentum is fixed to be $(250~\text{GeV})^2$ for $\hat{\kappa}_{W,Z}$ and $(500~\text{GeV})^2$ for $\hat{\kappa}_{t,h}$. 
Moreover, the masses of the extra Higgs bosons are taken to be 400~GeV in common, and the triplet VEV $v_\Delta$ is set to be 10~GeV (black), 20~GeV (blue) and 30~GeV (red). 
The other parameters $\mu_1$ and $\mu_2$ are scanned in ranges that are taken as large as possible so as to maximize the allowed region of $\alpha$ for a fixed value of $v_\Delta$. 
In addition, we impose the constraints from perturbative unitarity and vacuum stability, to be described in Section~\ref{sec:theory} in the parameter scan.  The dots and curves shown in this figure satisfy both of these constraints.

We see that larger values of $\hat{\kappa}_Z$ and $\hat{\kappa}_W$ are obtained for larger values of $v_\Delta^{}$ and $|\alpha|$. 
This behavior can roughly be explained by the dominant tree-level prediction shown by the green dashed curves, determined solely by $\alpha$ and $\beta$ (or equivalently $v_\Delta$) in Eq.~(\ref{c5_1}). 
That is, the term proportional to $2\sqrt{6}/3$ in $c_{hVV}^{}$ becomes important for larger $v_{\Delta}$ and $|\alpha|$. 
We also see that the quantum effect typically reduces the values of $\hat{\kappa}_Z$ and $\hat{\kappa}_W$ by a few percent with respect to the tree-level predictions, where the most important source of quantum corrections comes from the counterterm for the $h$ wave function $\delta Z_h$ in Eq.~(\ref{reno-hvv2}). 
In fact, the 5-plet Higgs boson loop contributions to $\delta Z_h$ provide a term proportional to $\lambda_{H_5H_5h}^2$ (see Eq.~(\ref{pihh}) and notice $\delta Z_h = -\Pi_{hh}'(m_h^2)$) 
defined in Eq.~(\ref{lam3p}), which can be significant depending on $\mu_1$ and $\mu_2$, 
and it determines the typical size of quantum corrections.

Similar to $\hat{\kappa}_{W}^{}$ and $\hat{\kappa}_{Z}^{}$, the behavior of $\hat{\kappa}_t$ is roughly determined by the tree-level prediction, {\it i.e.}, $c_{hff}$ given in Eq.~(\ref{chff}). 
In fact, it is seen that $\hat{\kappa}_t$ becomes small when we take larger values of $v_\Delta$ and $|\alpha|$. 
In addition, the quantum correction reduces $\hat{\kappa}_t$, mainly because of the effect of $\delta Z_h$. 
We note that the predictions for $\hat{\kappa}_b$ and $\hat{\kappa}_\tau$ are almost the same as that of $\hat{\kappa}_t$. 

For $\hat{\kappa}_h$, there are several features different from $\hat{\kappa}_{W}^{}$, $\hat{\kappa}_Z^{}$ and $\hat{\kappa}_t$. 
First of all, the tree-level prediction, shown by the green dots, is not a single-valued curve, but spreads over a region on the plane. 
This is because $\hat{\kappa}_h$ depends not only on $\alpha$ and $v_\Delta$ but also on $\mu_1$ and $\mu_2$ as seen in Eq.~(\ref{ren_hhh}). 
Secondly, $\hat{\kappa}_h$ can receive a large quantum correction at several $100\%$ level with respect to the tree-level prediction. 
This large correction can be ascribed to the $\lambda_{H_5H_5h}^3$ dependence in the 1PI diagram contribution, see Eq.~(\ref{1pi_hhh}), when the value of $|\alpha|$ is not close to its maximum for a given value of $v_\Delta$. 

While the predicted scale factors presented here are only for some special cases, we will show their behaviors in more generic cases in Section~\ref{sec:numana}. 

\begin{table}[t]
\begin{center}
\begin{tabular}{l||cccc||cccc}\hline\hline 
           & $\alpha$ [degree]  & $v_\Delta^{}$ [GeV] & $\mu_1$ [GeV] & $\mu_2$ [GeV] & ~~~$\hat{\kappa}_W$~~~ & ~~~$\hat{\kappa}_Z$~~~   & ~~~$\hat{\kappa}_t$~~~ & ~~~$\hat{\kappa}_h$~~~ \\\hline
BP1        & $-7.0$          & 10                 & $-100.3$      & $112.3$       & $1.01$           & $1.01$            & $1.00$           &  $1.08$ \\\hline
BP2        & $-8.0$          & 10                 & $7.1$         & $2789.5$      & $0.99$           & $0.98$            & $0.96$           &  $4.71$ \\\hline
BP3        & $-15.1$         & 20                 & $-180.0$      & $171.0$       & $1.04$           & $1.04$            & $0.99$           &  $1.34$ \\\hline
BP4        & $-16.1$         & 20                 & $18.7$        & $1338.3$      & $1.02$           & $1.01$            & $0.95$           &  $4.00$ \\\hline
BP5        & $-22.4$         & 30                 & $-325.2$      & $-53.3$       & $1.09$           & $1.09$            & $0.98$           &  $1.39$ \\\hline
BP6        & $-24.9$         & 30                 & $10.0$        & $755.0$       & $1.07$           & $1.06$            & $0.93$           &  $0.87$ 
\\\hline\hline
\end{tabular} 
\end{center}
\caption{Six benchmark points allowed by the perturbative unitarity and the vacuum stability. 
The masses of the extra Higgs bosons are taken to be $m_{H_5}^{} =m_{H_3}^{} = m_{H_1}^{} =400$ GeV. All the other input parameters are shown in the first four columns.  
The numbers given in the latter four columns show the output of the renormalized scale factors at $\sqrt{p^2}=250$ (500) GeV for $\hat{\kappa}_{W,Z}$ ($\hat{\kappa}_{t,h}$).} \label{benchmark}
\label{bps}
\end{table}

Finally, we discuss the momentum dependence of $\hat{\kappa}_{W}^{}$, $\hat{\kappa}_Z^{}$, $\hat{\kappa}_t$ and $\hat{\kappa}_h$. 
We provide six benchmark points (BP1--BP6), all of which are allowed by both the constraints of perturbative unitarity and vacuum stability. 
In Table~\ref{bps}, we show the input parameters of BP1--BP6 and the output values of the renormalized scale factors at $\sqrt{p^2}=250$ (500)~GeV for $\hat{\kappa}_{W,Z}$ ($\hat{\kappa}_{t,h}$).
BP1, BP3 and BP5 (BP2, BP4 and BP6) are chosen such that the predictions of one-loop corrected scale factors are close to (far from) the tree-level predictions for $v_\Delta^{}=10$, 20 and 30~GeV, respectively.

\begin{figure}[t]
\centering
\hspace{-8mm}
\includegraphics[scale=0.21]{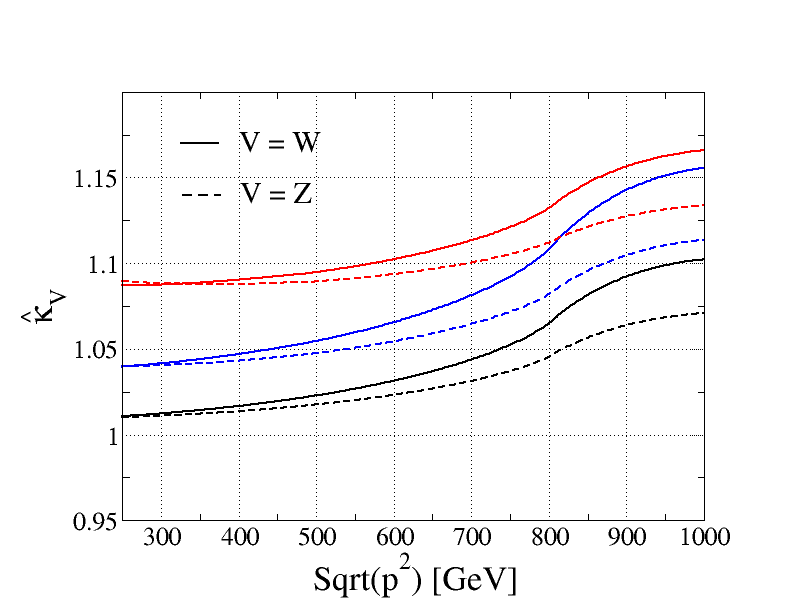}\hspace{-6mm}
\includegraphics[scale=0.21]{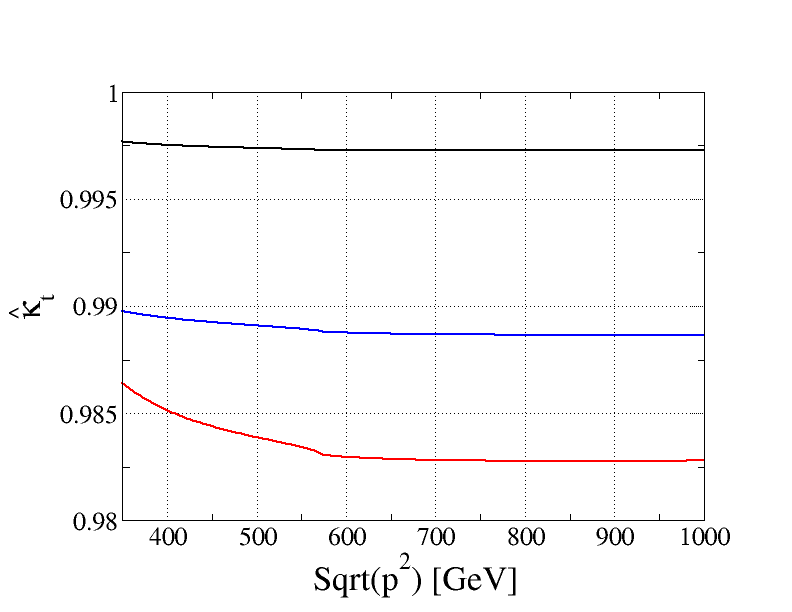}\hspace{-6mm}
\includegraphics[scale=0.21]{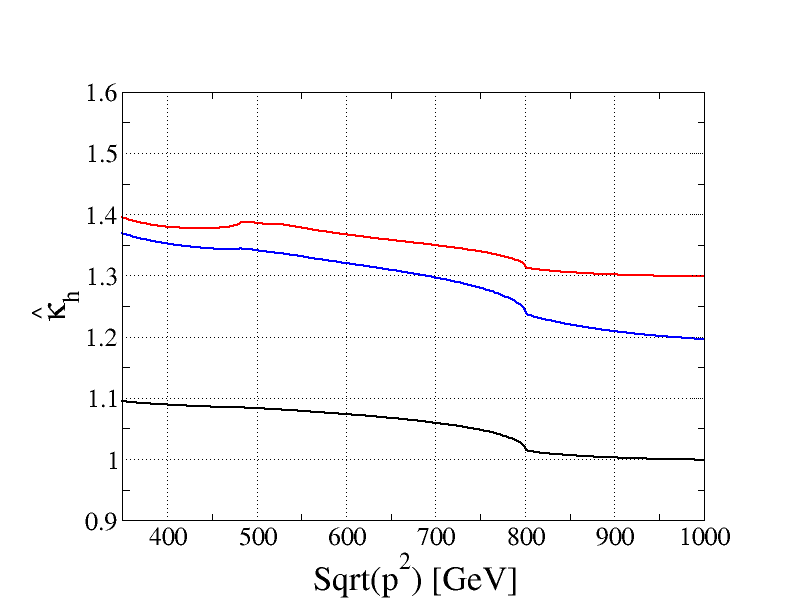}\\
\hspace{-8mm}
\includegraphics[scale=0.21]{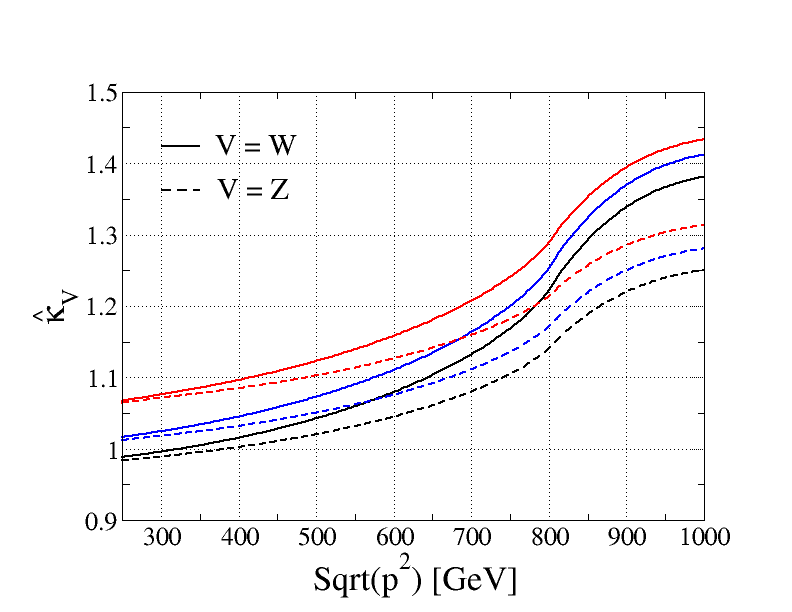}\hspace{-6mm}
\includegraphics[scale=0.21]{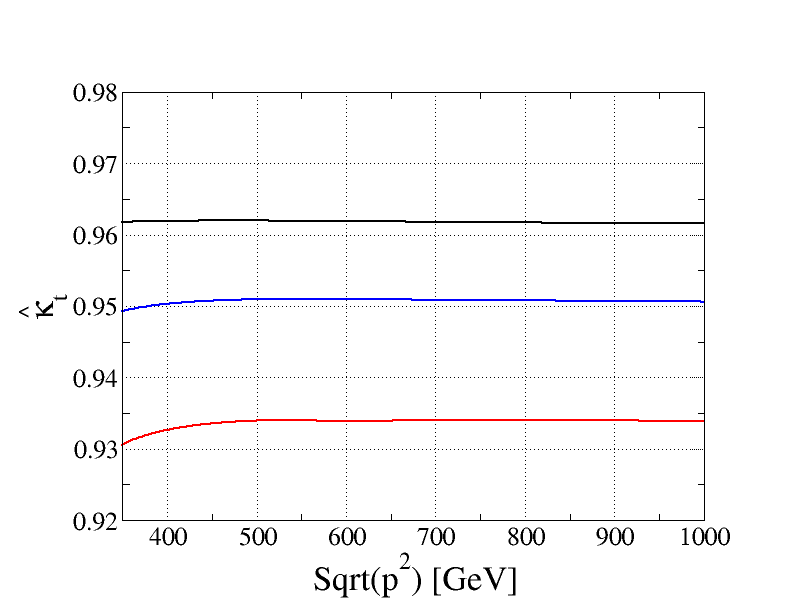}\hspace{-6mm}
\includegraphics[scale=0.21]{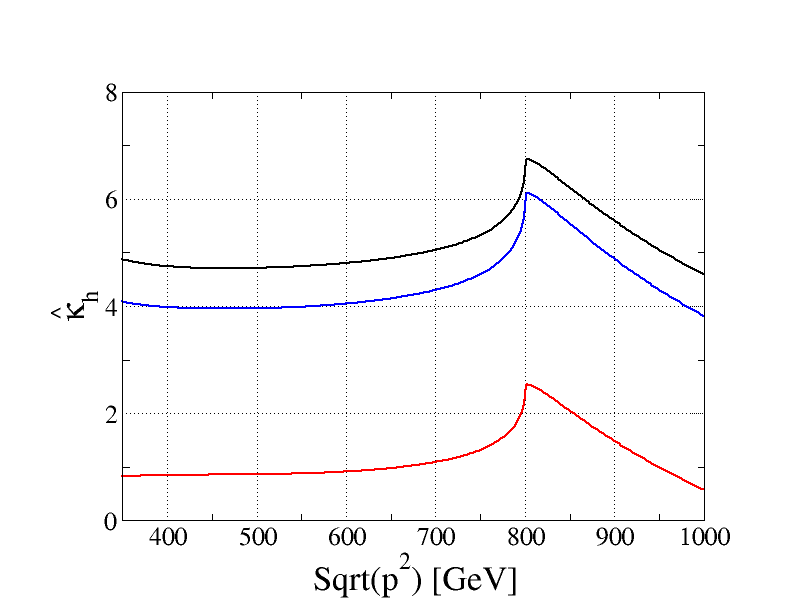}
\caption{Renormalized scale factors $\hat\kappa_{W,Z}$ (left), $\hat\kappa_t$ (middle) and $\hat\kappa_h$ (right) 
as functions of $\sqrt{p^2}$ for $m_{H_5}^{} =m_{H_3}^{} = m_{H_1}^{} =400$~GeV. 
The upper panels show the cases of BP1 (black), BP3 (blue) and BP5 (red), while the lower panels show the cases of
BP2 (black), BP4 (blue) and BP6 (red). }
\label{momentum-dep}
\end{figure}

In Fig.~\ref{momentum-dep}, the momentum dependence of $\hat\kappa_{V}^{}$, $\hat\kappa_t$ and $\hat\kappa_h$ are shown for the six benchmark points. 
For $\hat\kappa_V^{}$, both $\hat\kappa_W^{}$ and $\hat\kappa_Z^{}$ monotonically increase with $\sqrt{p^2}$, where 
the increasing rates for BP2, BP4 and BP6 are more significant as compared to those for BP1, BP3 and BP5. 
We also observe that the increasing rates becomes slightly higher at $\sqrt{p^2}\gtrsim 800$~GeV because of the threshold effects of the extra Higgs bosons. 
In addition, the difference between $\hat\kappa_W^{}$ and $\hat\kappa_Z^{}$ is getting larger as $\sqrt{p^2}$ increases. 
On the other hand, the momentum dependence of $\hat\kappa_t$ is quite mild with compared to that of $\hat\kappa_V^{}$, where 
the results for BP3 and BP6, having larger values of $|\alpha|$ and $v_\Delta^{}$, show more significant momentum dependence as compared to the other benchmark points. 
For $\hat\kappa_h$, we see the sharp dip (peak) for BP1, BP3 and BP5 (BP2, BP4 and BP6) at around $\sqrt{p^2}=800$~GeV due to the threshold effects of the extra Higgs boson loops, where the appearance of the dip/peak can be explained by the destructive/constructive interference of the $\lambda_{H_5H_5h}^3$ and $\lambda_{H_5H_5h}\lambda_{H_5H_5hh}$ terms in Eq.~(\ref{1pi_hhh}), {\it i.e.}, 
the signs of these two contributions flip between (BP1, BP3, BP5) and (BP2, BP4, BP6).

\section{Constraints on Parameter Space \label{sec:constraints}}

In this section, we discuss both theoretical and experimental constraints that we impose on the model parameters.
A search of viable exotic Higgs boson mass spectra based upon similar constraints and prospects for detecting the doubly-charged Higgs boson at the 14-TeV LHC and a 100-TeV future $pp$ collider had been studied in Ref.~\cite{Chiang:2015amq}.

\subsection{Theoretical bounds \label{sec:theory}}

Two theoretical constraints are taken into account to constrain the dimensionless quartic couplings of the scalar potential at tree level: the stability of the electroweak vacuum and the unitarity of the perturbation theory.  These constraints on the quartic couplings can be translated into bounds on the physical parameters such as the masses and mixing angles of the Higgs bosons through the relations given in Appendix~\ref{sec:mass}.

The vacuum stability requires the scalar potential to be bounded from below and leads to the following constraints for the quartic couplings~\cite{Logan}:
\begin{align}
\begin{split}
&\l_1>0~,\\
&\lambda_2>
\left\{
\begin{array}{l}
-\frac{1}{3}\l_3 \ \text{ for }\l_3\geq0~,
\\
-\l_3  \ \ \ \text{ for }\l_3<0~,
\end{array}
\right.\\
&\l_4>
\left\{
\begin{array}{l}
-\frac{1}{2}\l_5-2\sqrt{\l_1(\frac{1}{3}\lambda_3+\lambda_2)} \qquad\qquad
\text{ for }\l_5\leq0\text{ and }\l_3\geq0~,
\\
-\omega_+(\zeta)\l_5-2\sqrt{\l_1(\zeta\l_3+\l_2)}  \qquad
\text{ for }\l_5\leq0\text{ and }\l_3<0~,\\
-\omega_-(\zeta)\l_5-2\sqrt{\l_1(\zeta\l_3+\l_2)}  \qquad
\text{ for }\l_5>0~,
\end{array}
\right.
\end{split}
\label{eq:stability}
\end{align}
where 
\begin{align}
\omega_\pm(\zeta)=\frac{1}{6}(1-B)\pm\frac{\sqrt2}{3}\left[\left(1-B\right)\left(\frac{1}{2}+B\right)\right]^{1/2}~,
\end{align}
with $B$ randomly varying between 0 and 1.

The bound from perturbative unitarity is obtained by requiring that the $s$-wave amplitude matrix, $a_0$, for elastic $2 \to 2$ scalar boson scatterings does not become too large to violate $S$-matrix unitarity. 
One can set the criteria for this requirement as that the magnitudes of all the eigenvalues of $a_0$ do not exceed $1/2$. 
In the high-energy limit, the matrix elements of $a_0$ are expressed by the scalar quartic couplings because only the diagrams involving scalar contact interactions are relevant. 
In this setup, one can obtain the following conditions~\cite{Logan,Aoki:2007ah}:
\begin{align}
\begin{split}
&
\left\vert \, 6 \l_1 + 7 \l_3 + 11\l_2\,  \pm \sqrt{(6\l_1-7\l_3-11\l_2)^2+36\l_4^2}\, \right\vert < 4\pi ~,
\\
&
\left\vert \, 2 \l_1 -  \l_3 + 2\l_2\,  \pm \sqrt{(2\l_1+\l_3-2\l_2)^2+\l_5^2}\, \right\vert < 4\pi ~
\\
&
\left\vert \,  \l_4 + \l_5 \, \right\vert < 2\pi ~,
~~~ \left\vert \, 2 \l_3 + \l_2  \, \right\vert < \pi ~,
\\
&\left\vert \,  2\lambda_2 + \lambda_3 \, \right\vert < 2\pi ~,
~~~ \left\vert \, 4 \lambda_4 + \lambda_5  \, \right\vert < 8\pi ~,~~~ \left\vert \, 2 \lambda_4 - \lambda_5  \, \right\vert < 4\pi ~.
\end{split}
\label{eq:unitarity}
\end{align}
We note that by combining the vacuum stability condition, the first 2 inequalities of (\ref{eq:unitarity}) can be simply replaced by 
\begin{align}
\begin{split}
&
\left\vert \, 6 \l_1 + 7 \l_3 + 11\l_2\, \right\vert  + \sqrt{(6\l_1-7\l_3-11\l_2)^2+36\l_4^2} < 4\pi ~,
\\
&
\left\vert \, 2 \l_1 -  \l_3 + 2\l_2\, \right\vert  + \sqrt{(2\l_1+\l_3-2\l_2)^2+\l_5^2} < 4\pi.
\end{split}
\end{align}

\subsection{Experimental bounds}

Next, we discuss the experimental constraints from the electroweak oblique $S$ parameter, 
the signal strengths for the 125 GeV Higgs boson, and the direct searches for extra Higgs bosons. 
We note that the oblique $T$ parameter is used as one of the inputs (see the discussion in section~\ref{sec:reno-gauge}), so that it cannot be applied to the constrain the GM model. 
We require that predictions of these observables in the model be within the $95\%$ confidence level (CL) region.  
In the following, we explain how these constraints from experimental data are imposed in our analysis, in order. 

The current electroweak data fit gives~\cite{pdg}
\begin{align}
S=0.07\pm0.08~,
\end{align}
by fixing $U=0$.

The Higgs signal strengths have been measured from 20 channels with different combinations of production and decay channels in Ref.~\cite{signalstrength}. 
Among these measurements, we do not include the signal strengths for the $Zh$ and $th$ productions with the $h \to WW^*$ decay in our study because the SM predictions for these two channels are excluded by the current data at $95\%$ CL.  
It should be noted that channels with $h$ decaying into a pair of photons provide effective constraints on the masses of extra Higgs bosons as their dependences appear in the charged Higgs boson ($H_3^\pm$, $H_5^\pm$ and $H_5^{\pm\pm}$) loop contributions to the $h \to \gamma\gamma$ decay.  In contrast, all the other channels depend on only two parameters: $\alpha$ and $\beta$.

The constraint from the direct search for the doubly-charged Higgs boson $H_5^{\pm\pm}$ is imposed.  A bound on the production cross section of $H_5^{\pm\pm}$ via the vector boson fusion mechanism times the branching fraction of the $H_5^{\pm\pm}\to W^\pm W^\pm$ decay has been set by the CMS Collaboration at the collision energy of 8~TeV and the integrated luminosity of $19.4~\text{fb}^{-1}$~\cite{directsearch}.  This constraint can be translated into the bound on the mass of the 5-plet Higgs bosons $m_{H_5}^{}$ and the triplet VEV $v_\Delta^{}$.

\subsection{Allowed parameter space \label{sec:allowed}}

We are now ready to present the allowed parameter space by imposing the constraints discussed in the previous subsections.

After fixing $v$ and $m_h$, there are totally seven independent free parameters  in the GM model as shown in Eq.~(\ref{para}), assuming the custodial symmetry at tree level.  
Instead of using the parameters given in Eq.~(\ref{para}), we choose four dimensionless quartic couplings $\l_{2-5}$ and three dimensionful parameters $\mu_{1,2}$ and $m_\Delta^{}$ in the Higgs potential as our inputs, with which all the other parameters are determined.  
We then perform a scan of the parameters in the following ranges:
\begin{align}
&-0.628 \leq\l_2\leq1.57~,\quad-1.57 \leq\l_3\leq1.88~,\quad-2.09 \leq\l_4\leq2.09~,\quad-8.38\leq\l_5\leq8.38~ \notag\\
&-650 \leq\mu_1\leq0 \ \text{GeV}~,\quad-400 \leq\mu_2\leq50 \ \text{GeV}~,\quad180 \leq m_\Delta\leq450 \ \text{GeV}~.
\label{eq:scanrange}
\end{align}
The ranges of $\l_{2,3}$ are determined by the constraints from perturbative unitarity and vacuum stability, while 
those of $\l_{4,5}$ are determined by the bounds from the perturbative unitarity only~\cite{Logan}.
The parameter scan is performed under two sets of constraints: Set-A takes into account the constraints of vacuum stability, 
perturbative unitarity and the $S$ parameter, and Set-B further considers the Higgs signal strengths and direct search of $H_5^{\pm\pm}$, all at $95\%$ CL.

\begin{figure}[!t]
\centering
\includegraphics[scale=0.55]{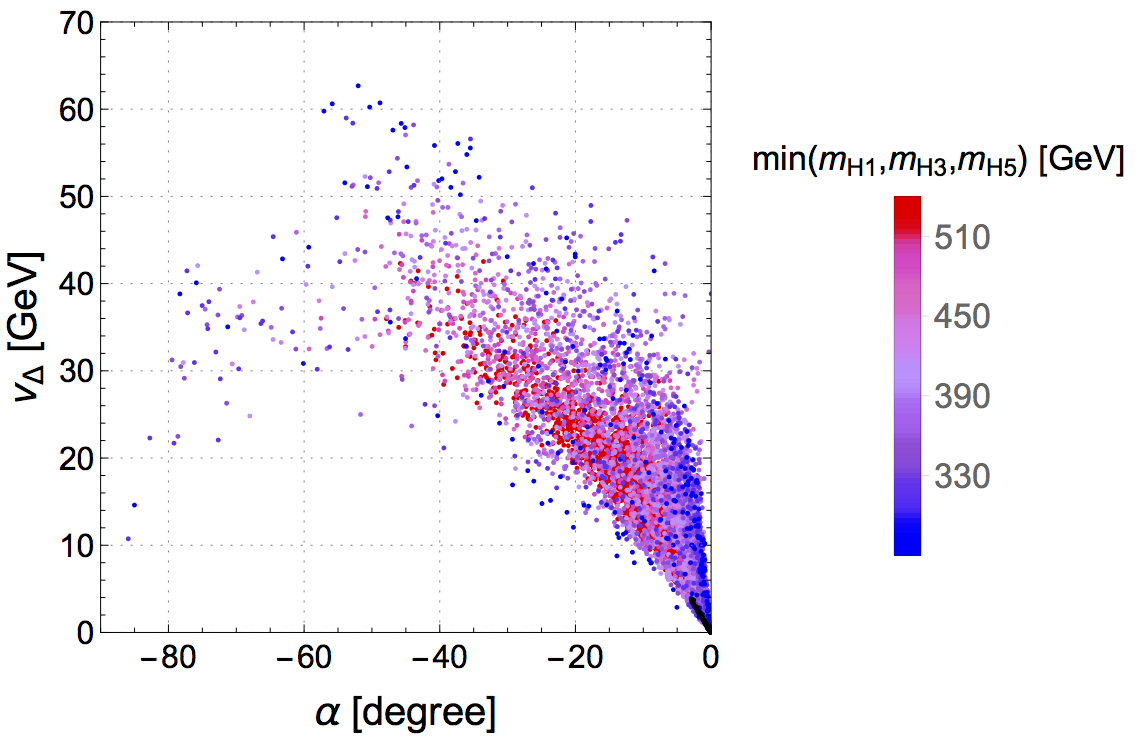}
\includegraphics[scale=0.55]{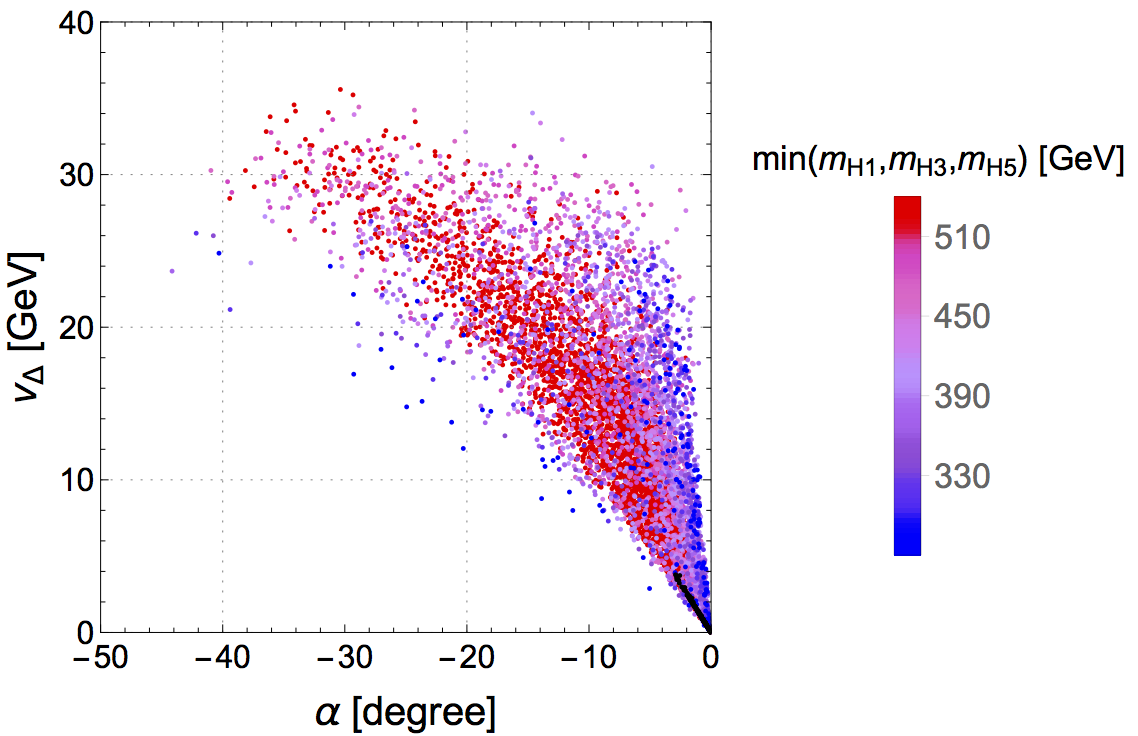}
\caption{Points allowed by Set-A constraints (left plot) and Set-B constraints (right plot). 
The color of the dots shows the value of min($m_{H_1}$, $m_{H_3}$, $m_{H_5}$).  The black dots are obtained by changing the scan range of $m_\Delta$ to $(1,2.5)$~TeV while keeping the scan ranges of all the other parameters as in Eq.~(\ref{eq:scanrange}).
}
\label{basis}
\end{figure}

In Fig.~\ref{basis}, points allowed by Set-A constraints (left plot) and Set-B constraints (right plot) are shown in the $\alpha$-$v_\Delta$ plane.  The color of the dots indicates the value of min($m_{H_1}$,$m_{H_3}$,$m_{H_5}$), the minimum of the exotic Higgs boson masses. 
It is seen that Set-B constraints exclude regions with larger values of $|\alpha|$ and $v_\Delta^{}$ in comparison with using only Set-A constraints. 
These two parameters are constrained to be $- 40^\circ \alt \alpha \alt 0^\circ$  and $v_\Delta^{} \alt 35$~GeV under Set-B constraints. 
We can also see from the right plot that for a fixed smaller value of $|\alpha|$ ({\it e.g.}, $|\alpha| \lesssim 10^\circ$), the maximally allowed value of min($m_{H_1}$,$m_{H_3}$,$m_{H_5}$) becomes smaller as $v_\Delta^{}$ increases. 
On the other hand, larger values of $v_\Delta^{}$ and min($m_{H_1}$,$m_{H_3}$,$m_{H_5}$) can be found for larger $|\alpha|$. 
In order to numerically check the decoupling behavior, we also add the black dots which are obtained by scanning 
$1\leq m_\Delta\leq 2.5$~TeV while keeping the scan ranges of all the other parameters as in Eq.~(\ref{eq:scanrange}). 
These black dots also correspond to min($m_{H_1}$,$m_{H_3}$,$m_{H_5}$) from 1.4 TeV to 3.5 TeV.
As expected, the decoupling limit $m_\Delta \gg v$ resides in the region where both $|\alpha|$ and $v_\Delta$ approach zero.

\section{Numerical Results for Renormalized Higgs Boson Couplings \label{sec:numana}}

In this section, we numerically show the deviations in the one-loop corrected $hVV$, $hf\bar{f}$ and $hhh$ couplings from the corresponding SM predictions under the scan defined in Eq.~(\ref{eq:scanrange}). 
These deviations are expressed in terms of the renormalized scale factors $\hat\kappa_X$ defined in Eqs.~(\ref{kappa-def1}) and (\ref{kappa-def2}). 
In the following analysis, we restrict ourselves to the parameter region allowed by Set-B constraints defined in Section~\ref{sec:allowed}.

\begin{figure}[t]
\centering
\includegraphics[scale=0.7]{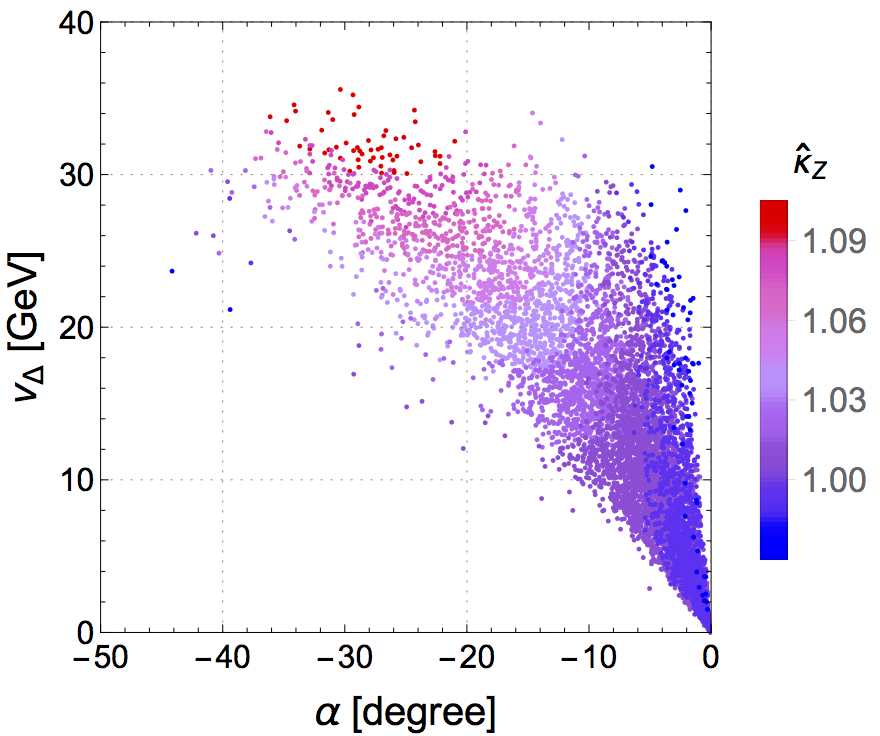}
\includegraphics[scale=0.7]{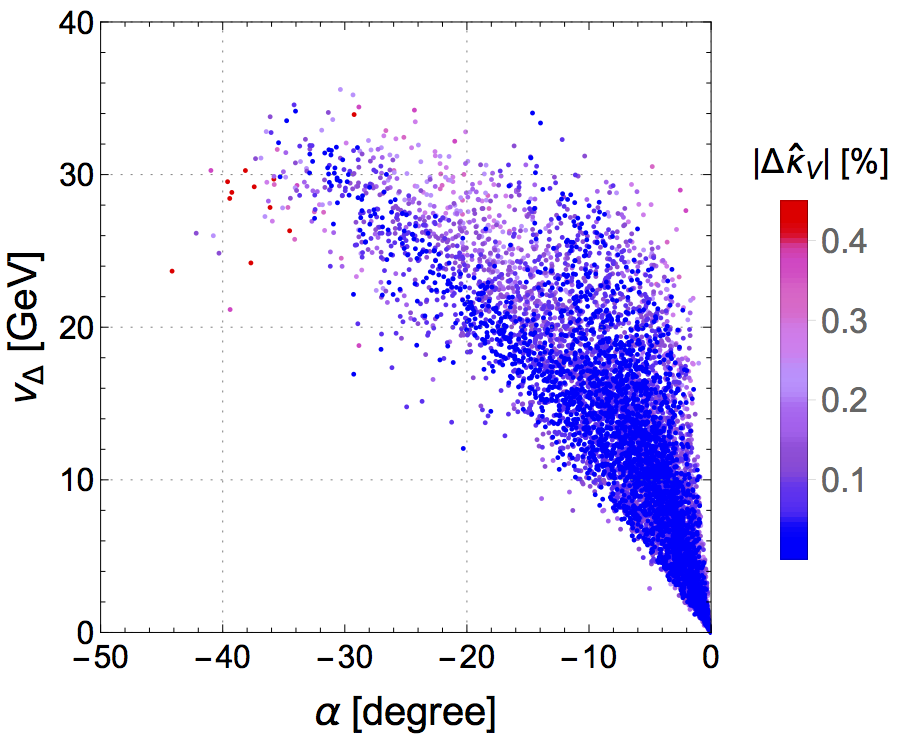}
\includegraphics[scale=0.7]{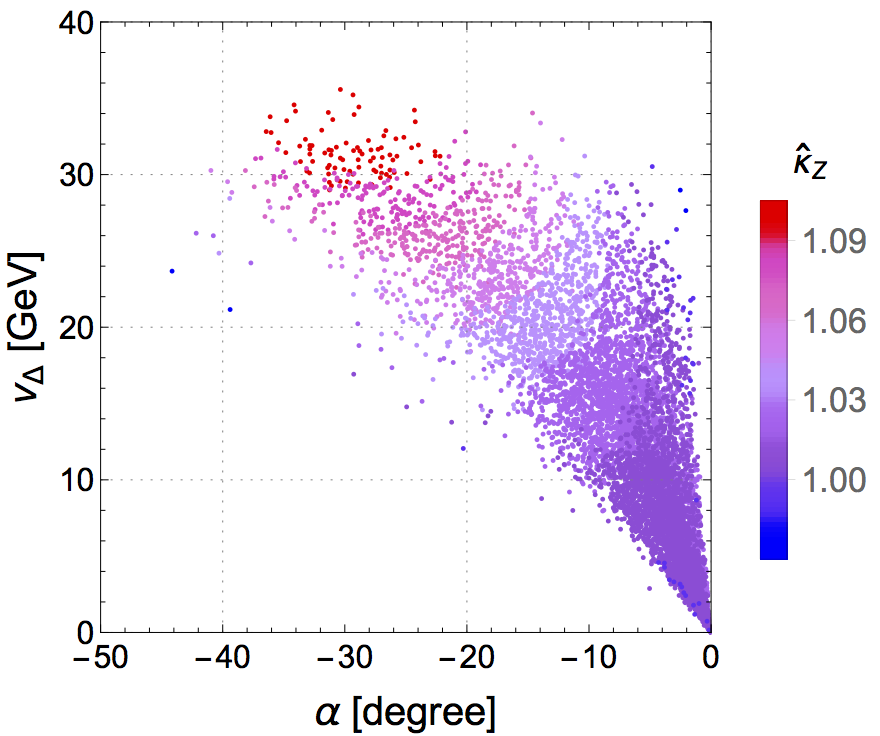}
\includegraphics[scale=0.7]{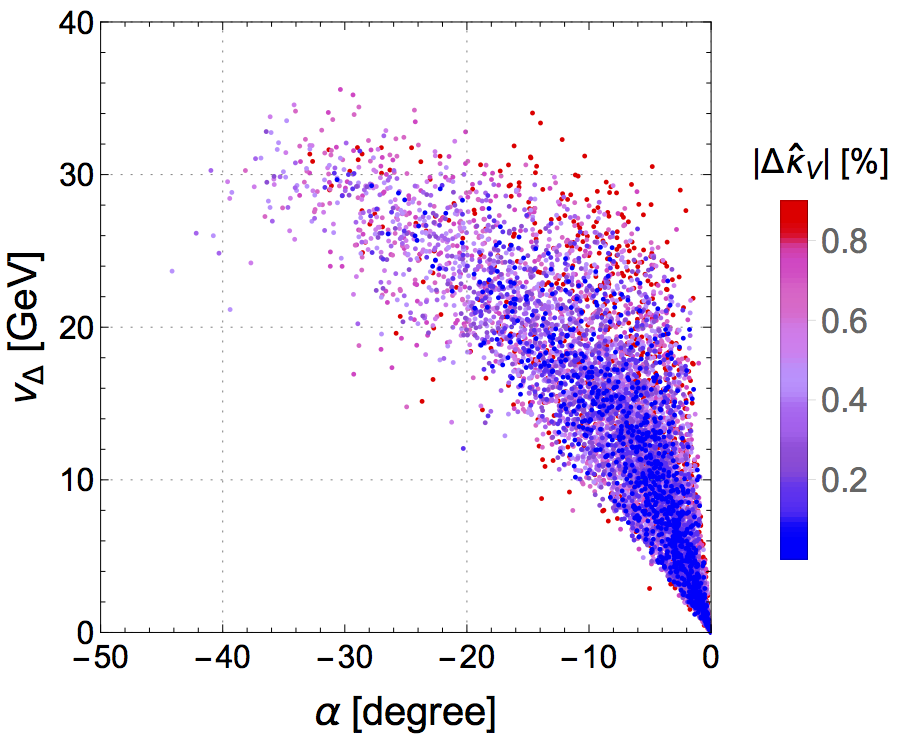}
\caption{Scatter plots of $\hat\kappa_Z^{}$ (left) and $\left|\Delta\hat\kappa_V\right|$ (right) for $\sqrt{p^2}=250$~GeV (upper) and $\sqrt{p^2}=500$~GeV (lower) in the $\alpha$--$v_\Delta$ plane.  }
\label{kvscatter}
\end{figure}

First, we show the behavior of the scale factors for the Higgs couplings with weak gauge bosons. 
Moreover, we define
\begin{align}
\Delta\hat\kappa_V(p^2)\equiv\hat\kappa_Z(p^2)-\hat\kappa_W(p^2). 
\end{align}
Fig.~\ref{kvscatter} shows the scatter plots of $\hat\kappa_Z$ (left plots) and $\left|\Delta\hat\kappa_V\right|$ (right plots) for $\sqrt{p^2}=250$~GeV (upper plots) and $500$~GeV (lower plots) in the $\alpha$--$v_\Delta$ plane.  
It is clear from the left plots that larger values of $\hat\kappa_Z$ are obtained in the region with larger values of $v_\Delta^{}$ and $\left|\alpha\right|$. 
The result for $\hat\kappa_Z$ does not change much as we change from $\sqrt{p^2} = 250$~GeV to 500~GeV, in agreement with the special case in Fig.~\ref{momentum-dep}. 
We note that within our parameter scan ranges, $\hat\kappa_Z$ varies from $0.88~(0.93)$ to $1.12~(1.13)$ for $\sqrt{p^2} = 250$ (500)~GeV. 
On the other hand, from the right plots we see that maximal $|\Delta\hat\kappa_V |$ is typically around 0.2\% for $\sqrt{p^2} = 250$~GeV, while the maximum becomes around 0.8\% for $\sqrt{p^2} = 500$~GeV.  It is also seen that $|\Delta\hat\kappa_V|$ does not depend on $v_\Delta$ and $\alpha$ so much. 
We note that $\Delta\hat\kappa_V$ can be either positive or negative, and it falls in the range of $-0.7\%$ to $0.4\%$ for $\sqrt{p^2}=250$~GeV and $0.0\%$ to $1.45\%$ for $\sqrt{p^2}=500$~GeV.

\begin{figure}[t]
\centering
\includegraphics[scale=0.7]{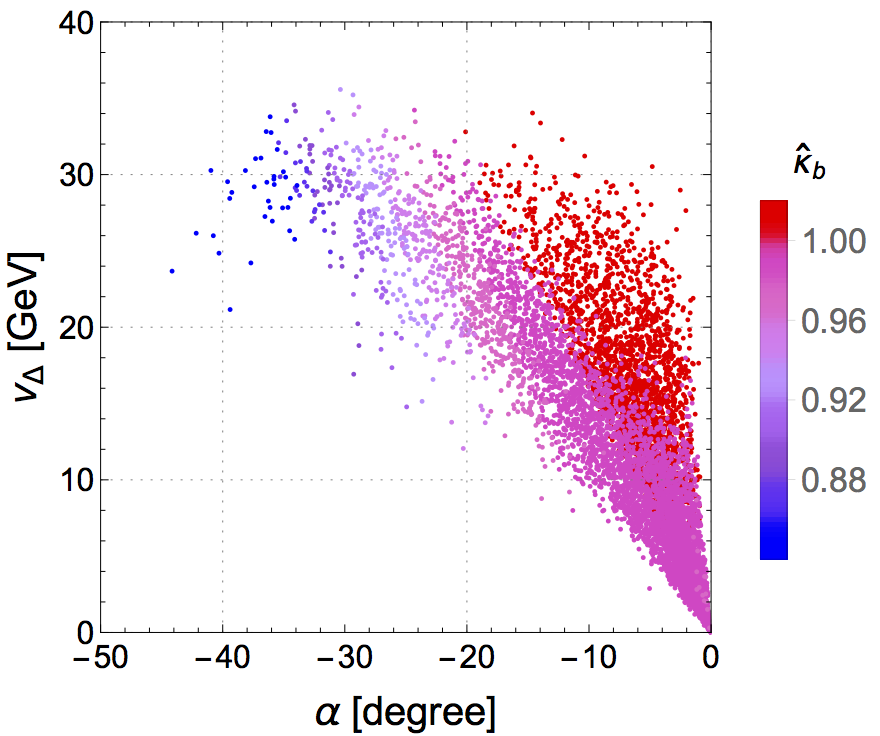}
\includegraphics[scale=0.7]{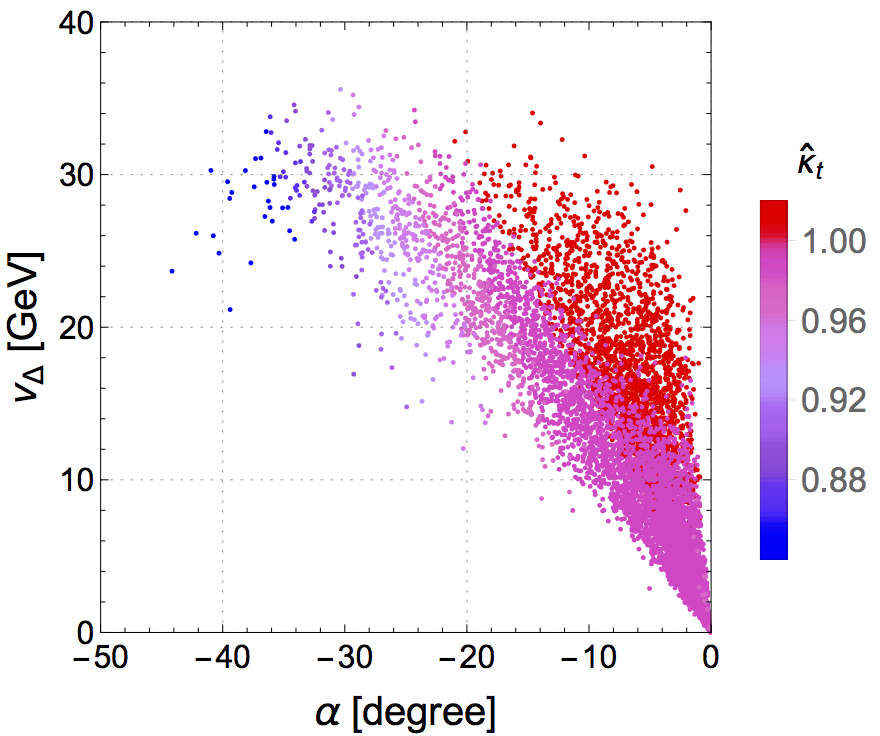}
\caption{
Scatter plots of $\hat\kappa_b^{}$ (left) and $\hat\kappa_t^{}$ (right) in the $\alpha$--$v_\Delta$ plane.
For $\hat\kappa_t^{}$, we take $\sqrt{p^2}=500$~GeV. 
}
\label{kbkhscatter}
\end{figure}

In Fig.~\ref{kbkhscatter}, we show the scatter plots of $\hat\kappa_b$ (left) and $\hat\kappa_t$ with $\sqrt{p^2} = 500$~GeV (right) in the $\alpha-v_\Delta$ plane. 
As shown, the behaviors of $\hat\kappa_b$ and $\hat\kappa_t$ are almost the same as each other. 
The result for $\hat\kappa_\tau$ is also very similar to that of $\hat\kappa_b$. 
In contrast to the case of $\hat\kappa_Z^{}$, the value of $\hat\kappa_b$ becomes smaller when $|\alpha|$ becomes larger.

\begin{figure}[t]
\centering
\includegraphics[scale=0.8]{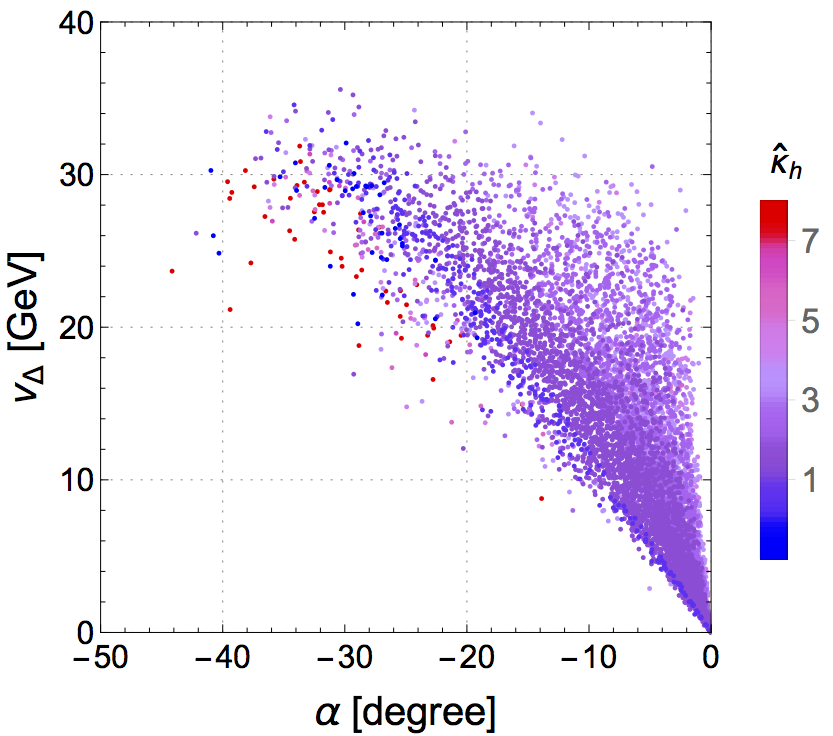}
\caption{
Scatter plot of $\hat\kappa_h^{}$ with $\sqrt{p^2}=500$~GeV. }
\label{contour_kh_500}
\end{figure}

Fig.~\ref{contour_kh_500} is a scatter plot of $\hat\kappa_h$ for $\sqrt{p^2} = 500$~GeV. 
While the variations from the SM predictions in Figs.~\ref{kvscatter} and \ref{kbkhscatter} are typically less than about 10\%, the magnitude of the deviation in the $hhh$ coupling, {\it i.e., $\hat\kappa_h -1$}, can be at a few 100\% level.   
In addition, $\hat\kappa_h$ does not depend much on $\alpha$ and $v_\Delta$ as compared to $\hat\kappa_Z$, $\hat\kappa_b$ and $\hat\kappa_t$.

\begin{figure}[t]
\centering
\includegraphics[scale=0.6]{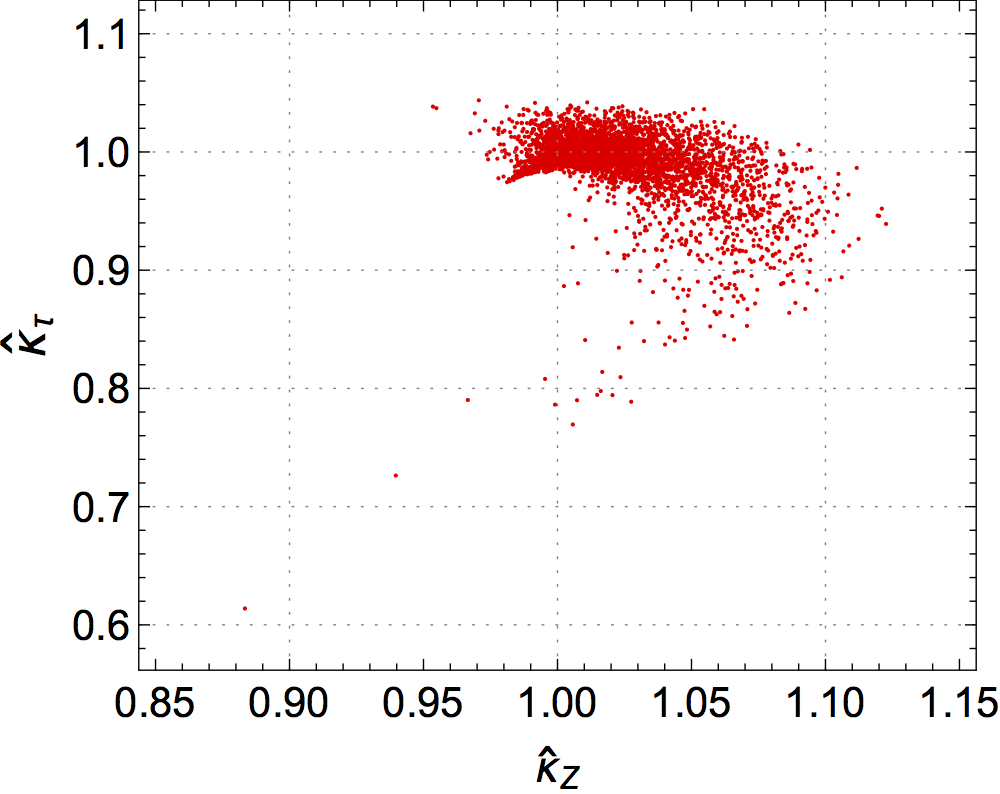}\hspace{5mm}
\includegraphics[scale=0.6]{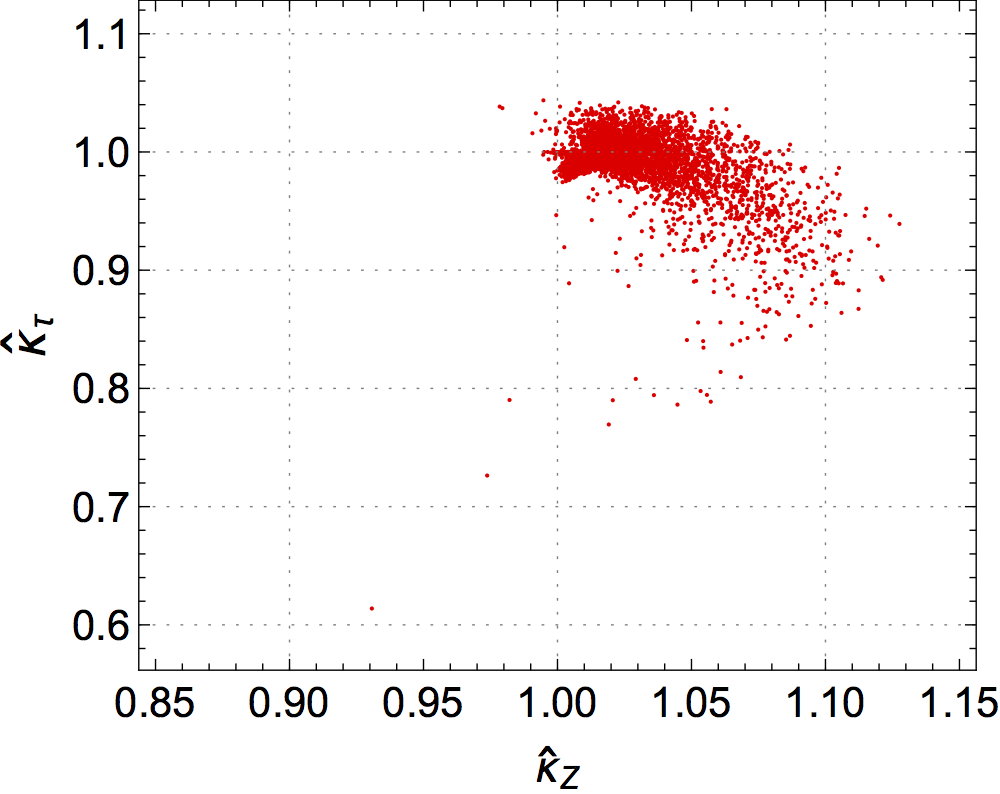}
\caption{Correlation between $\hat\kappa_Z$ and $\hat\kappa_\tau$, where $\sqrt{p^2}$ of $\hat\kappa_Z$ is taken to be 250~GeV (left) and 500~GeV (right). }
\label{kztokx}
\end{figure}

Finally, we show the correlation of the renormalized scale factors. 
Fig.~\ref{kztokx} shows the correlation between $\hat\kappa_Z$ and $\hat\kappa_\tau$, where the momentum $\sqrt{p^2}$ of $\hat\kappa_Z$ is set to be 250~GeV and 500~GeV in the left and right plots, respectively.  
We see that the distribution of the dots in the $\hat\kappa_Z$--$\hat\kappa_\tau$ plane for $\sqrt{p^2} = 500$~GeV is almost the same as that for $\sqrt{p^2} = 250$~GeV, except for slight shrinking in the range of $\hat\kappa_Z$ in the former case. 
It is also seen that the range of possible $\kappa_\tau$ gets restricted when $\hat\kappa_Z$ becomes larger. 
At $\hat\kappa_Z^{} \simeq 1.13$, $\hat\kappa_\tau$ is predicted to be about $0.95$.

\begin{figure}[t]
\centering
\includegraphics[scale=0.6]{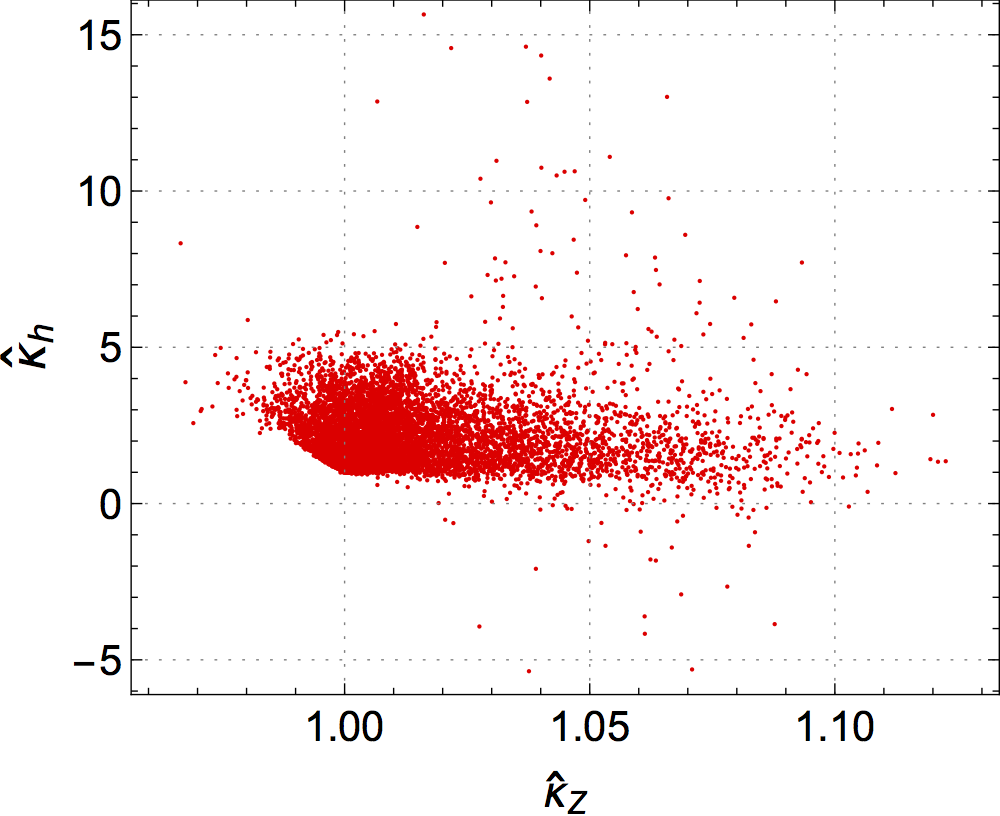}\hspace{5mm}
\includegraphics[scale=0.6]{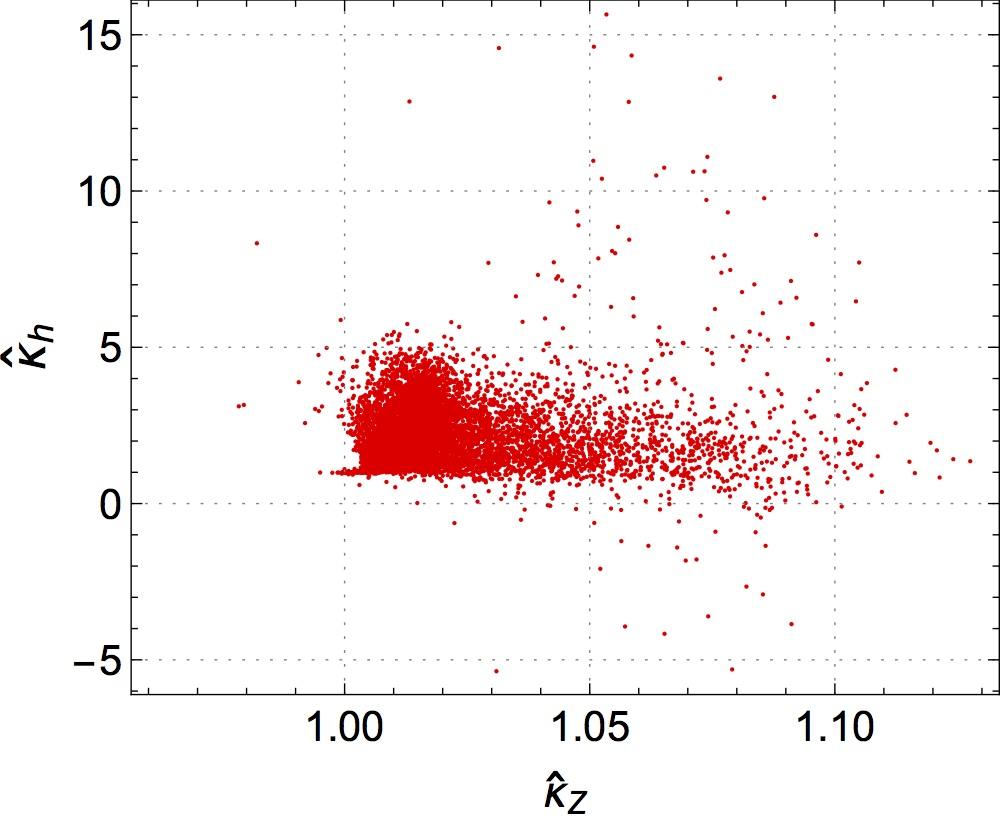}
\caption{
Correlation between $\hat\kappa_Z$ and $\hat\kappa_h$, where $\sqrt{p^2}$ of $\hat\kappa_Z$ is taken to be 250~GeV (left) and 500~GeV (right), while $\sqrt{p^2}$ of $\hat\kappa_h$ is fixed at 500~GeV for both plots. }
\label{kz_kh}
\end{figure}

Fig.~\ref{kz_kh} shows the correlation between $\hat\kappa_Z$ and $\hat\kappa_h$, where the momentum $\sqrt{p^2}$ of $\hat\kappa_Z$ is set to be 250~GeV and 500~GeV in the left and right plots, respectively, while that of $\hat\kappa_h$ is fixed at 500~GeV for both plots. 
Aside from some shifting in the dot distributions between the two plots, most of the predicted $\hat\kappa_h$ values are between 1 and 5. 
Again, the possible range of $\kappa_h$ is restricted when $\hat\kappa_Z$ becomes larger. 
In particular, $\hat\kappa_h$ is predicted to be about 1 when $\hat\kappa_Z^{} \simeq 1.13$. 
We also notice some of the predicted $\hat\kappa_h$ values are less than 1 or even negative for $\hat\kappa_Z \agt 1$.

\section{Conclusions \label{sec:summary}}

In this work, we have calculated the one-loop renormalized vertices  of the 125 GeV Higgs boson ($h$) with the weak gauge bosons ($hVV$), fermions ($hff$), and itself ($hhh$) in the GM model.  
We have chosen to work with the on-shell renormalization scheme and the minimal subtraction scheme, where the latter is only applied to the determination of the counterterms appearing in the renormalized $hhh$ coupling. 
Special care has been taken to check gauge dependence of the counterterms for the mixing parameters $\delta\alpha$ and $\delta\beta$. 
We have defined the gauge-independent counterterm $\delta\alpha$ by adding the pinch-term contributions to the mixing 2-point functions for the CP-even Higgs bosons. 
For $\delta \beta$, we have clarified that its gauge dependence cannot be removed completely even if we add the pinch-term contributions to the mixing 2-point functions for the CP-odd Higgs bosons, due to the gauge-philic and fermio-phobic nature of the 5-plet Higgs bosons. 
Such gauge dependence, however, is exactly cancelled in 2-to-2 fermion scattering processes.

We have numerically evaluated the renormalized Higgs boson couplings, subject to the theoretical bounds of perturbative unitarity and vacuum stability at tree level.  
We have further imposed the constraints from experimental data: the oblique $S$ parameter, the Higgs signal strengths and the direct search for doubly-charged Higgs boson. 
It has been found that the magnitudes of deviations in the one-loop corrected $hVV$ and $hf\bar{f}$ couplings from the SM predictions can be up to about $10\%$ level, where 
the signs of the deviations are typically positive and negative for the $hVV$ and $hf\bar{f}$ couplings, respectively. 
The one-loop corrected $hhh$ coupling, on the other hand, can be significantly larger than the SM prediction by several hundred percent.  Finally, we have studied and shown the correlations of renormalized scale factors for the Higgs boson couplings.

\newpage
\begin{appendix}

\section{Mass eigenstates of the scalar fields \label{sec:mass}}

In this appendix, we give the mass eigenstates of the scalar fields in the GM model and their masses as derived from the potential in Eqs.~(\ref{custb}) and (\ref{eq:pot}).

The mass eigenstates of the scalar fields are related to the original fields given in Eq.~(\ref{eq:Higgs_matrices}) by the following transformations:
\begin{align}
\begin{split}
\begin{pmatrix}
\chi_i\\
\phi_i
\end{pmatrix}
&=
\begin{pmatrix}
c_{\beta_{\text{odd}}} & -s_{\beta_{\text{odd}}} \\
s_{\beta_{\text{odd}}} & c_{\beta_{\text{odd}}}
\end{pmatrix}
\begin{pmatrix}
G^0\\
H_3^0
\end{pmatrix}, \\
\begin{pmatrix}
\phi^\pm \\
\xi^\pm \\
\chi^\pm
\end{pmatrix}
&= R_{H_5^\pm} R_{\beta_\pm}R_\gamma
\begin{pmatrix}
G^\pm\\
H_3^\pm\\
H_5^\pm
\end{pmatrix},~
\begin{pmatrix}
\xi_r\\
\phi_r\\
\chi_r
\end{pmatrix}
= R_{H_5^0} R_{\alpha}
\begin{pmatrix}
H_1\\
h\\
H_5^0
\end{pmatrix},  
\end{split}
\label{eigen2}
\end{align}
where $G^\pm$ and $G^0$ are the NG bosons to become the longitudinal components of $W^\pm$ and $Z$ bosons, respectively. 
The rotation matrices in Eq.~(\ref{eigen2})
\begin{align}
\begin{split}
R_{H_5^\pm} &=
\begin{pmatrix}
1&0&0 \\
0 & \frac{1}{\sqrt{2}} & -\frac{1}{\sqrt{2}} \\
0 & \frac{1}{\sqrt{2}} & \frac{1}{\sqrt{2}} 
\end{pmatrix},~
R_{\beta_\pm} = 
\left(
\begin{array}{ccc}
s_{\beta_1^\pm} & c_{\beta_1^\pm} & 0\\
c_{\beta_1^\pm} &-s_{\beta_1^\pm} & 0\\
0&0&1
\end{array}
\right)
\begin{pmatrix}
c_{\beta_2^\pm} & 0 & s_{\beta_2^\pm}\\
0&1&0 \\
-s_{\beta_2^\pm} & 0 & c_{\beta_2^\pm}
\end{pmatrix}, \\
R_\gamma &= 
\begin{pmatrix}
1 & 0 & 0\\
0 & c_\gamma& -s_\gamma \\
0 & s_{\gamma}  & c_{\gamma}
\end{pmatrix}, ~~
R_{H_5^0}=\begin{pmatrix}
\frac{1}{\sqrt{3}} &0& -\sqrt{\frac{2}{3}}\\
0 & 1 &0\\
\sqrt{\frac{2}{3}} & 0 & \frac{1}{\sqrt{3}}
\end{pmatrix}, \\
R_{\alpha} &= 
\begin{pmatrix}
1 & 0 & 0 \\
0 & c_{\alpha_{1}} & -s_{\alpha_{1}} \\
0 & s_{\alpha_{1}} &  c_{\alpha_{1}}
\end{pmatrix}
\begin{pmatrix}
c_{\alpha_{2}} &0& -s_{\alpha_{2}} \\
0 & 1 & 0 \\
s_{\alpha_{2}} &0&  c_{\alpha_{2}}
\end{pmatrix}
\begin{pmatrix}
c_{\alpha_{3}} & -s_{\alpha_{3}} & 0\\
s_{\alpha_{3}} &  c_{\alpha_{3}} & 0\\
0&0&1
\end{pmatrix}, 
\end{split}
\end{align}
with the mixing angles satisfying 
\begin{align}
\tan\beta_{\text{odd}} = \frac{v_\phi}{2\sqrt{2}v_\Delta},\quad
\tan\beta_1^\pm = \frac{v_\phi}{\sqrt{2}(2v_\Delta + \nu)},\quad
\tan\beta_2^\pm = \frac{\sqrt{2}\nu}{\sqrt{v_\phi^2 + 2(2v_\Delta + \nu)^2}}.  \label{betas}
\end{align}
The other mixing angles $\gamma$, $\alpha_1$, $\alpha_2$ and $\alpha_3$ generally have very complicated forms.  Nevertheless, an important thing is that 
in the  $\nu \to  0$ limit $\gamma$, $\alpha_1$ and $\alpha_2$ become zero, while $\alpha_3$ can be nonzero.

The squared masses of the physical Higgs bosons are given by
 \begin{align}
 \begin{split}
 m_{H_5^{\pm\pm}}^2 & =  m_{H_5}^2 -\frac{\nu}{v_\Delta}\left(\frac{v_\phi^2}{2}\lambda_5 +12\mu_2v_\Delta \right) , \\
 m_{H_5^{\pm}}^2   & =  m_{H_5}^2 +\frac{\nu}{8v_\Delta}\left(64v_\Delta^2\lambda_3 +2v_\phi^2 \lambda_5 +v_\phi^2\frac{\mu_1}{v_\Delta} \right) + {\cal O}(\nu^2), \\
 m_{H_5^{0}}^2    & =  m_{H_5}^2 +\frac{\nu}{6v_\Delta}\left(64v_\Delta^2\lambda_3 +3v_\phi^2 \lambda_5 + v_\phi^2\frac{\mu_1}{v_\Delta}+24\mu_2v_\Delta \right) + {\cal O}(\nu^2), \\
 m_{H_3^{\pm}}^2   & =  m_{H_3}^2 +\frac{\nu}{8v_\Delta}(v_\phi^2 - 8v_\Delta^2)\left(2\lambda_5 +\frac{\mu_1}{v_\Delta}\right) + {\cal O}(\nu^2) , \\
m_{H_3^{0}}^2    &  =  m_{H_3}^2 - \frac{\nu}{2v_\Delta}(v_\phi^2 + 8v_\Delta^2)\lambda_5  ,\\
m_{H_1}^2        &  = c_{\alpha_3}^2 M_{11}^2 + s_{\alpha_3}^2 M_{22}^2 + 2s_{\alpha_3}c_{\alpha_3} M_{12}^2 + {\cal O}(\nu^2), \\
m_{h}^2         &   = s_{\alpha_3}^2 M_{11}^2 + c_{\alpha_3}^2 M_{22}^2 - 2s_{\alpha_3}c_{\alpha_3} M_{12}^2 + {\cal O}(\nu^2), 
\label{massH}
\end{split}
\end{align}
where 
 \begin{align}
 \begin{split}
 m_{H_5}^2 &= 8v_\Delta^2\lambda_3  -\frac{3}{2}v_\phi^2\lambda_5   - \frac{v_\phi^2}{4}\frac{\mu_1}{v_\Delta} -12 v_\Delta\mu_2 , \\
 m_{H_3}^2 &= -\frac{1}{4}(v_\phi^2 + 8v_\Delta^2)\left(2\lambda_5 +\frac{\mu_1}{v_\Delta} \right), 
 \end{split}
 \label{m3m5}
 \end{align}
and 
 \begin{align}
 \begin{split}
 M^2_{11}&= 8v_\Delta^2(3\lambda_2 + \lambda_3) - v_\phi^2\frac{\mu_1}{4v_\Delta} + 6v_\Delta \mu_2 + \frac{\nu}{v_\Delta}\left[\frac{16}{3}v_\Delta^2(3\lambda_2 + \lambda_3) 
+ v_\phi^2\frac{\mu_1}{12v_\Delta} +2v_\Delta\mu_2\right], \\
 M^2_{22}&=8v_\phi^2\lambda_1 , \\
 M^2_{12}&=\frac{\sqrt{3}}{2}v_\phi \left[4v_\Delta(2\lambda_4+\lambda_5) + \mu_1 \right] + \frac{2\sqrt{3}}{3}\nu v_\phi(2\lambda_4 + \lambda_5). 
 \end{split}
\end{align}
It is observed that in the $\nu \to 0$ limit, the different charged states within each multiplet have the same mass as the consequence of the restoration of the custodial symmetry.

\section{Interaction terms of the Higgs bosons \label{sec:hcoup}}

We give expressions for the relevant 3-point and 4-point interaction terms of the Higgs bosons.  
The scalar-gauge-gauge interaction terms are given by 
\begin{align}
{\cal L}_{SVV} 
=& 
\sum_\varphi\left[\frac{2m_W^2}{v} c_{\varphi WW}^{}\, \varphi W^{+\mu}W_\mu^-  +\frac{m_Z^2}{v}c_{\varphi ZZ}^{}\,\varphi Z^{\mu}Z_\mu \right]\notag\\
& + \frac{2m_W^2}{v}\left(\frac{c_\beta}{\sqrt{2}} H_5^{++}W^{-\mu}W_\mu^-  - \frac{c_\beta}{c_W}H_5^+Z^\mu W_\mu - \frac{s_W^2}{c_W^{}} G^+Z^{\mu}W_\mu^-    + \text{H.c.} \right),  \label{sgg}
\end{align}
where
\begin{align}
\begin{split}
&c_{hVV}^{} =   c_\alpha s_\beta  - \frac{2\sqrt{6}}{3} s_\alpha c_\beta, \quad  c_{H_1VV}^{}  = s_\alpha s_\beta+ \frac{2\sqrt{6}}{3} c_\alpha c_\beta ,~ (V=W,Z), \\
&c_{H_5^0WW}   =   -\frac{c_\beta}{\sqrt{3}}, \quad\quad\quad\quad\quad c_{H_5^0ZZ}   = \frac{2c_\beta}{\sqrt{3}}, 
\end{split}
 \label{c5_1}
\end{align}
and 
\begin{align}
\sum_\varphi \equiv \sum_{\varphi = h, H_1, H_5^0}. 
\end{align}

The scalar-scalar-gauge interaction terms are given by 
\begin{align}
{\cal L}_{SSV} 
=& i\frac{g}{2} \Bigg[
\sum_\varphi c_{\varphi WW}^{}(\varphi \,\partial\, G^+)^\mu -  \sum_\varphi c_{\varphi H_3W}^{}(\varphi\,\partial\, H_3^+)^\mu \notag\\
&\quad\quad\quad -i(G^0\,\partial\, G^+)^\mu -i(H_3^0\,\partial\, H_3^+)^\mu\notag\\
&\quad\quad\quad-ic_\beta (G^0\,\partial\, H_5^+)^\mu +is_\beta(H_3^0\,\partial\, H_5^+)^\mu +\sqrt{3}(H_5^0\,\partial\, H_5^+)^\mu \notag\\
&\quad\quad\quad+\sqrt{2}c_\beta(G^-\,\partial\, H_5^{++})^\mu - \sqrt{2}s_\beta(H_3^-\,\partial\, H_5^{++})^\mu+\sqrt{2}(H_5^-\,\partial\, H_5^{++})^\mu   \Bigg]W_\mu^- + \text{H.c.}, \notag\\
&+\frac{g_Z^{}}{2} \Bigg\{ \sum_{\varphi} c_{\varphi ZZ}^{}(\varphi \,\partial\, G^0)^\mu 
 -\sum_{\varphi} c_{\varphi H_3Z}^{}(\varphi \,\partial\, H_3^0)^\mu    \notag\\
&\quad\quad\quad-ic_{2W}^{} \left[(G^+\,\partial\, G^-)^\mu 
+(H_3^+\,\partial\, H_3^-)^\mu 
+(H_5^+\,\partial\, H_5^-)^\mu 
+2(H_5^{++}\,\partial\, H_5^{--})^\mu   \right] \notag\\
&\quad\quad\quad +ic_\beta(H_5^+ \partial G^-)-is_\beta(H_5^+ \partial H_3^-)   \Bigg\} Z_\mu, 
\end{align}
where $(A\,\partial\, B)^\mu \equiv A (\partial^\mu B) - (\partial^\mu A) B$, $c_{2W}^{}\equiv  c_W^2 - s_W^2$ and 
\begin{align}
\begin{split}
&c_{hH_3V}^{}   =   -c_\alpha c_\beta- \frac{2\sqrt{6}}{3}s_\alpha s_\beta, \quad
c_{H_1H_3V}^{}  =    -s_\alpha c_\beta + \frac{2\sqrt{6}}{3} c_\alpha s_\beta, ~ (V=W,Z), \\
&c_{H_5^0 H_3W}^{} = -\frac{s_\beta}{\sqrt{3}},\quad\quad\quad\quad\quad\quad c_{H_5^0H_3Z}^{}  = \frac{2s_\beta}{\sqrt{3}}.   
\end{split}
 \label{c5_2}
\end{align}

The scalar-scalar-gauge-gauge interaction terms are given by 
\begin{align}
{\cal L}_{SSVV} &=\sum_\varphi \left(\frac{g^2}{4}   c_{h\varphi WW}  W^{+\mu}W_\mu^- h\varphi 
+\frac{g_Z^2}{8}  c_{h\varphi ZZ} Z^\mu Z_\mu h\varphi \right), 
\end{align}
where
\begin{align}
\begin{split}
&c_{hhVV}^{}    = \frac{11-5c_{2\alpha}}{6}, \quad
c_{hH_1VV}^{}  = -\frac{5}{3}s_{2\alpha},~ (V = W,Z) \notag\\
&c_{hH_5WW}^{}  =  \frac{4}{3}\sqrt{2}s_\alpha,\quad ~
c_{hH_5ZZ}^{}  =  -\frac{8}{3}\sqrt{2}s_\alpha , \\
&c_{H_1H_1VV}^{}  = \frac{11+5c_{2\alpha}}{6}, \quad
c_{H_5^0H_5^0WW}^{}  = \frac{10}{3}. \\
\end{split}
\end{align}

The Yukawa interaction terms for the third-generation fermions are given by
\begin{align}
{\cal L}_{ffS} 
=& -\sum_{f=t,b,\tau} \frac{m_f}{v}\left(c_{hff}\bar{f}f h + c_{H_1ff}\bar{f}f H_1  
-2iI_f \cot\beta \bar{f}\gamma_5f H_3^0  \right)\notag\\
&- \frac{\sqrt{2}}{v}\cot\beta \left[\bar{t}(m_b P_R - m_t P_L) bH_3^+ + \bar{\nu}_\tau \, m_\tau  P_R \,\tau H_3^+  +\text{h.c.}\right], \label{int_gm}
\end{align}
where $I_t = 1/2$ and $I_b = I_\tau = -1/2$, and 
\begin{align}
c_{hff} = \frac{c_\alpha}{s_\beta},\quad c_{H_1ff} = \frac{s_\alpha}{s_\beta}. \label{chff}
\end{align}

\section{1PI contributions \label{sec:1pi}}

We give the analytic expressions for 1PI diagram contributions that appear in the renormalized Higgs boson vertices. 
Section~\ref{sec:lf} defines the required loop functions. 
The formulas for the 1PI diagram contributions to 1-, 2- and 3-point functions are given in Sections~\ref{sec:1pf}, \ref{sec:2pf} and \ref{sec:3pf}, respectively. 
Calculations are performed in the 't~Hooft-Feynman gauge, where the masses of the NG bosons $m_{G^\pm}$ and $m_{G^0}$ become $m_W^{}$ and $m_{Z}^{}$, respectively.

\subsection{Loop functions \label{sec:lf}}

In order to systematically express all one-loop amplitudes, we introduce the Passarino-Veltman 1-, 2- and 3-point functions~\cite{PV} as follows: 
\begin{align}
\frac{i}{16\pi^2}A(m_1)&=\mu^{4-D}\int\frac{d^Dk}{(2\pi)^D}\frac{1}{D_1},\\
\frac{i}{16\pi^2}[B_0,B^\mu,B^{\mu\nu}](p_1^2;m_1,m_2)&=\mu^{4-D}\int\frac{d^Dk}{(2\pi)^D}\frac{[1,k^\mu,k^\mu k^\nu]}{D_1D_2},\\
\frac{i}{16\pi^2}[C_0,C^\mu,C^{\mu\nu}](p_1^2,p_2^2,(p_1+p_2)^2;m_1,m_2,m_3)&=\mu^{4-D}\int\frac{d^Dk}{(2\pi)^D}\frac{[1,k^\mu,k^\mu k^\nu]}{D_1D_2D_3}, 
\end{align}
where $D=4-2\epsilon$, and $\mu$ is a dimensionful parameter.
The functions in the denominators, $D_{1,2,3}$, are defined by 
\begin{align}
&D_1=k^2-m_1^2+i\varepsilon,\quad D_2=(k+p_1)^2-m_2^2+i\varepsilon,\quad D_3=(k+p_1+p_2)^2-m_3^2+i\varepsilon. 
\end{align}
The $B$ and $C$ tensor functions are decomposed into the following forms in terms of scalar coefficients $B_{1,21,22}$ and $C_{11,12,21,22,23,24}$:
\begin{align}
B^\mu&=p_1^\mu B_1,\label{b1}\\
B^{\mu\nu}&=p_1^\mu p_1^\nu B_{21}+g^{\mu\nu}B_{22},\label{b2} \\
C^\mu &=p_1^\mu C_{11}+p_2^\mu C_{12},\\
C^{\mu\nu} &=p_1^\mu p_1^\nu C_{21}+p_2^\mu p_2^\nu C_{22}
+(p_1^\mu p_2^\nu+p_1^\nu p_2^\mu) C_{23}+g^{\mu\nu}C_{24}.    
\label{cmunu} 
\end{align}

It is convenient to define coefficients $\lambda_{\phi_i\phi_j\phi_k}$ and $\lambda_{\phi_i\phi_j\phi_k\phi_l}$ respectively for the 3-point and 4-point scalar interaction terms as  
\begin{align}
{\cal L} = +\lambda_{\phi_i\phi_j\phi_k}\phi_i\phi_j\phi_k + \lambda_{\phi_i\phi_j\phi_k\phi_l}\phi_i\phi_j\phi_k\phi_l + \cdots. \label{self}
\end{align}
As some of these coefficients are proportional to each other, we thus define the following quantities: 
\begin{align}
\begin{split}
\lambda_{H_5 H_5 \phi} &\equiv \lambda_{H_5^{0} H_5^{0} \phi} = \frac{1}{2}\lambda_{H_5^{+} H_5^{-} \phi} = \frac{1}{2}\lambda_{H_5^{++} H_5^{--} \phi}, \\
\lambda_{H_3 H_3 \phi} &\equiv \lambda_{H_3^0 H_3^0\phi}    = \frac{1}{2}\lambda_{H_3^{+} H_3^{-} \phi}, \\
\lambda_{GG\phi}      &\equiv \lambda_{G^{0}G^{0}\phi}      =  \frac{1}{2}\lambda_{G^+ G^-\phi}, \\
\lambda_{GGH_5^0}   &\equiv \lambda_{G^{0}G^{0} H_5^0}  = - \lambda_{G^+ G^- H_5^0}, \\
\lambda_{H_3G\phi}    &\equiv \lambda_{H_3^{0}G^{0}\phi}    =   \lambda_{H_3^\pm G^\mp \phi}, \\
\lambda_{H_3GH_5^0} &\equiv \lambda_{H_3^{0}G^{0} H_5^0} = -2\lambda_{H_3^\pm G^\mp H_5^0}, \\
\end{split}
\label{lam3p}
\end{align}
and 
\begin{align}
\begin{split}
&\lambda_{H_5 H_5 hh}\equiv \lambda_{H_5^{++} H_5^{--} hh} = \lambda_{H_5^{+} H_5^{-} hh} = 2\lambda_{H_5^{0} H_5^{0} hh}, \\
&\lambda_{H_3 H_3 hh}\equiv \lambda_{H_3^{+} H_3^{-} hh} = 2\lambda_{H_3^{0} H_3^{0} hh}, \\
&\lambda_{G G hh}\equiv \lambda_{G^{+} G^{-} hh} = 2\lambda_{G^{0} G^{0} hh}, \\
&\lambda_{H_3 G hh}\equiv \lambda_{H_3^{+} G^{-} hh} = \lambda_{H_3^{0} G^{0} hh}, 
\end{split}
\end{align}
with $\phi = h$ or $H_1$.

\subsection{1-point functions \label{sec:1pf}}

The 1PI diagram contributions to 1-point functions for $h$, $H_1$ and $H_5^0$ are given by 
\begin{align}
T_h^{1\text{PI}} 
=& -\sum_{f = t,b,\tau}\frac{N_c^f}{16\pi^2}\frac{4m_f^2}{v}c_{hff}A(m_f) \notag\\
& -\frac{1}{16\pi^2}\Bigg\{ 
            5\l_{H_5H_5h}A(m_{H_5})+3\l_{H_3H_3h}A(m_{H_3})+\l_{H_1H_1h}A(m_{H_1}) \notag\\
& \qquad\qquad  +3\l_{hhh}A(m_h) -\l_{GGh}\left[2A(m_{G^\pm}) + A(m_{G^0})\right] \notag\\
& \qquad\qquad -3c_{hVV}^{}\left[gm_WA(m_W)+\frac{g_Z}{2}m_ZA(m_Z) -\frac{2}{3}gm_W^3 -\frac{g_Z^{}}{3}m_Z^3\right]\Bigg\}~, \\[10pt]
T_{H_1}^{1\text{PI}} 
=& -\sum_{f = t,b,\tau}\frac{N_c^f}{16\pi^2}\frac{4m_f^2}{v}c_{H_1ff}^{}A(m_f) \notag\\ 
& -\frac{1}{16\pi^2}\Bigg\{ 
           5\l_{H_5H_5H_1}A(m_{H_5})+3\l_{H_3H_3H_1}A(m_{H_3})+3\l_{H_1H_1H_1}A(m_{H_1}) \notag\\
& \qquad\qquad +\l_{H_1hh}A(m_h) -\l_{GGH_1}\left[2A(m_{G^\pm}) + A(m_{G^0})\right] \notag\\
& \qquad\qquad -3c_{H_1VV}^{}\left[gm_WA(m_W)+\frac{g_Z}{2}m_ZA(m_Z) -\frac{2}{3}gm_W^3 -\frac{g_Z^{}}{3}m_Z^3\right]\Bigg\}~,\\[10pt]
T_{H_5^0}^{1\text{PI}}
=& \frac{1}{16\pi^2}\Big\{ \l_{GGH_5^0}\left[A(m_{G^\pm}) - A(m_{G^0}) \right] \notag\\
& \qquad +gm_Wc_{H_5^0WW}^{} 
\left[ 3A(m_W)-2m_W^2\right]+\frac{g_Z^{}}{2}m_Z^{}c_{H_5^0ZZ}^{}\left[3A(m_Z^{})-2m_Z^2 \right]\Big\}, 
\end{align}
where $N_c^f = 3~(1)$ for $f=t,b~ (\tau)$. 
We note that for $T_{H_5^0}^{1\text{PI}}$, the 5-plet and 3-plet Higgs boson loop contributions are cancelled among themselves.

\subsection{2-point functions \label{sec:2pf}}

The 1PI diagram contributions to 2-point functions for CP-even Higgs bosons are given by 
\begin{align}
&\Pi_{hh}^{\text{1PI}}(q^2) = \Pi_{hh}^{ff}(q^2) + \Pi_{hh}^{SV+VV}(q^2) \notag\\
&\quad+\frac{1}{16\pi^2}\Big\{
10\lambda_{H_5H_5h}^2B_0(q^2;m_{H_5}^{},m_{H_5}^{}) + 6\lambda_{H_3H_3h}^2B_0(q^2;m_{H_3}^{},m_{H_3}^{}) \notag\\
&\quad +18\lambda_{hhh}^2B_0(q^2;m_h,m_h) +2\lambda_{H_1H_1h}^2B_0(q^2;m_{H_1}^{},m_{H_1}^{}) +4\lambda_{H_1hh}^2B_0(q^2;m_{H_1}^{},m_h) \notag\\
&\quad +\lambda_{H_3Gh}^2 \left[2B_0(q^2;m_{G^\pm}^{},m_{H_3}) + B_0(q^2;m_{G^0}^{},m_{H_3}^{}) \right]\notag\\
&\quad +2\lambda_{GGh}^2 \left[2B_0(q^2;m_{G^\pm}^{},m_{G^\pm}^{}) + B_0(q^2;m_{G^0}^{},m_{G^0}^{})   \right]\Big\} \notag\\
&\quad -\frac{2}{16\pi^2}\sum_{X=\text{scalars}}(1 + 5\delta_{Xh})\lambda_{XX^*hh} A(m_{X}^{}),  \label{pihh}  \\[10pt]
&\Pi_{H_1h}^{\text{1PI}}(q^2) = \Pi_{H_1 h}^{ff}(q^2) + \Pi_{H_1h}^{SV+VV}(q^2)  \notag\\
&  \quad +\frac{1}{16\pi^2}\Big[
10\lambda_{H_5H_5h}\lambda_{H_5H_5H_1} B_0(q^2;m_{H_5}^{},m_{H_5}^{})+6\lambda_{H_3H_3h}\lambda_{H_3H_3H_1}B_0(q^2;m_{H_3}^{},m_{H_3}^{}) \notag\\
&\quad +6\lambda_{hhh}\lambda_{H_1hh}B_0(q^2;m_h,m_h) +6\lambda_{H_1H_1h}\lambda_{H_1H_1H_1}B_0(q^2;m_{H_1}^{},m_{H_1}^{}) \notag\\
&\quad+4\lambda_{H_1hh}\lambda_{H_1H_1h}B_0(q^2;m_{H_1}^{},m_h) \notag\\
&\quad+\lambda_{H_3Gh}\lambda_{H_3GH_1} \left[2B_0(q^2;m_{G^\pm}^{},m_{H_3}^{})+B_0(q^2;m_{G^0}^{},m_{H_3}^{}) \right]\notag\\
&\quad +2\lambda_{GGh}\lambda_{GGH_1}\left[2B_0(q^2;m_{G^\pm}^{},m_{G^\pm}^{}) + B_0(q^2;m_{G^0}^{},m_{G^0}^{})\right] \notag\\
&\quad -\frac{1}{16\pi^2}\sum_{X=\text{scalars}}(1 + 2\delta_{Xh}+2\delta_{XH_1})\lambda_{XX^*H_1h} A(m_X^{}), \\[10pt]
&\Pi_{H_5^0h}^{\text{1PI}}(q^2) =\Pi_{H_5^0h}^{SV+VV}(q^2) \notag\\
& \quad -\frac{1}{16\pi^2}\Big\{\lambda_{H_3Gh}\lambda_{H_3GH_5^0}\left[B_0(q^2;m_{G^\pm}^{},m_{H_3}^{}) -B_0(q^2;m_{G^0}^{},m_{H_3}^{})\right]\notag\\
& \quad +2\lambda_{GGh}\lambda_{GGH_5^0}\left[B_0(q^2;m_{G^\pm}^{},m_{G^\pm}^{}) - B_0(q^2;m_{G^0}^{},m_{G^0}^{})\right]  \Big\} \notag\\
& \quad -\frac{1}{16\pi^2}\sum_{X=\text{scalars}}(1 + 2\delta_{Xh} + 2\delta_{XH_1})\lambda_{XX^*H_3^0 G^0} A(m_X^{}),
\end{align}
where 
\begin{align}
\Pi_{\varphi h}^{ff}(q^2) &=  -\frac{1}{16\pi^2}\sum_{f=t,b,\tau}\frac{4m_f^2N_c^f}{v^2}c_{h ff}c_{\varphi ff}
\left[A(m_f) + \left(2m_f^2 -\frac{q^2}{2}B_0(q^2;m_f,m_f) \right) \right], 
\end{align}
\begin{align}
&\Pi_{\varphi h}^{SV+VV}(q^2) =  -\frac{1}{16\pi^2}\frac{2m_W^2}{v^2}\Bigg\{
c_{hWW}^{}c_{\varphi WW}^{} \Big[(2q^2 - 6m_W^2)B_0(q^2;m_W^{},m_W^{}) +A(m_W^{}) + 4m_W^2\Big]\notag\\
&\quad +\frac{c_{hZZ}^{}c_{\varphi ZZ}^{}}{2c_W^2}\Big[(2q^2 - 6m_Z^2 )B_0(q^2;m_Z^{},m_Z^{}) +A(m_Z^{}) +4m_Z^2\Big] \notag\\
&\quad+c_{hH_3W}^{}c_{\varphi H_3W}^{}
\Big[(2q^2+2m_{H_3}^2 -m_W^2)B_0(q^2;m_{H_3}^{},m_W^{}) +2A(m_W^{})-A(m_{H_3}^{})\Big]\notag\\
&\quad +\frac{c_{hH_3Z}^{}c_{\varphi H_3Z}^{}}{2c_W^2}\Big[(2q^2+2m_{H_3}^2 -m_Z^2)B_0(q^2;m_{H_3}^{},m_Z^{}) +2A(m_Z^{})-A(m_{H_3}^{})\Big]\Bigg\}\notag\\
&\quad +(1+\delta_{\varphi h})\frac{1}{16\pi^2} \left\{ g^2c_{h\varphi WW}^{}\left[A(m_W^{}) -\frac{m_W^2}{2}\right] 
+\frac{g_Z^2}{2}c_{h\varphi ZZ}^{}\left[A(m_Z^{}) -\frac{m_Z^2}{2} \right]\right\}. 
\end{align}
That for the $H_3^0$--$G^0$ mixing is expressed as 
\begin{align}
&\Pi^{1\text{PI}}_{H_3^0G^0}(q^2)  = 
-\frac{1}{16\pi^2}\sum_{f = t,b,\tau}\frac{4m_f^2N_c^f}{v^2}\cot\beta\left[A(m_f) -\frac{q^2}{2}B_0(q^2;m_f,m_f)  \right] \notag\\
  &\quad +\frac{1}{16\pi^2}\Big\{\frac{g^2}{2}\sb\cb\left[(2q^2 +2m_{H_5}^2-m_W^2)B_0(q^2;m_{H_5}^{},m_W^{}) +2A(m_W^{})-A(m_{H_5}^{})\right]\notag\\
  &\quad +\frac{g_Z^2}{4}\sum_{\varphi}c_{\varphi H_3Z}c_{\varphi ZZ}\left[(2q^2+2m_\varphi^2-m_Z^2)B_0(q^2;m_\varphi,m_Z^{})+2A(m_Z^{})-A(m_\varphi)\right] \Big\}\notag\\
  &\quad -\frac{1}{16\pi^2}\sb\cb\left[2g^2 A(m_W^{})+3g_Z^2A(m_Z^{})-g^2m_W^2 - \frac{3g_Z^2}{2}m_Z^2 \right] \notag\\
  &\quad +\frac{1}{16\pi^2}\Big\{ 3\l_{H_3GH_5^0}[\l_{H_5^0H_3^0H_3^0}B_0(q^2;m_{H_5}^{},m_{H_3}^{}) + \l_{GGH_5^0}B_0(q^2;m_{H_5}^{},m_{G^\pm}^{} )] \notag\\
  &\quad +2\sum_{\varphi}\left[\l_{ H_3H_3\varphi}\l_{H_3G\varphi}B_0(q^2;m_\varphi,m_{H_3}^{})
 +\l_{GG\varphi}\l_{ H_3G\varphi}B_0(q^2;m_\varphi,m_{G^0}^{})      \right]\Big\}\notag\\
  &\quad -\frac{1}{16\pi^2}\sum_{X=\text{scalar}}(1 + 2\delta_{XG^0} + 2\delta_{XH_3^0})\lambda_{XX^*H_3^0 G^0} A(m_X^{}). 
\end{align}

Next, fermion 2-point functions can be decomposed into the following three parts:
\begin{align}
\Pi_{ff}^{\text{1PI}}(q^2) =  
 q\hspace{-2mm}/\Pi_{ff,V}^{\text{1PI}}(q^2) - q\hspace{-2mm}/\gamma_5\Pi_{ff,A}^{\text{1PI}}(q^2)  
+m_f\Pi_{ff,S}^{\text{1PI}}(q^2).   \label{piffdef}
\end{align}
Each part subtracted by the SM contribution is calculated as
\begin{align}
\Delta \Pi_{ff,V}^{\text{1PI}}(q^2) 
=& -\frac{1}{16\pi^2}\frac{m_f^2}{v^2}
\left[ (c_{hff}^2 -1 )B_1(q^2;m_f,m_h)+c_{H_1ff}^2B_1(q^2;m_f,m_{H_1}^{})  \right. \notag\\
& \qquad +\cot^2\beta B_1(q^2;m_f,m_{H_3}^{})
+ \left. \left( 1 + \frac{m_{f'}^2}{m_f^2} \right)\cot^2\beta B_1(q^2;m_{f'},m_{H_3}^{}) \right] , \\[10pt]
\Delta \Pi_{ff,A}^{\text{1PI}}(q^2) 
=& \frac{1}{16\pi^2}
\frac{m_f^2-m_{f'}^2}{v^2}\cot^2\beta B_1(q^2;m_{f'},m_{H_3}^{}) ,\\[10pt]
\Delta \Pi_{ff,S}^{\text{1PI}}(q^2) 
=& \frac{1}{16\pi^2}\frac{m_f^2}{v^2}
\left[(c_{hff}^2 - 1)B_0(q^2;m_f,m_h)+c_{H_1ff}^2 B_0(q^2;m_f,m_{H_1}^{})  \right. \notag\\
& \qquad \left. -\cot^2\beta B_0(q^2;m_f,m_{H_3}^{})-2\frac{m_{f'}^2}{m_f^2}\cot^2\beta B_0(q^2;m_{f'},m_{H_3}^{}) \right], 
\end{align}
where $v_f$ and $a_f$ are the coefficients of the vector coupling and axial-vector coupling of the $Zf\bar{f}$ vertex, given by
\begin{align}
v_f = \frac{I_f}{2}-s_W^2Q_f,\quad a_f = \frac{I_f}{2}, 
\end{align}
with $Q_f$ being the electric charge of the fermion $f$. 
In addition, $m_{f'}$ is the mass of fermion $f'$ with an opposite weak isospin to $f$.

Finally, we present the expressions for the transverse components of the gauge boson 2-point functions. 
Each of the functions subtracted by the SM contribution is given by 
\begin{align}
\begin{split}
\Delta \Pi_{WW}^{\text{1PI}}(q^2) 
=& \frac{g^2}{64\pi^2}\Big\{ 5B_5(q^2;m_{H_5},m_{H_5})+ 3\sb^2B_5(q^2;m_{H_5},m_{H_3}) +B_5(q^2;m_{H_3},m_{H_3})  
\\
&\qquad\qquad 
+2\cb^2B_5(q^2;m_{H_5},m_{G^\pm})+\cb^2B_5(q^2;m_{H_5},m_{G^0}) 
\\
&\qquad\qquad
+\sum_\varphi \left[c_{\varphi H_3W}^2B_5(q^2;m_{H_3},m_\varphi) + c_{\varphi WW}^2B_5(q^2;m_{G^\pm},m_\varphi) \right]
\\
&\qquad\qquad
-B_5(q^2;m_{G^\pm},m_h) \Big\}
\\
&\quad+\frac{g^2m_W^2}{16\pi^2}\Big[2\cb^2B_0(q^2;m_{H_5},m_W)+\frac{\cb^2}{c_W^2}B_0(q^2;m_{H_5},m_Z) 
\\
&\qquad\qquad
+\sum_\varphi c_{\varphi WW}^2B_0(q^2;m_\varphi,m_W) -B_0(q^2;m_h,m_W)\Big], ~
\\
\Delta\Pi_{ZZ}^{\text{1PI}}(q^2)
=& 
\frac{g_Z^2}{64\pi^2}\Big\{5c_{2W}^2B_5(q^2;m_{H_5},m_{H_5}) + c_{2W}^2B_5(q^2;m_{H_3},m_{H_3}) 
\\
&\qquad\qquad 
+ 2\sb^2B_5(q^2;m_{H_5},m_{H_3})+2\cb^2B_5(q^2;m_{H_5},m_{G^\pm})
\\
&\qquad\qquad 
+\sum_\varphi \left[c_{\varphi H_3^0Z}^2B_5(q^2;m_{H_3},m_\varphi) + c_{\varphi ZZ}^2B_5(q^2;m_\varphi,m_{G^0})\right] 
\\
&\qquad\qquad
-B_5(q^2;m_h,m_{G^0}) \Big\}
\\
&
+\frac{g_Z^2m_Z^2}{16\pi^2}\Big[2\cb^2c_W^2B_0(q^2;m_{H_5},m_W) 
+\sum_\varphi c_{\varphi ZZ}^2B_0(q^2;m_\varphi,m_Z)
\\
& \qquad\qquad
-B_0(q^2;m_h,m_Z)\Big], 
\\
\Delta\Pi_{Z\gamma}^{\text{1PI}}(q^2) 
=& 
\frac{eg_Z}{32\pi^2}c_{2W}^{}\Big[5B_5(q^2;m_{H_5},m_{H_5})+B_5(q^2;m_{H_3},m_{H_3}) \Big], 
\\
\Delta\Pi_{\gamma\gamma}^{\text{1PI}}(q^2) 
=& 
\frac{e^2}{16\pi^2}\Big[5B_5(q^2;m_{H_5},m_{H_5})+B_5(q^2;m_{H_3},m_{H_3})\Big], 
\end{split}
\end{align}
where $B_5(q^2;m_1,m_2)\equiv  A(m_1) + A(m_2) - 4B_{22}(q^2;m_1,m_2)$.

\subsection{3-point functions \label{sec:3pf}}

For the 1PI diagram contributions to 3-point functions, we use a shorthand notation for the Passarino-Veltman's $C$ functions: 
\begin{align}
C_{i,ik}(A^{},B^{},C^{}) \equiv  C_{i,ik}(p_1^2,p_2^2,q^2;m_A^{},m_B^{},m_C^{}), 
\end{align}
where $p_1^\mu$ and $p_2^\mu$ are incoming 4-momenta of gauge bosons, fermions and (on-shell) Higgs bosons for the $hVV$, $hff$ and $hhh$ vertices, respectively, and $q^\mu$ is the outgoing momentum of the Higgs boson. 
For the $hVV$ and $hff$ vertices, we show the expressions corresponding to the first form factor defined in Eqs.~(\ref{form_factor}) and (\ref{form_factor2}), respectively. 
First, the 1PI diagram contributions to the $hVV$ vertices are 
\begin{align}
&\!\!\!\! \Gamma_{hWW}^{1,\text{1PI}}(p_1^2,p_2^2,q^2) \notag\\
=& 
-\frac{3g^2m_t^2}{16\pi^2 v}c_{hff}
\Big[4C_{24}(t,b,t)  -\frac{1}{2}B_0(p_2^2;m_t,m_b)-B_0(q^2;m_t,m_t)-\frac{1}{2}B_0(p_1^2;m_t,m_b) \notag\\
 & \qquad\qquad\qquad -\frac{1}{2}(2m_t^2 + 2m_b^2 -p_1^2 - p_2^2)C_0(t,b,t) \Big] + (m_t \leftrightarrow m_b)\notag\\
&+ \frac{g^3m_W}{16\pi^2}\Bigg\{c_{hWW}^{}[C_{hVV}^{VVV}(Z,W,Z)+c_W^2C_{hVV}^{VVV}(W,Z,W)+s_W^2C_{hVV}^{VVV}(W,\gamma,W)]\notag\\
&\quad -\frac{s_W^2}{2}c_{hWW}^{}[C_{hVV}^{SVV}(G^\pm ,Z,W)+C_{hVV}^{VVS}(W,Z,G^\pm)-C_{hVV}^{SVV}(G^\pm,\gamma,W)-C_{hVV}^{VVS}(W,\gamma,G^\pm )]\notag\\
&\quad-m_W^2c_{hWW}\Big[\sum_{\varphi} c_{\varphi VV}^2C_0(W,\varphi,W)  +  t_W^4C_0(Z,G^\pm,Z) \notag\\
& \quad\quad\quad\quad\quad\quad +\frac{c_\beta^2}{c_W^4}C_0(Z,H_5^\pm,Z)+2c_\beta^2C_0(W,H_5^{\pm\pm},W) \Big]\notag\\
&\quad +\frac{1}{2} \sum_{\varphi}\left[ c_{hWW}c_{\varphi WW}^2 \tilde{C}_{24}(G^\pm,\varphi,W)
 + c_{hH_3W}^{}c_{\varphi H_3W}^{}\,c_{\varphi WW}^{} \tilde{C}_{24}(H_3^\pm,\varphi,W) \right]\notag\\
&\quad  + \,c_\beta^2 c_{hWW}^{}\tilde{C}_{24}(G^\pm,H_5^{\pm\pm},W)
   + s_{\beta}c_{\beta}c_{hH_3V}^{} \tilde{C}_{24}(H_3^\pm,H_5^{\pm\pm},W)\notag\\
&\quad + \frac{t_W^2}{2}c_{hVV} \tilde{C}_{24}(G^0,G^\pm,Z)
  +\frac{c_\beta^2}{2c_W^2} c_{hVV}\tilde{C}_{24}(G^0,H_5^\pm,Z)
  +\frac{s_\beta c_\beta}{2c_W^2} c_{hH_3V}^{} \tilde{C}_{24}(H_3^0,H_5^\pm,Z)\notag\\
&\quad -c_{hVV}^{}[3B_0(q^2,m_W^{},m_W^{}) + 3B_0(q^2,m_Z^{},m_Z^{}) -4]\notag\\
&\quad -\frac{1}{4}\sum_{\varphi}(1+\delta_{h\varphi})c_{h\varphi WW}^{}c_{\varphi WW}^{}\tilde{B}_0(W,\varphi)-\frac{s_W^2}{2}c_{hWW} \left[t_W^2\tilde{B}_0(Z,G^\pm) + \tilde{B}_0(\gamma,G^\pm)\right] \notag\\
&\quad +\frac{\sqrt{3}}{2} c_\beta c_{hH_5WW}[\tilde{B}_0(W^{},H_5^{\pm\pm}) +\frac{1}{2c_W^2}\tilde{B}_0(Z,H_5^{\pm}) ] \Bigg\}\notag\\
&\quad+\frac{g^2m_W^2}{16\pi^2}\Bigg\{
6\lambda_{hhh}c_{hWW}^2C_0(h,W,h)
+2\lambda_{H_1H_1h}c_{H_1WW}^2C_0(H_1,W,H_1) \notag\\
& \quad\quad\quad\quad\quad\quad +2\lambda_{H_1hh}c_{hWW}^{}c_{H_1WW}^{}\tilde{C}_0(h,W,H_1)\notag\\
&\quad+2\lambda_{H_5H_5h} \left[c_{H_5^0WW}^2C_0(H_5^0,W,H_5^0)+2c_\beta^2C_0(H_5^{\pm\pm},W,H_5^{\pm\pm})+\frac{c_\beta^2}{c_W^2}  C_0(H_5^\pm,Z,H_5^\pm) \right]\notag\\
&\quad+2\lambda_{GGh} s_W^2  \left[C_0(G^\pm,\gamma,G^\pm)+t_W^2  C_0(G^\pm,Z,G^\pm) \right]\Bigg\}\notag
\end{align}
\begin{align}
&-\frac{g^2}{16\pi^2}\Bigg\{
2\lambda_{GGh}\Big[\sum_{\varphi}c_{\varphi WW}^2C_{24}(G^\pm,\varphi,G^\pm) + C_{24}(G^\pm,G^0,G^\pm) +2c_\beta^2C_{24}(G^\pm,H_5^{\pm\pm},G^\pm) \notag\\
&\quad\quad\quad\quad\quad\quad\quad                +C_{24}(G^0,G^\pm,G^0)+c_\beta^2C_{24}(G^0,H_5^\pm,G^0) \Big] \notag\\
&\quad + 2\lambda_{H_3H_3h}\Big[\sum_{\varphi}c_{\varphi H_3 W}^2C_{24}(H_3^\pm,\varphi,H_3^\pm) + 2C_{24}(H_3,H_3,H_3)+3s_\beta^2C_{24}(H_3,H_5,H_3)   \Big]\notag\\
&\quad - \lambda_{H_3Gh}\left[\sum_{\varphi}c_{\varphi WW}c_{\varphi H_3 W}\tilde{C}_{24}(G^\pm,\varphi,H_3^\pm) +s_{2\beta}\tilde{C}_{24}(G^\pm,H_5^{\pm\pm},H_3^\pm)
+ \frac{s_{2\beta}}{2}\tilde{C}_{24}(G^0,H_5^\pm,H_3^0)\right] \notag\\
&\quad + 2\lambda_{H_5H_5h}\Big[10C_{24}(H_5,H_5,H_5) + \frac{10s_\beta^2}{3} C_{24}(H_5,H_3,H_5)  \notag\\
&\quad\quad\quad\quad\quad                       + c_\beta^2 C_{24}(H_5,G^0,H_5) + \frac{7c_\beta^2}{3} C_{24}(H_5,G^\pm,H_5) \Big]\notag\\
&\quad+6\lambda_{hhh}\left[c_{hWW}^2C_{24}(h,G^\pm,h)+c_{hH_3W}^2C_{24}(h,H_3^\pm,h)\right]\notag\\
&\quad + 2\lambda_{H_1H_1h}\left[c_{H_1WW}^2C_{24}(H_1,G^\pm,H_1)+c_{H_1H_3W}^2C_{24}(H_1,H_3^\pm,H_1)\right]\notag\\
&\quad + 2\lambda_{H_1hh}\left[c_{hWW}^{}c_{H_1WW}^{}\tilde{C}_{24}(h,G^\pm,H_1)+c_{hH_3W}^{}c_{H_1H_3W}^{}\tilde{C}_{24}(h,H_3^\pm,H_1)\right] \Bigg\} \notag\\
&+\frac{g^2}{64\pi^2}\Bigg\{
\frac{80}{3}\lambda_{H_5H_5h}B_0(q^2;m_{H_5}^{},m_{H_5}^{})+(6+10s_\beta^2)\lambda_{H_3H_3h} B_0(q^2;m_{H_3}^{},m_{H_3}^{})\notag\\
&\quad+6\lambda_{hhh}\,c_{hhWW}^{}B_0(q^2;m_h,m_h)+2\lambda_{H_1H_1h}\,c_{H_1H_1WW}^{}B_0(q^2;m_{H_1}^{},m_{H_1}^{})\notag\\
&\quad+2\lambda_{H_1hh}\,c_{H_1hWW}^{}B_0(q^2;m_{H_1}^{},m_h)\notag\\
&\quad+2\lambda_{GGh}[2(2+c_{2\beta}) B_0(q^2;m_{G^\pm}^{},m_{G^\pm}^{}) + (1+c_\beta^2)B_0(q^2;m_{G^0}^{},m_{G^0}^{})]\notag\\
&\quad-\lambda_{H_3Gh} s_{2\beta}[4B_0(q^2;m_{H_3}^{},m_{G^\pm}^{})+ B_0(q^2;m_{H_3}^{},m_{G^0}^{})] \Bigg\}, 
\end{align}
\begin{align}
&\!\!\!\! \Gamma_{hZZ}^{1,\text{1PI}}(p_1^2,p_2^2,q^2)\notag\\
=&
-\sum_{f=t,b,\tau}\frac{4m_f^2g_Z^2N_c^f}{16\pi^2 v}
 \Bigg\{(v_f^2 - a_f^2)\Big[B_0(p_1^2;m_f,m_f)+B_0(p_2^2; m_f,m_f) +(4m_f^2-q^2)C_0(f,f,f)\Big] \notag\\
& \qquad\qquad\qquad\qquad\qquad -(v_f^2 + a_f^2)\Big[B_0(p_1^2;m_f,m_f)+B_0(p_2^2;m_f,m_f)+2B_0(q^2;m_f,m_f) \notag\\
& \qquad\qquad\qquad\qquad\qquad +(4m_f^2-p_1^2-p_2^2)C_0(f,f,f)-8C_{24}(f,f,f)\Big]\Bigg\}\notag\\
&+\frac{g^3m_W^{}}{16\pi^2}\Big\{2c_W^2c_{hVV}^{}C_{VVV}^{hVV}(W,W,W)+s_W^2c_{hVV}^{}\left[C_{hVV}^{SVV}(G^\pm,W,W)+C_{hVV}^{VVS}(W,W,G^\pm)\right]\notag\\
&\quad -m_W^2c_{hVV}^{}\left[2s_W^2t_W^2  C_0(W,G^\pm,W)+2\frac{c_\beta^2}{c_W^2}C_0(W,H_5^\pm,W)
+\frac{1}{c_W^6} \sum_\varphi  c_{\varphi ZZ}^2 C_0(Z,\varphi,Z)\right] \notag\\
&\quad -c_{2W}^{}t_W^2c_{hVV} \tilde{C}_{24}(W,G^\pm,G^\pm)
+\frac{c_\beta^2}{c_W^2}c_{hVV}\tilde{C}_{24}(W,H_5^\pm,G^\pm)+\frac{s_\beta c_\beta}{c_W^2}c_{hH_3V} \tilde{C}_{24}(W,H_5^\pm,H_3^\pm)\notag\\
&\quad +\frac{c_{hZZ}}{2c_W^4}\sum_\varphi c_{\varphi ZZ}^2\tilde{C}_{24}(Z,\varphi,G^0)
+\frac{c_{hH_3Z}}{2c_W^4}\sum_\varphi c_{\varphi H_3Z}c_{\varphi ZZ}\tilde{C}_{24}(Z,\varphi,H_3^0)\notag\\
&\quad -6c_W^2c_{hVV}\left[B_0(q^2,m_W^{},m_W^{})-\frac{2}{3}\right]\notag\\
&\quad -s_W^2t_W^2c_{hVV}^{}\tilde{B}_0(W,G^\pm)+\frac{2\sqrt{6}}{3c_W^2}c_\beta s_\alpha \tilde{B}_0(W,H_5^\pm)-\frac{1}{4c_W^4}\sum_\varphi (1+\delta_{h\varphi})c_{\varphi VV}c_{h\varphi VV}\tilde{B}_0(\varphi,Z)\Big\}\notag\\
&+\frac{g_Z^2m_Z^2}{16\pi^2} \Big\{6\lambda_{hhh}c_{hVV}^2C_0(h,Z,h)
+2\lambda_{H_1H_1h}c_{H_1VV}^2C_0(H_1,Z,H_1) \notag\\
&\quad +2\lambda_{H_1hh}c_{hVV}^{}c_{H_1VV}^{}\tilde{C}_0(h,Z,H_1)\notag\\
&\quad +2\lambda_{H_5H_5h}\left[c_{H_5ZZ}^2C_0(H_5^0,Z,H_5^0)+2c_W^2c_\beta^2 C_0(H_5^\pm,W,H_5^\pm)\right] \notag\\
&\quad +4\lambda_{GGh}s_W^4c_W^2C_0(G^\pm,W,G^\pm)  \Big\} \notag
\end{align}
\begin{align}
&-\frac{g_Z^2}{16\pi^2}\Bigg\{\sum_\varphi\Big[2\lambda_{GGh}  c_{\varphi ZZ}^2 C_{24}(G^0,\varphi,G^0)
 +2\lambda_{H_3H_3h}  c_{\varphi H_3Z}^2 C_{24}(H_3^0,\varphi,H_3^0) \notag\\
&\quad\quad\quad\quad -\lambda_{H_3Gh}  c_{\varphi ZZ}c_{\varphi H_3 Z} \tilde{C}_{24}(G^0,\varphi,H_3^0) \Big]\notag\\
&\quad+6\lambda_{hhh}c_{hVV}^2C_{24}(h,G^0,h)
 +2\lambda_{H_1H_1h}c_{H_1VV}^2C_{24}(H_1,G^0,H_1)\notag\\
&\quad +2\lambda_{H_1hh}c_{hVV}^{}c_{H_1VV}^{}\tilde{C}_{24}(H_1,G^0,h)\notag\\
&\quad+6\lambda_{hhh}c_{hH_3V}^2C_{24}(h,H_3^0,h)
 +2\lambda_{H_1H_1h}c_{H_1H_3V}^2C_{24}(H_1,H_3^0,H_1)\notag\\
&\quad +2\lambda_{H_1hh}c_{hH_3V}^{}c_{H_1H_3V}^{}\tilde{C}_{24}(H_1,H_3^0,h)\notag\\
&\quad +2\lambda_{H_5H_5h} \Big[\frac{10}{3}C_{24}(H_5,H_3,H_5) +10c_{2W}^2C_{24}(H_5,H_5,H_5)  \notag\\
&\quad\quad\quad\quad                      + c_{H_5ZZ}^2C_{24}(H_5,G^0,H_5)+2c_\beta^2C_{24}(H_5,G^\pm,H_5)\Big]\notag\\
&\quad +4\lambda_{H_3H_3h}[s_\beta^2C_{24}(H_3,H_5,H_3) + c_{2W}^2C_{24}(H_3,H_3,H_3)] -\lambda_{H_3Gh}s_{2\beta}\tilde{C}_{24}(G^\pm,H_5^\pm,H_3^\pm) \notag\\
&\quad +4\lambda_{GGh}[c_{2W}^2C_{24}(G^\pm,G^\pm,G^\pm)+c_\beta^2C_{24}(G^\pm,H_5^\pm,G^\pm)]\Bigg\} \notag\\
&+\frac{g_Z^2}{16\pi^2}\Bigg\{
5\lambda_{H_5H_5h}\left(\frac{1}{3} + c_{2W}^2 \right)B_0(q^2;m_{H_5}^{},m_{H_5}^{})  \notag\\
& \quad +\frac{\lambda_{H_3H_3h}}{2}\left(5s_\beta^2 + 2c_{2W}^2 +1\right)B_0(q^2;m_{H_3}^{},m_{H_3}^{}) \notag\\
&\quad +\lambda_{GGh}\left[(c_{2W}^2+c_\beta^2)B_0(q^2,G^\pm,G^\pm)+\frac{1+3c_\beta^2}{2}B_0(q^2,G^0,G^0) \right]\notag\\
&\quad-\frac{s_{2\beta}}{4} \lambda_{H_3 G h}\left[3B_0(q^2;G^0,H_3^0) + 2B_0(q^2;G^\pm,H_3^\pm) \right]\notag\\
&\quad+\frac{3}{2}\lambda_{hhh}c_{hhVV}B_0(q^2,m_h,m_h)
+\frac{1}{2}\lambda_{H_1H_1h}c_{H_1H_1VV}B_0(q^2,m_{H_1}^{},m_{H_1}^{}) \notag\\
&\quad+\frac{1}{2}\lambda_{H_1hh}c_{H_1hVV}B_0(q^2,m_{H_1}^{},m_h) \Bigg\}. 
\end{align}
In the above expressions, we have introduced 
\begin{align}
\begin{split}
\tilde{B}_{0}(A,B) &\equiv B_0(p_1^2;m_A^{},m_B^{}) +  B_0(p_2^2;m_A^{},m_B^{}), \\
\tilde{C}_{24}(A,B,C) &\equiv C_{24}(A,B,C) +C_{24}(C,B,A) , \\
\tilde{C}_{0}(A,B,C) &\equiv C_0(A,B,C) +C_0(C,B,A), 
\end{split}
\end{align}
and 
\begin{align}
\begin{split}
&C_{hVV}^{VVV}(X,Y,Z)\equiv \Big[17C_{24}+p_1^2(2C_{21}+3C_{11}+C_{0})+p_2^2(2C_{22}+C_{12}) \\
                        &\quad\quad\quad\quad\quad\quad\quad\quad +p_1\cdot p_2(4C_{23}+3C_{12}+C_{11}-4C_0)\Big]
(X,Y,Z)-3,\\
&C_{hVV}^{SVV}(X,Y,Z)\equiv \Big[3C_{24}+p_1^2(C_{21}-C_0)+p_2^2(C_{22}-2C_{12}+C_0)  \\
&\quad\quad\quad\quad\quad\quad\quad\quad +2p_1\cdot p_2 (C_{23}-C_{11}) \Big](X,Y,Z)-\frac{1}{2},\\
&C_{hVV}^{VVS}(X,Y,Z)\equiv \Big[3C_{24}+p_1^2(C_{21}+4C_{11}+4C_0)+p_2^2(C_{22}+2C_{12}) \\
&\quad\quad\quad\quad\quad\quad\quad\quad  +2p_1\cdot p_2 (C_{23}+2C_{12}+C_{11}+2C_0)\Big](X,Y,Z)-\frac{1}{2}.
\end{split}
\end{align}
The 1PI diagram contribution to the $hff$ vertex is calculated as 
\begin{align}
&\Gamma^{S,\text{1PI}}_{hff}(p_1^2,p_2^2,q^2) = \frac{m_f}{16\pi^2 v}\Bigg\{ -2g_Z^4v^2(v_f^2-a_f^2)c_{hVV}C_0(Z,f,Z)\notag\\
&\quad  -4c_{hff}\left[ e^2Q_f^2 C_{hff}^{FVF}(f,\gamma,f)+g_Z^2(v_f^2-a_f^2)C_{hff}^{FVF}(f,Z,f) \right]\notag\\
&\quad +c_{hff}
\frac{m_f^2}{v^2}\Big[
c_{hff}^2C_{hff}^{FSF}(f,h,f) + c_{H_1ff}^2C_{hff}^{FSF}(f,H_1,f) \notag\\
& \quad\quad\quad\quad\quad\quad -C_{hff}^{FSF}(f,G^0,f)-\cot^2\beta C_{hff}^{FSF}(f,H_3^0,f) \Big]  \notag\\
& \quad -c_{hff}\frac{2m_{f'}^2}{v^2}\Big[C_{hff}^{FSF}(f',G^\pm,f')+\cot^2\beta C_{hff}^{FSF}(f',H_3^\pm,f')\Big]\notag\\
& \quad -\frac{m_f^2}{v}\Big\{6c_{hff}^2\lambda_{hhh} C_0(h,f,h)+2c_{H_1ff}^2\lambda_{HHh}C_0(H_1,f,H_1) 
 +2c_{hff}c_{H_1ff}\lambda_{H_1hh}\tilde{C}_0(h,f,H_1)\notag\\
&\quad\quad\quad-2\lambda_{GGh}C_0(G^0,f,G^0) -2\cot^2\beta\lambda_{H_3H_3h}C_0(H_3^0,f,H_3^0)-\cot\beta\lambda_{H_3Gh}\tilde{C}_0(H_3^0,f,G^0)  \Big\}\notag\\
&\quad+\frac{2m_{f'}^2}{v}\Big\{
2\lambda_{GGh}C_0(G^\pm,f',G^\pm) +2\cot^2\beta \lambda_{H_3H_3h}C_0(H_3^\pm,f',H_3^\pm)\notag\\
&\quad\quad\quad\quad+\lambda_{H_3Gh}\cot\beta[C_0({G^\pm},{f'},{H_3^\pm})+C_0({H_3^\pm},{f'},{G^\pm})]\Big\}\notag\\
&\quad-\frac{g^2}{4}c_{hVV}\Big[
C_{hff}^{VFS}(W,{f'},{G^\pm})+C_{hff}^{SFV}(G^\pm,f',W)\Big]\notag\\
&\quad+\frac{g^2}{4}\cot\beta c_{hH_3V}\Big[C_{hff}^{VFS}(W,f',H_3^\pm)+C_{hff}^{SFV}(H_3^\pm,f',W)\Big]\notag\\
&\quad-\frac{g_Z^2}{8}c_{hVV}\Big[C_{hff}^{VFS}(Z,f,G^0)
+C_{hff}^{SFV}({G^0},f,Z) \Big]\notag\\
&\quad+\frac{g_Z^2}{8}\cot\beta c_{hH_3V}\Big[C_{hff}^{VFS}(Z,f,H_3^0)+C_{hff}^{SFV}(H_3^0,f,Z)\Big] \Bigg\},
\end{align}
where
\begin{align}
\begin{split}
C_{hff}^{FVF}(X,Y,Z) &\equiv \Big[m_f^2C_0+p_1^2(C_{11}+C_{21})+p_2^2(C_{12}+C_{22})\\
& \qquad                    +p_1\cdot p_2(2C_{23}-C_0)+4C_{24}-1\Big](X,Y,Z) \\
C_{hff}^{FSF}(X,Y,Z) &\equiv \Big[m_f^2C_0+p_1^2(C_{11}+C_{21})+p_2^2(C_{12}+C_{22}) \\
& \qquad       +2p_1\cdot p_2(C_{12}+C_{23})+4C_{24}\Big](X,Y,Z)-\frac{1}{2}, \\
C_{hff}^{VFS}(X,Y,Z) &\equiv \Big[p_1^2(2C_0+3C_{11}+C_{21})+p_2^2(2C_{12}+C_{22})  \\
& \qquad +2p_1\cdot p_2(2C_0+2C_{11}+C_{12}+C_{23})+4C_{24}\Big](X,Y,Z)-\frac{1}{2}, \\
C_{hff}^{SFV}(X,Y,Z) &\equiv \Big[p_1^2(C_{21}-C_0)+p_2^2(C_{22}-C_{12})  \\
& \qquad +2p_1\cdot p_2(C_{23}-C_{12})+4C_{24}\Big](X,Y,Z)-\frac{1}{2}. \\
\end{split}
\end{align}
The 1PI diagram contribution to the $hhh$ vertex is given by
\begin{align}
&\!\!\!\!
\Gamma_{hhh}^{\text{1PI}}(p_1^2,p_2^2,q^2) \notag\\
=& 
-\sum_{f=t,b,\tau}\frac{8m_f^4N_c^f}{16\pi^2 v^3}
\Big[\bar{B}_0(f,f) +(4m_f^2-q^2+p_1 \cdot p_2)C_0(f,f,f)\Big] \notag\\
&+\frac{1}{16\pi^2}\Bigg\{
g^3m_W^3c_{hVV}^3\left[ \frac{15}{2}C_0(W,W,W) + \frac{15}{4c_W^6}C_0(Z,Z,Z)\right] \notag\\
&\quad -\frac{g^3m_W}{2}c_{hVV}^3\left[C_{SVV}^{hhh}(G^\pm,W,W)+\frac{1}{2c_W^4}C_{SVV}^{hhh}(G^0,Z,Z)\right] \notag\\
&\quad-\frac{g^3m_W}{2}c_{hVV}c_{hH_3V}^2\left[C_{SVV}^{hhh}(H_3^\pm,W,W)+\frac{1}{2c_W^4}C_{SVV}^{hhh}(H_3^0,Z,Z)\right] \notag\\
&\quad+ g^2\lambda_{GGh}c_{hVV}^2 \left[C_{VSS}^{hhh}(W,G^\pm,G^\pm)+\frac{1}{2c_W^2}C_{VSS}^{hhh}(Z,G^0,G^0)\right] \notag\\
&\quad+g^2\lambda_{H_3H_3h}c_{hH_3V}^2 \left[C_{VSS}^{hhh}(W,H_3^\pm,H_3^\pm)+\frac{1}{2c_W^2}C_{VSS}^{hhh}(Z,H_3^0,H_3^0) \right]\notag\\
&\quad-g^2\lambda_{H_3Gh}c_{hVV}c_{hH_3V} \left[C_{VSS}^{hhh}(W,H_3^\pm,G^\pm)+\frac{1}{2c_W^2}C_{VSS}^{hhh}(Z,H_3^0,G^0) \right]\notag\\
&\quad-40\lambda_{H_5H_5h}^3 C_0(H_5,H_5,H_5) -24\lambda_{H_3H_3h}^3C_0(H_3,H_3,H_3) \notag\\
&\quad -8\lambda_{GGh}^3\left[2C_0(G^\pm,G^\pm,G^\pm)+C_0(G^0,G^0,G^0)\right]\notag\\
&\quad-2\lambda_{H_3H_3h}\lambda_{H_3Gh}^2[2\bar{C}_0(H_3^\pm,H_3^\pm,G^\pm)+\bar{C}_0(H_3^0,H_3^0,G^0)]\notag\\
&\quad-2\lambda_{GGh}\lambda_{H_3Gh}^2[2\bar{C}_0(G^\pm,G^\pm,H_3^\pm)+\bar{C}_0(G^0,G^0,H_3^0)]\notag\\
&\quad-8\lambda_{H_1H_1h}^3C_0(H_1,H_1,H_1)-216\lambda_{hhh}^3C_0(h,h,h) \notag\\
&\quad -24\lambda_{H_1hh}^2\lambda_{hhh}\bar{C}_0(H_1,h,h)-8\lambda_{H_1hh}^2\lambda_{H_1H_1h}\bar{C}_0(H_1,H_1,h)\notag\\
&\quad+c_{hVV}c_{hhVV}\left[2g^3m_W\bar{B}_0(W,W) +g_Z^3m_Z\bar{B}_0(Z,Z) -\left(3g^3m_W+\frac{3}{2}g_Z^3m_Z\right)\right]\notag\\
&\quad+4\lambda_{H_1H_1h}\lambda_{H_1H_1hh}\bar{B}_0(H_1,H_1)
+12\lambda_{H_1hh}\lambda_{H_1hhh}\bar{B}_0(H_1,h)
+72\lambda_{hhh}\lambda_{hhhh}\bar{B}_0(h,h) \notag\\
&\quad+ 10\lambda_{H_5H_5h}\lambda_{H_5H_5hh}\bar{B}_0(H_5,H_5)+6\lambda_{H_3H_3h}\lambda_{H_3H_3hh}\bar{B}_0(H_3,H_3) \notag\\
&\quad  +2\lambda_{GGh}\lambda_{GGhh}[2\bar{B}_0(G^\pm,G^\pm) + \bar{B}_0(G^0,G^0)] \notag\\
&\quad+2\lambda_{H_3Gh}\lambda_{H_3Ghh}[2\bar{B}_0(H_3^{\pm},G^{\pm})+\bar{B}_0(H_3^{0},G^{0})] \Bigg\},   \label{1pi_hhh}
\end{align}
where
\begin{align}
\begin{split}
C_{hhh}^{SVV}(A,B,C)
\equiv&  \Big[p_1^2(C_{21}-2C_{11}+C_0)+p_2^2(C_{22}-C_{12})  \\
& \qquad +p_1\cdot p_2(2C_{23}-C_{11}-2C_{12}+C_0)+4C_{24}-\frac{1}{2}\Big](A,B,C)\\
&+\Big[p_1^2(C_{21}+3C_{11}+2C_0) +p_2^2(C_{22}-C_{12})  \\
& \qquad +p_1\cdot p_2(2C_{23}+3C_{12}-C_{11}-2C_0)+4C_{24}-\frac{1}{2} \Big](C,A,B)\\
&+\Big[p_1^2(C_{21}+3C_{11}+2C_0)+p_2^2(C_{22}+4C_{12}+4C_0) \\
& \qquad +p_1\cdot p_2(2C_{23}+3C_{12}+4C_{11}+6C_0)+4C_{24} \\
& \qquad -\frac{1}{2}C_0\Big](B,C,A),\\
C_{hhh}^{VSS}(A,B,C) 
\equiv& \Big[p_1^2(C_{21}+4C_{11}+4C_0)+p_2^2(C_{22}+2C_{12}) \\&\qquad +p_1\cdot p_2(2C_{23}+4C_{12}+2C_{11}+4C_0 ) +4C_{24}-\frac{1}{2}\Big](A,B,C)\\
&+\Big[p_1^2(C_{21}-C_0)+p_2^2(C_{22}+2C_{12})\\&\qquad+p_1\cdot p_2(2C_{23}+2C_{11}-2C_0)+4C_{24}-\frac{1}{2}\Big](C,A,B)\\
&+\Big[p_1^2(C_{21}-C_0) + p_2^2(C_{22}-2C_{12}+C_0)\\&\qquad+ p_1\cdot p_2 (2C_{23}-2C_{11})+4C_{24}-\frac{1}{2}\Big](B,C,A), \\
\bar{C}_0(A,B,C)    
\equiv&  C_0(A,B,C)+ C_0(C,A,B)+ C_0(B,C,A), \\ 
\bar{B}_0(A,B)      
\equiv&  B_0(p_1^2,m_A,m_B)+B_0(p_2^2,m_A,m_B)+B_0(q^2,m_A,m_B). 
\end{split}
\end{align}

\end{appendix}


\end{document}